\documentclass{myelsart}

\journal{Information and Computation}

\mathchardef\Omega="700A

\usepackage{ifthen}

\newboolean{TR}
\setboolean{TR}{true}
\usepackage{amsmath}
\usepackage{amssymb}
\usepackage{stmaryrd}
\usepackage{url}
\usepackage{hhline}
\usepackage{varioref}

\allowdisplaybreaks

\newenvironment{program}
               {
                 \ttfamily
                 \upshape
                 \begin{tabular}{l}
               }
               {
                 \end{tabular}
               }
\newenvironment{centeredprogram}
               {
                 \begin{center}
                   \begin{program}
               }
               {
                   \end{program}
                 \end{center}
               }

\newcommand{\summary}[1]{\textrm{\textbf{\textup{#1}}}}

\providecommand*{\Nset}{\mathbb{N}}            % Naturals

\newcommand*{\cD}{\ensuremath{\mathcal{D}}}

\newcommand*{\cR}{\ensuremath{\mathcal{R}}}
\newcommand*{\cS}{\ensuremath{\mathcal{S}}}

\renewcommand{\emptyset}{\mathord{\varnothing}}
\newcommand{\card}{\mathop{\#}\nolimits}

\newcommand*{\wpf}{\mathop{\wp_\mathrm{f}}\nolimits}

\newcommand*{\sseq}{\subseteq}

\newcommand*{\sslt}{\subset}
\newcommand*{\Sseq}{\supseteq}

\newcommand*{\union}{\cup}
\newcommand*{\bigunion}{\bigcup}
\newcommand*{\inters}{\cap}
\newcommand*{\biginters}{\bigcap}
\newcommand*{\setdiff}{\setminus}

\newcommand{\sset}[2]{{\renewcommand{\arraystretch}{1.2}
                      \left\{\,#1 \,\left|\,
                               \begin{array}{@{}l@{}}#2\end{array}
                      \right.   \,\right\}}}

\newcommand*{\reld}[3]{\mathord{#1}\subseteq#2\times#3}
\newcommand*{\fund}[3]{\mathord{#1}\colon#2\rightarrow#3}

\newcommand{\st}{\mathrel{.}}
\newcommand{\itc}{\mathrel{:}}

\newcommand{\lambdadot}{\mathbin{.}}

\newcommand*{\Vars}{\mathord{\mathrm{Vars}}}
\newcommand*{\Sig}{\mathord{\mathrm{Sig}}}

\newcommand*{\Terms}{\mathord{\mathrm{Terms}}}
\newcommand*{\GTerms}{\mathord{\mathrm{GTerms}}}
\newcommand*{\LTerms}{\mathord{\mathrm{LTerms}}}
\newcommand*{\HTerms}{\mathord{\mathrm{HTerms}}}

\newcommand*{\vars}{\mathop{\mathrm{vars}}\nolimits}
\newcommand*{\mvars}{\mathop{\mathrm{mvars}}\nolimits}
\newcommand*{\size}{\mathop{\mathrm{size}}\nolimits}

\newcommand*{\HT}{\ensuremath{\mathcal{FT}}}
\newcommand*{\RT}{\ensuremath{\mathcal{RT}}}

\newcommand*{\entails}{\mathrel{\vdash}}
\newcommand*{\bigland}{\mathop{\bigwedge}\limits}

\newcommand*{\piff}{\mathrel{\leftrightarrow}}
\newcommand*{\pimplies}{\mathrel{\rightarrow}}

\newcommand{\defrel}[1]{\mathrel{\buildrel \mathrm{def} \over {#1}}}
\newcommand{\defeq}{\defrel{=}}
\newcommand{\defiff}{\defrel{\Longleftrightarrow}}
\newcommand*{\dom}{\mathop{\mathrm{dom}}\nolimits}

\newcommand*{\compose}{\mathbin{\circ}}

\newcommand*{\Bind}{\mathrm{Bind}}
\newcommand*{\RSubst}{\mathrm{RSubst}}
\newcommand*{\VSubst}{\mathrm{VSubst}}
\newcommand*{\ISubst}{\mathrm{ISubst}}

\newcommand{\Sstep}{\mathrel{\buildrel{\cS}\over\longmapsto}}
\newcommand{\Sstepstar}{\mathrel{\buildrel{\mkern-10mu\cS}\over{\longmapsto^{\smash{\mkern-2.2mu\ast}}}}}

\newcommand{\rt}{\mathop{\mathrm{rt}}\nolimits}

\newcommand*{\Eqs}{\mathrm{Eqs}}
\newcommand*{\mgs}{\mathop{\mathrm{mgs}}\nolimits}

\newcommand*{\domainD}{{\cD^\flat}}
\newcommand*{\projectD}{\mathop{\exists\kern-.4em\exists}\nolimits}

\newcommand*{\amgu}{\mathop{\mathrm{amgu}}\nolimits}
\newcommand*{\proj}{\mathop{\mathrm{proj}}\nolimits}

\newcommand*{\VI}{\mathrm{VI}}
\newcommand*{\SG}{\mathrm{SG}}
\newcommand*{\SH}{\mathrm{SH}}
\newcommand*{\sh}{\mathrm{sh}}
\newcommand{\sg}{\mathop{\mathrm{sg}}\nolimits}
\newcommand{\occ}{\mathop{\mathrm{occ}}\nolimits}

\newcommand*{\irel}{\mathop{\overline{\mathrm{rel}}}\nolimits}
\newcommand*{\rel}{\mathop{\mathrm{rel}}\nolimits}
\newcommand*{\bin}{\mathop{\mathrm{bin}}\nolimits}
\newcommand*{\cyclicreduce}{\mathop{\mathrm{cyclic}}\nolimits}
\newcommand*{\projSH}{\proj_\SH}

\newcommand*{\SFL}{\mathrm{SFL}}
\newcommand*{\leqSFL}{\mathrel{\leq_S}}
\newcommand*{\botSFL}{\bot_S}
\newcommand{\gvars}{\mathop{\mathrm{gvars}}\nolimits}

\newcommand{\nonlinvars}{\mathop{\mathrm{nlvars}}\nolimits}

\newcommand*{\sfl}{\mathrm{d}}
\newcommand*{\amguSFL}{\amgu_S}
\newcommand*{\projSFL}{\proj_S}
\newcommand{\hvars}{\mathop{\mathrm{hvars}}\nolimits}
\newcommand*{\abstrH}{\alpha_H}
\newcommand*{\concrH}{\gamma_H}
\newcommand*{\amguH}{\amgu_H}
\newcommand*{\projH}{\proj_H}
\newcommand*{\hterm}{\mathrm{hterm}}

\newcommand*{\concrP}{\gamma_P}
\newcommand*{\amguP}{\amgu_P}
\newcommand*{\projP}{\proj_P}
\newcommand*{\ground}{\mathrm{ground}}
\newcommand*{\free}{\mathrm{free}}
\newcommand*{\gfree}{\mathrm{gfree}}
\newcommand*{\lin}{\mathrm{lin}}
\newcommand*{\orlin}{\mathrm{or\_lin}}
\newcommand*{\ind}{\mathrm{ind}}
\newcommand*{\sharelin}{\mathrm{share\_lin}}
\newcommand*{\occlin}{\mathrm{occ\_lin}}
\newcommand*{\sharewith}{\mathrm{share\_with}}
\newcommand*{\sharesamevar}{\mathrm{share\_same\_var}}

\newcommand*{\PSD}{\mathrm{PSD}}
\newcommand*{\PSDFL}{\SFL_2}
\newcommand*{\ASub}{\mathsf{ASub}}
\newcommand*{\ScozzariShPSh}{\mathsf{Sh}^\mathsf{PSh}}

\newcommand*{\Pattern}{\mathop{\mathrm{Pattern}}\nolimits}

\newcommand*{\Bool}{\mathrm{Bool}}
\newcommand*{\Bval}{\mathord{\mathrm{Bval}}}
\newcommand*{\Bvalzero}{\mathbf{0}}
\newcommand*{\Bvalone}{\mathbf{1}}

\newcommand*{\Bfun}{\mathord{\mathrm{Bfun}}}
\newcommand*{\Pos}{\mathord{\mathrm{Pos}}}
\newcommand*{\Def}{\mathord{\mathrm{Def}}}

\newcommand{\pos}{\mathop{\mathrm{pos}}\nolimits}

\newcommand{\truev}{\mathop{\mathrm{true}}\nolimits}
\newcommand{\falsev}{\mathop{\mathrm{false}}\nolimits}

\newcommand*{\coanote}[1]{}

\newcommand*{\FD}{F}
\newcommand*{\hval}{\mathop{\mathrm{hval}}\nolimits}

\newcommand*{\concrFD}{\gamma_\FD}
\newcommand*{\amguFD}{\amgu_\FD}

\newcommand*{\GD}{G}
\newcommand*{\gval}{\mathop{\mathrm{gval}}\nolimits}

\newcommand*{\concrGD}{\gamma_\GD}
\newcommand*{\amguGD}{\amgu_\GD}

\newtheorem{theorem}{Theorem}
\newtheorem{definition}[theorem]{Definition}
\newtheorem{example}[theorem]{Example}
\newtheorem{lemma}[theorem]{Lemma}
\newtheorem{proposition}[theorem]{Proposition}
\newtheorem{corollary}[theorem]{Corollary}

\newcommand*{\down}{\mathop{\downarrow}\nolimits}

\newcommand{\china}{\textmd{\textsc{China}}}

\newcommand*{\CLPR}{CLP($\cR$)}

\newcommand*{\concat}{\mathbin{.}}

\newcommand*{\checked}[1]{}

\newcommand*{\law}[1]{\text{[#1]}}

\begin{document}
\begin{frontmatter}

\ifthenelse{\boolean{TR}}{%%
\title{%%
  Finite-Tree Analysis\\
  for Constraint Logic-Based Languages:\\
  The Complete Unabridged Version\thanksref{th}
}
}{%%\ifthenelse{\boolean{TR}}{%%
\title{%%
  Finite-Tree Analysis\\
  for Constraint Logic-Based Languages\thanksref{th}
}
}%%\ifthenelse{\boolean{TR}}{%%

\thanks[th]{This work has been partly supported by MURST projects
``Automatic Program Certification by Abstract Interpretation'',
``Abstract Interpretation, Type Systems and Control-Flow Analysis'', and
``Constraint Based Verification of Reactive Systems.''
Some of this work was done during visits of the fourth
author to Leeds, funded by EPSRC under grant M05645.}

\author[Parma]{Roberto Bagnara},
\ead{bagnara@cs.unipr.it}
\author[Pisa]{Roberta Gori},
\ead{gori@di.unipi.it}
\author[Leeds]{Patricia M. Hill},
\ead{hill@comp.leeds.ac.uk}
\author[Parma]{Enea Zaffanella}
\ead{zaffanella@cs.unipr.it}
                                                                                
\address[Parma]{Department of Mathematics, University of Parma, Italy}
\address[Pisa]{Department of Computer Science, University of Pisa, Italy}
\address[Leeds]{School of Computing, University of Leeds, UK}

\begin{abstract}
Logic languages based on the theory of rational, possibly
infinite, trees have much appeal in that rational trees allow for
faster unification (due to the safe omission of the occurs-check) and
increased expressivity (cyclic terms can provide very efficient
representations of grammars and other useful objects).
Unfortunately, the use of infinite rational trees has problems.
For instance, many of the built-in and library predicates
are ill-defined for such trees and need to be supplemented
by run-time checks whose cost may be significant.
Moreover, some widely-used program analysis and manipulation techniques are
correct only for those parts of programs working over finite trees.
It is thus important to obtain, automatically, a knowledge of the
program variables (the \emph{finite variables}) that, at the program
points of interest, will always be bound to finite terms.
For these reasons, we propose here a new data-flow analysis,
based on abstract interpretation, that captures such information.
We present a parametric domain where a simple component for recording
finite variables is coupled, in the style of the \emph{open product}
construction of Cortesi et al., with a generic domain (the parameter of
the construction) providing sharing information.
The sharing domain is abstractly specified so as to guarantee the
correctness of the combined domain and the generality of the
approach.
This finite-tree analysis domain is further enhanced
by coupling it with a domain of Boolean functions,
called \emph{finite-tree dependencies},
that precisely captures how the finiteness of some
variables influences the finiteness of other variables.
We also summarize our experimental results showing how finite-tree analysis,
enhanced with finite-tree dependencies, is a practical means of obtaining
precise finiteness information.
\end{abstract}


\end{frontmatter}

\section{Introduction}
\label{sec:introduction}

The intended computation domain
of most logic-based languages\footnote{That is,
ordinary logic languages, (concurrent) constraint logic languages,
functional logic languages and variations of the above.}
includes the algebra (or structure) of \emph{finite trees}.
Other (constraint) logic-based languages,
such as Prolog~II and its successors \cite{Colmerauer82,Colmerauer90},
SICStus Prolog \cite{SICStusManual-3_9},
and Oz \cite{SmolkaT94},
refer to a computation domain of \emph{rational trees}.\footnote{Support
for rational trees is also provided as an option
by the YAP Prolog system \cite{YAPManual-4_3_20}.}
A rational tree is a possibly infinite tree with a finite number
of distinct subtrees and
%, as is the case for finite trees,
where each node has a finite number of immediate descendants.
These properties ensure that rational trees,
even though infinite in the sense that they admit paths
of infinite length, can be finitely represented.
One possible representation makes use of connected, rooted, directed
and possibly cyclic graphs
where nodes are labeled with variable and function symbols
as is the case of finite trees.

Applications of rational trees in logic programming include
graphics \cite{EggertC83},
parser generation and grammar manipulation \cite{Colmerauer82,GiannesiniC84},
and computing with finite-state automata \cite{Colmerauer82}.
Rational trees also constitute the basis of the abstract domain of
\emph{rigid type graphs}, which is used for type analysis of logic programs
\cite{CousotC95a,JanssensB92,VanHentenryckCLC95a}.
Other applications are described in \cite{Filgueiras84}
and \cite{HaridiS84}.
Very recently, Manuel Carro has described a nice application of
rational trees where they are used to represent imperative programs
within interpreters.  Taking a continuation-passing style approach,
each instruction is coupled with a data structure representing the
remaining part of the program to be executed so that sequences of
instructions for realizing (backward) jumps, iterations and recursive
calls give rise to cyclic structures in the form of rational trees.
Compared to a naive interpreter for the same language, this threaded
interpreter is faster and uses less memory, at the cost of a simple
preliminary ``compilation pass'' to generate the rational tree
representation for the program \cite{Carro04}.

Going from Prolog to CLP, in \cite{Mukai91th} K.~Mukai
has combined constraints on rational trees and record structures,
while the logic-based language
\emph{Oz} allows constraints over rational and feature trees
\cite{SmolkaT94}.
The expressive power of rational trees is put to use, for instance,
in several areas of natural language processing.
Rational trees are used in implementations of
the HPSG formalism (Head-driven Phrase Structure Grammar) \cite{PollardS94},
in the ALE system (Attribute Logic Engine) \cite{Carpenter92},
and in the ProFIT system (Prolog with Features, Inheritance and Templates)
\cite{Erbach95}.

While rational trees allow for increased expressivity,
they also come equipped with a surprising number of problems.
As we will see, some of these problems are so serious
that rational trees must be used in a very controlled way,
disallowing them in any context where they are ``dangerous.''
This, in turn, causes a secondary problem: in order to disallow
rational trees in selected contexts one must first detect them,
an operation that may be expensive.

The first thing to be aware of is that almost any
semantics-based program manipulation technique developed
in the field of logic programming ---whether it be an analysis,
a transformation, or an optimization--- assumes a computation
domain of \emph{finite trees}.
Some  of these techniques might work with rational trees
but their correctness has only been proved in the case of
finite trees.
Others are clearly inapplicable.
Let us consider a very simple Prolog program:
\begin{centeredprogram}
  list([]). \\
  list([\_|T]) :- list(T).
\end{centeredprogram}
Most automatic and semi-automatic tools for proving program
termination\footnote{Such as TerminWeb \cite{CodishT97,CodishT99},
TermiLog \cite{LindenstraussSS97}, cTI \cite{MesnardR05TPLP},
and LPTP \cite{Stark96,Stark98}.}
and for complexity analysis\footnote{Systems like
GAIA \cite{CortesiLCR97},
CASLOG \cite{DebrayL93},
and the Ciao-Prolog preprocessor \cite{HermenegildoBPL99}.}
agree on the fact that \verb+list/1+ will terminate
when invoked with a ground argument.
Consider now the query
\begin{centeredprogram}
  ?- X = [a|X], list(X).
\end{centeredprogram}
and note that, after the execution of the first rational unification,
the variable \verb+X+ will be bound to a rational term
containing no variables, i.e., the predicate \verb+list/1+
will be invoked with \verb+X+ ground.
However, if such a query is given to, say, SICStus Prolog, then
the only way to get the prompt back is by interrupting the program.
The problem stems from the fact that the analysis techniques
employed by these tools are only sound for finite trees:
as soon as they are applied to a system where the creation of cyclic
terms is possible, their results are inapplicable.
The situation can be improved by combining these termination
and/or complexity analyses with a finiteness analysis providing
the precondition for the applicability of the other techniques.

The implementation of built-in predicates is another problematic issue.
Indeed, it is widely acknowledged that, for the implementation
of a system that provides real support for rational trees,
the biggest effort concerns proper handling of built-ins.
Of course, the meaning of `proper' depends on the actual built-in.
Built-ins such as \verb+copy_term/2+ and \verb+==/2+
maintain a clear semantics when passing from finite to rational trees.
For others, like \verb+sort/2+, the extension can be
questionable:\footnote{Even though \texttt{sort/2} is not required to
be a built-in by the ISO Prolog standard, it is offered as such by several
implementations.}  failing, raising an exception,
answering \verb+Y = [a]+ (if duplicates are deleted)
and answering \verb+Y = [a|Y]+ (if duplicates are kept)
can all be argued to be ``the right reaction'' to the query
\begin{centeredprogram}
  ?- X = [a|X], sort(X, Y).
\end{centeredprogram}
Other built-ins do not tolerate infinite trees in some argument
positions.
A good implementation should check for finiteness of the
corresponding arguments and make sure ``the right thing''
---failing or raising an appropriate exception---
always happens.
However, such behavior appears to be uncommon.
A small experiment we conducted on six Prolog implementations with queries
like
\begin{centeredprogram}
  ?- X = 1+X, Y is X. \\
  ?- X = [97|X], name(Y, X). \\
  ?- X = [X|X], Y =.. [f|X]. \\
\end{centeredprogram}
resulted in
infinite loops,
memory exhaustion and/or system thrashing,
segmentation faults or other fatal errors.
One of the implementations tested, SICStus Prolog, is a professional one and
implements run-time checks to avoid most cases where built-ins
can have catastrophic effects.\footnote{SICStus 3.11
still loops on \texttt{?- X = [97|X], name(Y, X).}}
The remaining systems are a bit more than research prototypes,
but will clearly have to do the same if they evolve to the stage
of production tools.
Again, a data-flow analysis aimed at the detection of those variables
that are definitely bound to finite terms could be used to avoid
a (possibly significant) fraction of the useless run-time checks.
Note that what has been said for built-in predicates applies
to libraries as well.
Even though it may be argued that it is enough for programmers to know
that they should not use a particular library predicate with infinite
terms, it is clear that the use of a ``safe'' library,
including automatic checks ensuring that such a predicate
is never called with an illegal argument,
will result in a robuster system.
With the appropriate data-flow analyses,
safe libraries do not have to be inefficient libraries.

Another serious problem is the following:
the standard term ordering dictated by ISO Prolog
\cite{ISO-Prolog-part-1} cannot be extended to rational trees
[M.~Carlsson, Personal communication, October 2000].
Consider the rational trees defined by
$\texttt{A = f(B, a)}$ and $\texttt{B = f(A, b)}$.
Clearly, $\texttt{A == B}$ does not hold.
Since the standard term ordering is total, we must have either
$\texttt{A @< B}$ or $\texttt{B @< A}$.
Assume $\texttt{A @< B}$.
Then $\texttt{f(A, b) @< f(B, a)}$, since the ordering of terms having the
same principal functor is inherited by the ordering of subterms
considered in a left-to-right fashion.
Thus $\texttt{B @< A}$ must hold, which is a contradiction.
A dual contradiction is obtained by assuming $\texttt{B @< A}$.
As a consequence, applying any Prolog term-ordering predicate
to terms where one or both of them is infinite may cause inconsistent results,
giving rise to bugs that are exceptionally difficult to diagnose.
For this reason, any system that extends ISO Prolog
with rational trees ought to detect such situations
and make sure they are not ignored
(e.g., by throwing an exception
or aborting execution with a meaningful message).
However, predicates such as the term-ordering ones
are likely to be called a significant number of times,
since they are often used to maintain structures implementing
ordered collections of terms.
This is another instance of the efficiency issue mentioned above.

Still on efficiency, it is worth noting that even for built-ins whose
definition on rational trees is not problematic, there is often a
performance penalty in catering for the possibility of infinite
trees. Thus, for such predicates, which include rational unification
provided by \verb+=/2+, a compile-time knowledge of term finiteness
can be beneficial.
For instance, rational-tree implementations of the built-ins
\verb+ground/1+, \verb+term_variables/2+, \verb+copy_term/2+,
\verb+subsumes/2+, \verb+variant/2+ and \verb+numbervars/3+ need
more expensive marking techniques to ensure they do not enter an
infinite loop.  With finiteness information it is possible to avoid
this overhead.

In this paper,
we present a parametric abstract domain for finite-tree analysis,
denoted by $H \times P$.
This domain combines a simple component $H$
(written with the initial of \emph{Herbrand}
and called the \emph{finiteness} component)
recording the set of definitely finite variables,
with a generic domain $P$
(the parameter of the construction)
providing sharing information.
The term ``sharing information'' is to be understood
in its broader meaning, which includes variable aliasing, groundness,
linearity, freeness and any other kind of information that can improve
the precision on these components, such as explicit structural information.
Several domain combinations and abstract operators,
characterized by different precision/complexity trade-offs,
have been proposed to capture these properties
(see~\cite{BagnaraZH00,BagnaraZH05TPLP} for an account of some of them).
By giving a generic specification for this parameter component,
in the style of the \emph{open product} construct
proposed in~\cite{CortesiLCVH00},
it is possible to define and establish the correctness
of abstract operators on the finite-tree domain
independently from any particular domain for sharing analysis.

The information encoded by $H$
is \emph{attribute independent} \cite{CousotC92LP},
which means that each variable is considered in isolation.
What this lacks is information about how finiteness of one variable
affects the finiteness of other variables.
This kind of information, usually called \emph{relational information},
is not captured at all by $H$ and is only partially captured
by the composite domain $H \times P$.
Moreover, $H \times P$ is designed to capture the ``negative'' aspect
of term-finiteness, that is, the circumstances under which finiteness
can be lost.
However, term-finiteness has also a ``positive'' aspect:
there are cases where a variable is granted to be bound to a finite term and
this knowledge can be propagated to other variables.
Guarantees of finiteness are provided by several built-ins like
\verb+unify_with_occurs_check/2+, \verb+var/1+, \verb+name/2+,
all the arithmetic predicates, besides those explicitly
provided to test for term-finiteness
such as the \verb+acyclic_term/1+ predicate
of SICStus Prolog.
For these reasons $H \times P$ is coupled with a domain of Boolean functions
that precisely captures how the finiteness of some variables
influences the finiteness of other variables.
This domain of \emph{finite-tree dependencies} provides relational
information that is important for the precision of the overall
finite-tree analysis.
It also combines obvious similarities, interesting differences
and somewhat unexpected connections with classical domains for
\emph{groundness dependencies}.
Finite-tree and groundness dependencies are similar in that they both
track \emph{covering} information
(a term $s$ covers $t$ if all the variables in $t$ also occur in~$s$)
and share several abstract operations.
However, they are different because covering does not tell the whole story.
Suppose $x$ and $y$ are free variables before either the unification
$x = f(y)$ or the unification $x = f(x, y)$ are executed.
In both cases, $x$ will be ground if and only if $y$ will be so.
However, when $x = f(y)$ is the performed unification,
this equivalence will also carry over to finiteness.
In contrast, when the unification is $x = f(x, y)$,
$x$ will never be finite and will be totally independent,
as far as finiteness is concerned, from $y$.
Among the unexpected connections is the fact that
finite-tree dependencies can improve the groundness information
obtained by the usual approaches to groundness analysis.

The paper is structured as follows.
The required notations and preliminary concepts are given
in Section~\ref{sec:preliminaries}.
The concrete domain for the analysis is presented
in Section~\ref{sec:concrete-domain}.
The finite-tree domain is then introduced
in Section~\ref{sec:finite-tree-domain}:
Section~\ref{subsec:P} provides the specification
of the parameter domain $P$;
Section~\ref{subsec:operators-on-RSubst} defines some computable operators
that extract, from substitutions in rational solved form, properties of
the denoted rational trees;
Section~\ref{subsec:abstrH} defines the abstraction function
for the finiteness component $H$;
Section~\ref{subsec:amguH} defines the abstract unification
operator for $H \times P$.
Section~\ref{sec:finite-tree-dependencies}
introduces the use of Boolean functions for tracking
finite-tree dependencies,
whereas Section~\ref{sec:groundness-dependencies}
illustrates the interaction between groundness
and finite-tree dependencies.
Our experimental results are presented in Section~\ref{sec:exp-results}.
We conclude the main body of the paper in Section~\ref{sec:conclusion}.
\ifthenelse{\boolean{TR}}{%%

Appendix~\ref{sec:implementing-P} specifies the sharing domain
$\SFL$ defined in~\cite{HillZB04TPLP,Zaffanella01th} as
a possible instance of the parameter $P$.
All the results are then proved in Appendix~\ref{sec:proofs}.

This paper is a combined and improved version of
\cite{BagnaraGHZ01} and \cite{BagnaraZGH01}.
}{%%\ifthenelse{\boolean{TR}}{%%
Appendix~\ref{sec:implementing-P} specifies the sharing domain
$\SFL$ defined in~\cite{HillZB04TPLP,Zaffanella01th} as
a possible instance of the parameter $P$.

This paper is a combination and improvement of
\cite{BagnaraGHZ01} and \cite{BagnaraZGH01}.
As a result of an editorial requirement,
the proofs of the stated results
have been omitted from this version of the paper;
the referees did check the proofs which were part of the submitted version.
The proofs can be found in~\cite{BagnaraGHZ04TR},
the unabridged version of this paper.
}%%\ifthenelse{\boolean{TR}}{%%

\section{Preliminaries}
\label{sec:preliminaries}

\subsection{Infinite Terms and Substitutions}

The cardinality of a set $S$ is denoted by $\card S$;
$\wp(S)$ is the powerset of $S$, whereas $\wpf(S)$ is
the set of all the \emph{finite} subsets of $S$.
Let $\Sig$ denote a possibly infinite set of function symbols, ranked
over the set of natural numbers.
It is assumed that $\Sig$ contains at least
one function symbol having rank $0$ and one having rank greater than $0$.
Let $\Vars$ denote a denumerable set of variables disjoint from $\Sig$
and $\Terms$ denote the free algebra of all (possibly infinite)
terms in the signature $\Sig$ having variables in $\Vars$.
Thus a term can be seen as an ordered labeled tree, possibly having
some infinite paths and possibly containing variables: every non-leaf
node is labeled with a function symbol in $\Sig$ with a rank matching
the number of the node's immediate descendants, whereas every leaf is
labeled by either a variable in $\Vars$ or a function symbol in $\Sig$
having rank $0$ (a constant).

If $t \in \Terms$ then $\vars(t)$ and $\mvars(t)$ denote the set and
the multiset of variables occurring in $t$, respectively.
We will also write $\vars(o)$ to denote the set of variables occurring
in an arbitrary syntactic object $o$.

Suppose $s,t \in \Terms$:
$s$ and $t$ are \emph{independent} if
$\vars(s) \inters \vars(t) = \emptyset$;
$t$ is said to be \emph{ground} if $\vars(t) = \emptyset$;
$t$ is \emph{free} if $t \in \Vars$;
if $y \in \vars(t)$ occurs exactly once in $t$,
then we say that variable $y$ \emph{occurs linearly in $t$},
more briefly written using the predication $\occlin(y, t)$;
$t$ is \emph{linear} if we have
$\occlin(y, t)$ for all $y \in \vars(t)$;
finally, $t$ is a \emph{finite term} (or \emph{Herbrand term})
if it contains a finite number of occurrences of function symbols.
The sets of all ground, linear and finite terms are denoted by
$\GTerms$, $\LTerms$ and $\HTerms$, respectively.
As we have specified that $\Sig$ contains function symbols of rank $0$
and rank greater than $0$,
$\GTerms \inters \HTerms \neq \emptyset$ and
$\GTerms \setdiff \HTerms \neq \emptyset$.

A \emph{substitution} is a total function
$\fund{\sigma}{\Vars}{\HTerms}$ that is the identity almost
everywhere; in other words, the \emph{domain} of~$\sigma$,
\[
  \dom(\sigma)
    \defeq
      \bigl\{\,
        x \in \Vars
      \bigm|
        \sigma(x) \neq x
      \,\bigr\},
\]
is finite.
Given a substitution $\fund{\sigma}{\Vars}{\HTerms}$,
we overload the symbol `$\sigma$' so as to denote also the function
$\fund{\sigma}{\HTerms}{\HTerms}$ defined as follows, for each term
$t \in \HTerms$:
\[
  \sigma(t)
    \defeq
      \begin{cases}
        t,                       &\text{if $t$ is a constant symbol;} \\
        \sigma(t),               &\text{if $t \in \Vars$;} \\
        f\bigl(\sigma(t_1), \ldots, \sigma(t_n)\bigr),
                                 &\text{if $t = f(t_1, \ldots, t_n)$.}
      \end{cases}
\]
If $t \in \HTerms$, we write $t\sigma$ to denote $\sigma(t)$
and $t\sigma\tau$ to denote $(t\sigma)\tau$.

If $x \in \Vars$ and $t \in \HTerms \setdiff \{x\}$,
then $x \mapsto t$ is called a \emph{binding}.
The set of all bindings is denoted by $\Bind$.
Substitutions are denoted by the set of their bindings,
thus a substitution $\sigma$ is identified with the (finite) set
\[
  \bigl\{\, x \mapsto x\sigma \bigm| x \in \dom(\sigma) \,\bigr\}.
\]
We denote by $\vars(\sigma)$ the set of variables occurring in the
bindings of $\sigma$.

A substitution is said to be \emph{circular} if, for $n > 1$, it has
the form
\[
  \{ x_1 \mapsto x_2, \ldots, x_{n-1} \mapsto x_n, x_n \mapsto x_1\},
\]
where $x_1$, \dots,~$x_n$ are distinct variables.
A substitution is in \emph{rational solved form}
if it has no circular subset.
The set of all substitutions in rational solved form
is denoted by $\RSubst$.

The composition of substitutions is defined in the usual way.
Thus $\tau \compose \sigma$ is the substitution such that,
for all terms $t \in \HTerms$,
\[
  (\tau \compose \sigma)(t)
    = \tau\bigl(\sigma(t)\bigr)
    = t\sigma\tau
\]
and has the formulation
\begin{equation*}
  \tau\compose \sigma
    =
      \bigl\{\,
        x \mapsto x\sigma\tau
      \bigm|
        x \in \dom(\sigma) \union \dom(\tau),
        x \neq x\sigma\tau
      \,\bigr\}.
\end{equation*}
As usual, $\sigma^0$ denotes the identity function
(i.e., the empty substitution)
and, when $i > 0$, $\sigma^i$ denotes the substitution
$(\sigma\circ\sigma^{i-1})$.

Consider an infinite sequence of terms
$t_0, t_1, t_2, \ldots$ with $t_i \in \HTerms$ for each $i \in \Nset$.
Suppose there exists $t \in \Terms$ such that, for each $n \in \Nset$,
there exists $m_0 \in \Nset$ such that,
for each $m \in \Nset$ with $m \geq m_0$,
the trees corresponding to the terms $t$ and $t_m$
coincide up to the first $n$ levels.
Then we say that \emph{the sequence $t_0, t_1, t_2, \ldots$ converges to $t$}
and we write $t = \lim_{i \rightarrow \infty} t_i$
\cite{BerarducciVZ93}.

For each $\sigma \in \RSubst$ and $t \in \HTerms$,
the sequence of finite terms
\[
  \sigma^0(t), \sigma^1(t), \sigma^2(t), \ldots
\]
converges~\cite{BerarducciVZ93,King00}.
Therefore, the function $\fund{\rt}{\HTerms \times \RSubst}{\Terms}$
such that
\[
  \rt(t, \sigma) \defeq \lim_{i \rightarrow \infty} \sigma^i(t)
\]
is well defined.

\subsection{Equations}

An \emph{equation} is a statement of the form $s = t$
where $s,t \in \HTerms$.
$\Eqs$ denotes the set of all equations.
As usual, a system of equations (i.e., a conjuction of elements in $\Eqs$)
will be denoted by a subset of $\Eqs$.
A substitution $\sigma$ may be regarded as a finite set of equations,
that is, as the set $\{\, x = t \mid x \mapsto t \in \sigma \,\}$.
A set of equations $e$ is in \emph{rational solved form}
if $\bigl\{\, s \mapsto t \bigm| (s = t) \in e \,\bigr\} \in \RSubst$.
In the rest of the paper, we will
often write a substitution $\sigma \in \RSubst$ to denote a set of
equations in rational solved form (and vice versa).

Languages such as Prolog~II, SICStus and Oz are
based on $\RT$, the theory of rational
trees~\cite{Colmerauer82,Colmerauer84}.
This is a syntactic equality theory (i.e., a theory where the function
symbols are uninterpreted), augmented with a \emph{uniqueness axiom}
for each substitution in rational solved form.
Informally speaking these axioms state that, after assigning a ground
rational tree to each non-domain variable, the substitution uniquely
defines a ground rational tree for each of its domain variables.
Thus, any set of equations in rational solved form is, by definition,
satisfiable in $\RT$.
\ifthenelse{\boolean{TR}}{%%
Equality theories and, in particular, $\RT$ are presented in more
detail in Appendix~\ref{subsec:equality-theories}.
}{%% \ifthenelse{\boolean{TR}}{%%
}%% \ifthenelse{\boolean{TR}}{%%
Note that being in rational solved form is a very weak property.
Indeed, unification algorithms returning a set of equations in
rational solved form are allowed to be much more ``lazy'' than one
would usually expect.
For instance, $\{ x = y, y = z \}$ and
$\bigl\{ x = f(y), y = f(x) \bigr\}$ are in rational
solved form.
We refer the interested reader to \cite{JaffarLM87,Keisu94th,Maher88}
for details on the subject.

Given a set of equations $e \in \wpf(\Eqs)$ that is satisfiable in $\RT$,
a substitution $\sigma \in \RSubst$
is called a \emph{solution for $e$ in $\RT$}
if $\RT \entails \forall(\sigma \pimplies e)$, i.e.,
if theory $\RT$ entails the first order
formula $\forall(\sigma \pimplies e)$.
If in addition $\vars(\sigma) \sseq \vars(e)$,
then $\sigma$ is said to be a \emph{relevant} solution for $e$.
Finally, $\sigma$ is a \emph{most general solution for $e$ in $\RT$}
if $\RT \entails \forall(\sigma \piff e)$.
In this paper, the set of all the relevant most general solutions for
$e$ in $\RT$ will be denoted by $\mgs(e)$.

In the sequel, in order to model the constraint accumulation process
of logic-based languages, we will need to characterize those sets of
equations that are stronger than (that can be obtained by adding
equations to) a given set of equations.
\begin{definition} \summary{($\down(\cdot)$)}
\label{def:downarrow}
The function
$\fund{\down(\cdot)}{\RSubst}{\wp(\RSubst)}$
is defined, for each $\sigma \in \RSubst$, by
\[
  \down\sigma
    \defeq
      \bigl\{\,
        \tau \in \RSubst
      \bigm|
        \exists \sigma' \in \RSubst
          \st \tau \in \mgs(\sigma \union \sigma')
      \,\bigr\}.
\]
\end{definition}
The next result shows that $\down(\cdot)$ corresponds to the closure
by entailment in $\RT$.
\begin{proposition}
\label{prop:down-entailment-equivalence}
Let $\sigma \in \RSubst$.
Then
\[
  \down \sigma
    =
      \bigl\{\,
        \tau \in \RSubst
      \bigm|
        \RT \entails \forall(\tau \pimplies \sigma)
      \,\bigr\}.
\]
\end{proposition}
%Thus, as entailment is a transitive relation,
%we have, for all $\sigma, \tau, \upsilon \in \RSubst$,
%\[
%  \upsilon \in \down\tau
%    \;\land\;
%      \tau \in \down \sigma
%        \quad\implies\quad
%          \upsilon \in \down\sigma.
%\]

\subsection{Boolean Functions}

Boolean functions have already been extensively used for data-flow
analysis of logic-based languages.
An important class of these functions used for tracking groundness dependencies
is $\Pos$ \cite{ArmstrongMSS98}.
This domain was introduced in \cite{MarriottS89} under the name \textit{Prop}
and further refined and studied in \cite{CortesiFW91,MarriottS93}.

The formal definition of the set of Boolean functions over
a finite set of variables is based on the notion of Boolean valuation.
Note that in all the following definitions we abuse notation by assuming
that the finite set of variables $V$ is clear from context, so as to
avoid using it as a suffix everywhere.

\ifthenelse{\boolean{TR}}{%%

\begin{definition} \summary{(Boolean valuation.)}
\label{def:boolean-valuations}
Let $V \in \wpf(\Vars)$
and $\Bool \defeq \{ 0, 1 \}$.
The set of \emph{Boolean valuations over $V$}
is given by
\[
  \Bval \defeq V \rightarrow \Bool.
\]
For each $a \in \Bval$, each $x \in V$, and each $c \in \Bool$
the valuation $a[c/x] \in \Bval$ is given, for each $y \in V$, by
\[
  a[c/x](y)
    \defeq
      \begin{cases}
        c,    & \text{ if $x = y$;} \\
        a(y), & \text{ otherwise.}
      \end{cases}
\]
If $X = \{ x_1, \ldots, x_k \} \sseq V$,
then $a[c/X]$ denotes $a[c/x_1]\cdots[c/x_k]$.
\end{definition}
The distinguished elements
$\Bvalzero, \Bvalone \in \Bval$ are given by
\begin{align*}
  \Bvalzero &\defeq \lambda x \in V \lambdadot 0, \\
  \Bvalone  &\defeq \lambda x \in V \lambdadot 1.
\end{align*}

\begin{definition} \summary{(Boolean function.)}
\label{def:boolean-functions}
The set of \emph{Boolean functions over $V$} is
\[
  \Bfun \defeq \Bval \rightarrow \Bool.
\]
$\Bfun$ is partially ordered by the relation $\models$
where, for each $\phi,\psi \in \Bfun$,
\[
  \phi \models \psi
    \quad\defiff\quad
      \bigl(
        \forall a \in \Bval
          \itc
            \phi(a) = 1 \implies \psi(a) = 1
      \bigr).
\]
\end{definition}
For $\phi \in \Bfun$, $x \in V$, and $c \in \Bool$,
the Boolean function
$\phi[c/x] \in \Bfun$ is given, for each $a \in \Bval$, by
\begin{align*}
  \phi[c/x](a) &\defeq \phi\bigl(a[c/x]\bigr). \\
\intertext{%
When $X \sseq V$, $\phi[c/X]$ is defined in the expected way.
If $\phi \in \Bfun$ and $x, y \in V$ the function
$\phi[y/x] \in \Bfun$ is given, for each $a \in \Bval$, by
}
  \phi[y/x](a) &\defeq \phi\Bigl(a\bigl[a(y)/x\bigr]\Bigr).
\end{align*}

}{%%\ifthenelse{\boolean{TR}}{%%

\begin{definition} \summary{(Boolean valuation and function.)}
\label{def:boolean-valuations-and-functions}
Let $V \in \wpf(\Vars)$
and $\Bool \defeq \{ 0, 1 \}$.
The set of \emph{Boolean valuations over $V$}
is given by
\[
  \Bval \defeq V \rightarrow \Bool.
\]
The set of \emph{Boolean functions over $V$} is
\[
  \Bfun \defeq \Bval \rightarrow \Bool.
\]
$\Bfun$ is partially ordered by the relation $\models$
where, for each $\phi,\psi \in \Bfun$,
\[
  \phi \models \psi
    \quad\defiff\quad
      \bigl(
        \forall a \in \Bval
          \itc
            \phi(a) = 1 \implies \psi(a) = 1
      \bigr).
\]
\end{definition}
}%%\ifthenelse{\boolean{TR}}{%%

\noindent
Boolean functions are constructed from the elementary functions
corresponding to variables and by means of the usual logical
connectives. Thus, for each $x \in V$, $x$ also denotes
the Boolean function $\phi$ such that, for each $a \in \Bval$,
$\phi(a) = 1$ if and only if $a(x) = 1$;
for $\phi \in \Bfun$, we write $\neg \phi$ to denote
the function $\psi$ such that, for each $a \in \Bval$, $\psi(a) = 1$
if and only if $\phi(a) = 0$;
for $\phi_1, \phi_2 \in \Bfun$, we write $\phi_1 \lor \phi_2$ to
denote the function $\phi$ such that, for each $a \in \Bval$, $\phi(a) = 0$
if and only if both $\phi_1(a) = 0$ and $\phi_2(a) = 0$.
A variable is restricted away using Schr\"{o}der's elimination
principle \textup{\cite{Schroeder1877}}:
\ifthenelse{\boolean{TR}}{%%
\[
  \exists x \st \phi
    \defeq
      \phi[1/x] \lor \phi[0/x].
\]
}{%%\ifthenelse{\boolean{TR}}{%%
\begin{align*}
  \exists x \st \phi
    &\defeq
      \phi[1/x] \lor \phi[0/x] \\
\intertext{%
where, for each $c \in \Bool$ and each $a \in \Bval$,
}
  \phi[c/x](a)
    &\defeq
      \phi\bigl(a[c/x]\bigr), \\
  a[c/x](y)
    &\defeq
      \begin{cases}
        c,    & \text{ if $x = y$;} \\
        a(y), & \text{ otherwise.}
      \end{cases}
\end{align*}
}%%\ifthenelse{\boolean{TR}}{%%
Note that existential quantification
is both monotonic and extensive on $\Bfun$.
The other Boolean connectives and quantifiers are handled similarly.
The distinguished elements $\bot, \top \in \Bfun$ are
the functions defined by
\begin{align*}
  \bot &\defeq \lambda a \in \Bval \lambdadot 0, \\
  \top &\defeq \lambda a \in \Bval \lambdadot 1.
\end{align*}
For notational convenience, when $X \sseq V$, we inductively define
\[
  \bigland X
    \defeq
      \begin{cases}
        \top,
          &\text{if $X = \emptyset$;} \\
	x \land \bigland \bigl(X \setdiff \{ x \}\bigr),
          &\text{if $x \in X$.}
      \end{cases}
\]
\ifthenelse{\boolean{TR}}{%%
}{%%\ifthenelse{\boolean{TR}}{%%
The distinguished valuation
\(
  \Bvalone \defeq \lambda x \in V \lambdadot 1
\)
is also called the \emph{everything-is-true} assignment.
}%%\ifthenelse{\boolean{TR}}{%%
$\Pos \sslt \Bfun$ consists precisely of those functions assuming
the true value under the \emph{everything-is-true} assignment,
i.e.,
\[
  \Pos \defeq \bigl\{\, \phi \in \Bfun \bigm| \phi(\Bvalone) = 1 \,\bigr\}.
\]
For each $\phi \in \Bfun$, the \emph{positive part of $\phi$},
denoted $\pos(\phi)$, is the strongest $\Pos$ formula that is
entailed by $\phi$. Formally,
\[
  \pos(\phi) \defeq \phi \lor \bigland V.
\]

For each $\phi \in \Bfun$,
the set of \emph{variables necessarily true for $\phi$} and
the set of \emph{variables necessarily false for $\phi$}
are given, respectively, by
\begin{align*}
  \truev(\phi) &\defeq
    \bigl\{\,
      x \in V
    \bigm|
      \forall a \in \Bval \itc \phi(a) = 1 \implies a(x) = 1
    \,\bigr\}, \\
  \falsev(\phi) &\defeq
    \bigl\{\,
      x \in V
    \bigm|
      \forall a \in \Bval \itc \phi(a) = 1 \implies a(x) = 0
    \,\bigr\}.
\end{align*}

\section{The Concrete Domain}
\label{sec:concrete-domain}

A knowledge of the basic concepts
of abstract interpretation theory~\cite{CousotC77,CousotC92fr} is assumed.
In this paper, the concrete domain consists of pairs of the form
$(\Sigma, V)$, where $V$ is a finite set of
\emph{variables of interest}~\cite{CortesiFW91}
and $\Sigma$ is a (possibly infinite) set of substitutions
in rational solved form.
\begin{definition}\summary{(The concrete domain.)}
\label{def:concrete-domain}
Let
\(
  \domainD \defeq \wp(\RSubst) \times \wpf(\Vars)
\).
If $(\Sigma, V) \in \domainD$, then $(\Sigma, V)$ represents the
(possibly infinite) set of first-order formulas
\(
  \bigl\{\,
    \exists \Delta \st \sigma
  \bigm|
    \sigma \in \Sigma,
    \Delta = \vars(\sigma) \setdiff V
  \,\bigr\}
\)
where $\sigma$ is interpreted as the logical conjunction of the
equations corresponding to its bindings.

The operation of projecting $x \in \Vars$\
away from $(\Sigma, V) \in \domainD$
is defined as follows:
\[
\projectD x \st (\Sigma, V)
    \defeq
      \sset{
        \sigma' \in \RSubst
      }{
        \sigma \in \Sigma,
        \overline{V} = \Vars \setdiff V, \\
         \RT
           \entails
             \forall\bigl(
                      \exists \overline{V} \st
                        (\sigma' \piff  \exists x \st \sigma)
                    \bigr)
      }.
\]
\end{definition}

Concrete domains for constraint languages would be similar.
If the analyzed language allows the use of constraints on various domains
to restrict the values of the variable leaves of rational trees,
the corresponding concrete domain would have one or more extra components
to account for the constraints (see \cite{BagnaraHZ00} for an example).

The concrete element
$\bigl(\bigl\{\{ x \mapsto f(y) \}\bigr\}, \{ x, y \}\bigr)$
expresses a dependency between $x$ and $y$.
In contrast,
$\bigl(\bigl\{\{ x \mapsto f(y) \}\bigr\}, \{ x \}\bigr)$
only constrains $x$.
The same concept can be expressed by saying that in
the first case the variable name `$y$' matters, but it does not in the
second case.  Thus, the set of variables of interest is crucial for
defining the meaning of the concrete and abstract descriptions.
Despite this, always specifying the set of variables of interest would
significantly clutter the presentation.
Moreover, most of the needed functions on concrete and abstract descriptions
preserve the set of variables of interest.
For these reasons, we assume the existence of a
set $\VI \in \wpf(\Vars)$ that contains, at each stage of the
analysis, the current variables of interest.\footnote{This parallels
what happens in the efficient implementation of data-flow analyzers.
In fact, almost all the abstract domains currently in use do not need
to represent explicitly the set of variables of interest.  In
contrast, this set is maintained externally and in a unique copy,
typically by the fixpoint computation engine.}
As a consequence, when the context makes it clear,
we will write $\Sigma \in \domainD$
as a shorthand for $(\Sigma, \VI) \in \domainD$.

\section{An Abstract Domain for Finite-Tree Analysis}
\label{sec:finite-tree-domain}

Finite-tree analysis applies to logic-based languages computing over a
domain of rational trees where cyclic structures are allowed.
In contrast, analyses aimed at occurs-check reduction
\cite{CrnogoracKS96,Sondergaard86} apply to programs that are meant to
compute on a domain of finite trees only, but have to be executed over
systems that are either designed for rational trees or
intended just for the finite trees but
omit the occurs-check for efficiency reasons.
Despite their different objectives, finite-tree and occurs-check
analyses have much in common:
in both cases, it is important to detect all program points where
cyclic structures can be generated.

Note however that, when performing occurs-check reduction,
one can take advantage of the following invariant:
all data structures generated so far are finite.
This property is maintained by transforming the program so as
to force finiteness whenever it is possible that a cyclic structure could
have been built.\footnote{%
Such a requirement is typically obtained by replacing
the unification with a call to the standard
predicate \texttt{unify\_with\_occurs\_check/2}.
As an alternative, in some systems based on rational trees it is
possible to insert, after each problematic unification,
a finiteness test for the generated term.}
In contrast, a finite-tree analysis has to deal with
the more general case when some of the data structures
computed so far may be cyclic.
It is therefore natural to consider an abstract domain
made up of two components.
The first one simply represents the set of variables
that are guaranteed not to be bound to infinite terms.
We will denote this \emph{finiteness component} by $H$
(from \emph{Herbrand}).
\begin{definition} \summary{(The finiteness component.)}
The \emph{finiteness component} is the set $H \defeq \wp(\VI)$
partially ordered by reverse subset inclusion.
\end{definition}
The second component of the finite-tree domain should maintain
any kind of information that may be useful for computing
finiteness information.

It is well-known that sharing information as a whole,
therefore including possible variable aliasing,
definite linearity, and definite freeness,
has a crucial role in occurs-check reduction so that,
as observed before, it can be exploited for finite-tree analysis too.
Thus, a first choice for the second component
of the finite-tree domain would be to consider
one of the standard combinations of sharing,
freeness and linearity as defined, e.g.,
in~\cite{BagnaraZH00,BagnaraZH05TPLP,BruynoogheCM94,HansW92}.
However, this would tie our specification to a particular
sharing analysis domain, whereas the overall approach
is inherently more general.
For this reason,
we will define a finite-tree analysis based on
the abstract domain schema $H \times P$,
where the generic \emph{sharing component} $P$
is a parameter of the abstract domain construction.
This approach can be formalized as an application
of the \emph{open product} operator~\cite{CortesiLCVH00},
where the interaction between the $H$ and $P$ components
is modeled by defining a suite of generic query operators:
thus, the overall accuracy of the finite-tree analysis will
heavily depend on the accuracy with which any specific instance
of the parameter $P$ is able to answer these queries.

\subsection{The parameter Component $P$}
\label{subsec:P}

Elements of $P$ can encode any kind of information.
We only require that substitutions that are equivalent
in the theory $\RT$ are identified in $P$.

\begin{definition} \summary{(The parameter component.)}
\label{def:P}
The parameter component $P$ is an abstract domain
related to the concrete domain $\domainD$
by means of the \emph{concretization function}
$\fund{\concrP}{P}{\wp(\RSubst)}$ such that,
for all $p \in P$,
\[
  \Bigl(
    \sigma \in \concrP(p)
      \land
        \bigl(\RT \entails \forall(\sigma \piff \tau)\bigr)
  \Bigr)
    \implies
      \tau \in \concrP(p).
\]
\end{definition}

The interface between $H$ and $P$ is provided
by a set of abstract operators that satisfy
suitable correctness criteria.
We only specify those
that are useful for defining abstract unification and projection
on the combined domain $H \times P$.
Other operations needed for a full description of the analysis,
such as renaming and upper bound,
are very simple and, as usual, do not pose any problems.

\begin{definition} \summary{(Abstract operators on $P$.)}
\label{def:predicates-and-functs-P}
Let $s,t \in \HTerms$ be finite terms.
For each $p \in P$,
we specify the following predicates:

\noindent
$s$ and $t$ are \emph{independent in $p$}
if and only if
$\fund{\ind_p}{\HTerms^2}{\Bool}$
holds for $(s, t)$, where
\begin{align*}
  \ind_p(s, t)
    &\implies
      \forall \sigma \in \concrP(p)
        \itc
          \vars\bigl(\rt(s, \sigma)\bigr)
            \inters
              \vars\bigl(\rt(t, \sigma)\bigr)
                = \emptyset; \\
\intertext{%
$s$ and $t$ \emph{share linearly in $p$}
if and only if
$\fund{\sharelin_p}{\HTerms^2}{\Bool}$
holds for $(s, t)$, where
}
  \sharelin_p(s, t)
    &\implies
      \forall \sigma \in \concrP(p)
        \itc \\
    &\qquad\qquad
          \forall y \in \vars\bigl(\rt(s, \sigma)\bigr)
                          \inters
                            \vars\bigl(\rt(t, \sigma)\bigr)
            \itc \\
    &\qquad\qquad\qquad
            \occlin\bigl(y, \rt(s, \sigma)\bigr)
              \land
                \occlin\bigl(y, \rt(t, \sigma)\bigr); \\
\intertext{%
$t$ is \emph{ground in $p$}
if and only if
$\fund{\ground_p}{\HTerms}{\Bool}$
holds for $t$, where
}
  \ground_p(t)
    &\implies
      \forall \sigma \in \concrP(p)
        \itc
          \rt(t, \sigma) \in \GTerms; \\
\intertext{%
$t$ is \emph{ground-or-free in $p$}
if and only if
$\fund{\gfree_p}{\HTerms}{\Bool}$
holds for $t$, where
}
  \gfree_p(t)
    &\implies
      \forall \sigma \in \concrP(p)
        \itc
          \rt(t, \sigma) \in \GTerms
            \lor
              \rt(t, \sigma) \in \Vars; \\
\intertext{%
$s$ is \emph{linear in $p$}
if and only if
$\fund{\lin_p}{\HTerms}{\Bool}$
holds for $s$, where
}
  \lin_p(s)
    &\implies
      \forall \sigma \in \concrP(p)
        \itc
          \rt(s, \sigma) \in \LTerms; \\
\intertext{%
$s$ and $t$ are \emph{or-linear in $p$}
if and only if
$\fund{\orlin_p}{\HTerms^2}{\Bool}$
holds for $(s, t)$, where
}
  \orlin_p(s,t)
    &\implies
      \forall \sigma \in \concrP(p)
        \itc
          \rt(s, \sigma) \in \LTerms
            \lor
              \rt(t, \sigma) \in \LTerms; \\
\end{align*}

For each $p \in P$, the following functions
compute subsets of the set of variables of interest:

\noindent
the function
$\fund{\sharesamevar_p}{\HTerms \times \HTerms}{\wp(\VI)}$
returns a set of variables that may share with the given terms
via the same variable. For each pair of terms $s,t \in \HTerms$,
\begin{align*}
  \sharesamevar_p(s,t)
    &\Sseq
      \sset
        {y \in \VI}
        {\exists \sigma \in \concrP(p) \st \\
           \quad \exists z \in \vars\bigl(\rt(y, \sigma)\bigr) \st \\
              \qquad z \in \vars\bigl(\rt(s, \sigma)\bigr)
                             \inters
                               \vars\bigl(\rt(t, \sigma)\bigr)
        }; \\
\intertext{%
the function $\fund{\sharewith_p}{\HTerms}{\wp(\VI)}$
yields a set of variables that may share with the given term.
For each $t \in \HTerms$,
}
  \sharewith_p(t)
    &\defeq
      \bigl\{\,
        y \in \VI
      \bigm|
        y \in \sharesamevar_p(y, t)
      \,\bigr\}.
\end{align*}

The function $\fund{\amguP}{P \times \Bind}{P}$
correctly captures the effects of a binding on an element of $P$.
For each $(x \mapsto t) \in \Bind$ and $p \in P$, let
\begin{equation*}
  p' \defeq \amguP(p, x \mapsto t);
\end{equation*}
for all $\sigma \in \concrP(p)$,
if $\tau \in \mgs\bigl(\sigma \union \{ x = t \}\bigr)$,
then $\tau \in \concrP(p')$.

The function $\fund{\projP}{P \times \VI}{P}$
correctly captures the operation of projecting away
a variable from an element of $P$.
For each $x \in \VI$, $p \in P$ and $\sigma \in \concrP(p)$,
if $\tau \in \projectD x \st \{\sigma\}$,
then $\tau \in \concrP\bigl(\projP(p, x)\bigr)$.
\end{definition}

As it will be shown in Appendix~\ref{sec:implementing-P},
some of these generic operators can be directly mapped to
the corresponding abstract operators defined for well-known
sharing analysis domains.
However, the specification given
in Definition~\ref{def:predicates-and-functs-P},
besides being more general than a particular implementation,
also allows for a modular approach when proving correctness results.

\subsection{Operators on Substitutions in Rational Solved Form}
\label{subsec:operators-on-RSubst}

There are cases when an analysis tries to capture properties of the
particular substitutions computed by a specific (ordinary or rational)
unification algorithm.
This is the case, for example, when the analysis needs to track
structure sharing for the purpose of compile-time garbage collection,
or provide upper bounds on the amount of memory needed to perform a
given computation.
More often the interest is on properties of the (finite or rational)
trees that are denoted by such substitutions.

When the concrete domain is based on the theory of finite trees,
idempotent substitutions provide a finitely computable \emph{strong
normal form} for domain elements, meaning that different substitutions
describe different sets of finite trees (as usual, this is modulo the
possible renaming of variables).
In contrast, when working on a concrete domain based on the theory of
rational trees, substitutions in rational solved form, while being
finitely computable, no longer satisfy this property: there can be an
infinite set of substitutions in rational solved form all describing
the same set of rational trees
(i.e., the same element in the ``intended'' semantics).
For instance, the substitutions
\[
  \sigma_n = \bigl\{ x \mapsto \overbrace{f( \cdots f(}^n x ) \cdots ) \bigr\}
\]
for $n = 1$, $2$,~\dots,
all map the variable $x$ to the same rational tree
(which is usually denoted by $f^\omega$).

Ideally, a strong normal form for the set of rational trees described by
a substitution $\sigma \in \RSubst$ can be obtained by computing
the limit function
\[
  \sigma^\infty \defeq \lambda t \in \HTerms \st \rt(t, \sigma),
\]
obtained by fixing the substitution parameter of `$\rt$'.
The problem is that, in general,
$\sigma^\infty$ is not a substitution: while having a finite domain,
its ``bindings'' $x \mapsto \lim_{i \rightarrow \infty} \sigma^i(x)$
can map a domain variable $x$ to an infinite rational term.
This poses a non-trivial problem when trying to define a ``good''
abstraction function, since it would be really desirable for this
function to map any two equivalent concrete elements to the same
abstract element.
Of course, it is important that the properties under investigation
are exactly captured, so as to avoid any unnecessary precision loss.
Pursuing this goal requires an ability to observe properties of
(infinite) rational trees while just dealing with one of their finite
representations.  This is not always an easy task since even simple
properties can be ``hidden'' when using non-idempotent substitutions.
For instance, when $\sigma^\infty$ maps variable $x$
to an infinite and ground rational tree
(i.e., when $\rt(x, \sigma) \in \GTerms \setdiff \HTerms$),
all of its finite representations in $\RSubst$
(i.e., all the $\tau \in \RSubst$ such that
$\RT \models \forall(\sigma \piff \tau)$)
will map the variable $x$ into a finite term that is not ground.
These are the motivations behind the introduction of the following
computable operators on substitutions.

The groundness operator `$\gvars$' captures the set of variables
that are mapped to ground rational trees by $\rt$.
We define it by means of the \emph{occurrence operator} `$\occ$'.
This was introduced in \cite{HillBZ02TPLP} as a replacement
for the sharing-group operator `$\sg$' of~\cite{JacobsL89}.
In \cite{HillBZ02TPLP} the `$\occ$' operator is used to define
a new abstraction function for set-sharing analysis that,
differently from the classical ones \cite{CortesiF99,JacobsL89},
maps equivalent substitutions in rational solved form
to the same abstract element.

\begin{definition} \summary{(Occurrence and groundness operators.)}
\label{def:groundness-op}
For each $n \in \Nset$,
the \emph{occurrence function}
$\fund{\occ_n}{\RSubst\times\Vars}{\wpf(\Vars)}$
is defined,
for each $\sigma \in \RSubst$ and each $v \in \Vars$, by
\begin{align*}
  \occ_n(\sigma, v)
    &\defeq
\begin{cases}
        \{v\}\setdiff \dom(\sigma),
            &\text{if $n = 0$;} \\
      \bigl\{\,
        y \in \Vars
      \bigm|
        \vars(y\sigma)\inters \occ_{n-1}(\sigma,v) \neq \emptyset
      \,\bigr\},
      &\text{if $n > 0$}.
\end{cases}
\end{align*}
The \emph{occurrence operator}
$\fund{\occ}{\RSubst\times\Vars}{\wpf(\Vars)}$
is given,
for each $\sigma \in \RSubst$ and $v \in \Vars$,
by
\(
  \occ(\sigma,v) \defeq \occ_\ell(\sigma,v)
\),
where $\ell = \card \sigma$.

The \emph{groundness operator}
$\fund{\gvars}{\RSubst}{\wpf(\Vars)}$
is given,
for each substitution $\sigma \in \RSubst$,
by
\[
 \gvars(\sigma)
   \defeq
     \bigl\{\,
       y \in \dom(\sigma)
     \bigm|
       \forall v \in \vars(\sigma) \itc y \notin \occ(\sigma, v)
     \,\bigr\}.
\]
\end{definition}

\begin{example}
Let
\begin{equation*}
  \sigma
    = \bigl\{
        x \mapsto f(y, z),
        y \mapsto g(z, x),
        z \mapsto f(a)
      \bigr\}.
\end{equation*}
Then $\gvars(\sigma) = \{ x, y, z \}$,
although $\vars(x\sigma^i) \neq \emptyset$
and $\vars(y\sigma^i) \neq \emptyset$,
for all $0 \leq i < \infty$.
\end{example}

The \emph{finiteness operator} is defined, like `$\occ$',
by means of a fixpoint construction.
\begin{definition} \summary{(Finiteness functions.)}
\label{def:finiteness-functs}
For each $n \in \Nset$,
the \emph{finiteness function}
$\fund{\hvars_n}{\RSubst}{\wp(\Vars)}$
is defined,
for each $\sigma \in \RSubst$, by
\begin{align*}
  \hvars_0(\sigma)
    &\defeq
      \Vars \setdiff \dom(\sigma) \\
\intertext{%
  and, for $n > 0$, by
}
  \hvars_n(\sigma)
    &\defeq
      \hvars_{n-1}(\sigma)
        \union
          \bigl\{\,
            y \in \dom(\sigma)
          \bigm|
            \vars(y\sigma) \sseq \hvars_{n-1}(\sigma)
          \,\bigr\}.
\end{align*}
\end{definition}

For each $\sigma \in \RSubst$ and each $i \geq 0$,
we have $\hvars_i(\sigma) \sseq \hvars_{i+1}(\sigma)$ and
also that $\Vars \setdiff \hvars_i(\sigma) \sseq \dom(\sigma)$
is a finite set.
By these two properties, the chain
$\hvars_0(\sigma) \sseq \hvars_1(\sigma) \sseq \cdots$
is stationary and finitely computable.
In particular, if $\ell = \card \sigma$, then,
for all $n \geq \ell$,
$\hvars_\ell(\sigma) = \hvars_n(\sigma)$.

\begin{definition} \summary{(Finiteness operator.)}
\label{def:finiteness-op}
For each $\sigma \in \RSubst$,
the \emph{finiteness operator}
$\fund{\hvars}{\RSubst}{\wp(\Vars)}$
is given by
\(
  \hvars(\sigma) \defeq \hvars_\ell(\sigma)
\)
where $\ell \defeq \ell(\sigma) \in \Nset$ is such that
$\hvars_\ell(\sigma) = \hvars_{n}(\sigma)$ for all $n \geq \ell$.
\end{definition}

The following proposition shows that the `$\hvars$' operator
precisely captures the intended property.
\begin{proposition}
\label{prop:rt-hvars-RSubst}
If $\sigma \in \RSubst$ and $x \in \Vars$ then
\[
  x \in \hvars(\sigma)
    \iff
      \rt(x, \sigma) \in \HTerms.
\]
\end{proposition}

\begin{example}
\label{ex:hvars}
Consider $\sigma \in \RSubst$, where
\begin{equation*}
  \sigma
     =
      \bigl\{
        x_1 \mapsto f(x_2),
        x_2 \mapsto g(x_5),
        x_3 \mapsto f(x_4),
        x_4 \mapsto g(x_3)
      \bigr\}.
\end{equation*}
Then,
\begin{align*}
  \hvars_0(\sigma) &= \Vars \setdiff \{x_1,x_2,x_3,x_4\}, \\
  \hvars_1(\sigma) &= \Vars \setdiff \{x_1,x_3,x_4\}, \\
  \hvars_2(\sigma) &= \Vars \setdiff \{x_3,x_4\} \\
                   &= \hvars(\sigma).
\end{align*}
Thus, $x_1 \in \hvars(\sigma)$,
although $\vars(x_1\sigma) \sseq \dom(\sigma)$.
\end{example}

The following proposition states how `$\gvars$' and `$\hvars$'
behave with respect to the further instantiation of variables.
\begin{proposition}
\label{prop:down-gvars-hvars}
Let $\sigma,\tau \in \RSubst$, where $\tau \in \down \sigma$.
Then
\begin{subequations}
\renewcommand{\theequation}{\ref{prop:down-gvars-hvars}\alph{equation}}
\begin{align}
\label{eq:down-gvars-hvars:hvars}
  \hvars(\sigma)
    &\Sseq
      \hvars(\tau), \\
\label{eq:down-gvars-hvars:gvars-inters-hvars}
  \gvars(\sigma) \inters \hvars(\sigma)
    &\sseq
      \gvars(\tau) \inters \hvars(\tau).
\end{align}
\end{subequations}
\end{proposition}

\subsection{The Abstraction Function for $H$}
\label{subsec:abstrH}

A Galois connection between the concrete domain $\wp(\RSubst)$
and the finiteness component $H = \wp(\VI)$
can now be defined naturally.

\begin{definition} \summary{(The Galois connection between
$\wp(\RSubst)$ and $H$.)}
\label{def:abstr-concr-H}
The abstraction function $\fund{\abstrH}{\RSubst}{H}$ is defined,
for each $\sigma \in \RSubst$, by
\begin{align*}
  \abstrH(\sigma)
    &\defeq
      \VI \inters \hvars(\sigma).
\intertext{%
The concrete domain $\domainD$ is related to $H$
by means of the \emph{abstraction function}
$\fund{\abstrH}{\domainD}{H}$
such that, for each $\Sigma \in \wp(\RSubst)$,
}
  \abstrH(\Sigma)
    &\defeq
      \bigcap
        \bigl\{\,
          \abstrH(\sigma)
        \bigm|
          \sigma \in \Sigma
        \,\bigr\}. \\
\intertext{%
Since the abstraction function $\abstrH$ is additive,
the concretization function is given
by its adjoint~\textup{\cite{CousotC77}}:
whenever $h \in H$,
}
  \concrH(h)
    &\defeq
      \bigl\{\,
        \sigma \in \RSubst
      \bigm|
        \abstrH(\sigma) \Sseq h
      \,\bigr\} \\
    &\defeq
      \bigl\{\,
        \sigma \in \RSubst
      \bigm|
        \hvars(\sigma) \Sseq h
      \,\bigr\}.
\end{align*}
\end{definition}

With these definitions, we have the desired result: equivalent
substitutions in rational solved form have the same finiteness
abstraction.

\begin{theorem}
\label{thm:equiv-RSubst-have-equiv-abstrH}
If $\sigma, \tau \in \RSubst$
and $\RT \entails \forall(\sigma \piff \tau)$,
then $\abstrH(\sigma) = \abstrH(\tau)$.
\end{theorem}

% Dual statement:
%
%\begin{proposition}
%\label{prop:concrH-equiv-RSubst}
%Let $\sigma,\tau \in \RSubst$, where $\sigma \in \concrH(h)$
%and $\RT \entails \forall(\sigma \piff \tau)$.
%Then $\tau \in \concrH(h)$.
%\end{proposition}

\subsection{Abstract Unification and Projection on $H \times P$}
\label{subsec:amguH}

The abstract unification for the combined domain $H \times P$
is defined by using the abstract predicates and functions
as specified for $P$
as well as a new finiteness predicate for the domain $H$.

\begin{definition} \summary{(Abstract unification on $H \times P$.)}
\label{def:abs-funcs-H}
A term $t \in \HTerms$ is a \emph{finite tree in $h \in H$}
if and only if the predicate
$\fund{\hterm_h}{\HTerms}{\Bool}$ holds for $t$,
where
\[
  \hterm_h(t)
    \defeq
      \bigl(\vars(t) \sseq h\bigr).
\]

The function $\fund{\amguH}{(H \times P) \times \Bind}{H}$
captures the effects of a binding on an $H$ element.
Let $\langle h, p \rangle \in H \times P$
and $(x \mapsto t) \in \Bind$.
Then
\[
  \amguH\bigl( \langle h, p \rangle, x \mapsto t \bigr) \defeq h',
\]
where $h'$ is given by the first case that applies in
\begin{align*}
  h'
    &\defeq
      \begin{cases}
        h \union \vars(t),
            &\text{if $\hterm_h(x)
                         \land
                           \ground_p(x)$;} \\
        h \union \{x\},
            &\text{if $\hterm_h(t)
                         \land
                           \ground_p(t)$;} \\
        h,
            &\text{if $\hterm_h(x)
                         \land
                           \hterm_h(t)$} \\
            &\text{$\quad\mathord{}
                         \land
                           \ind_p(x,t)
                             \land
                               \orlin_p(x,t)$;} \\
        h,
            &\text{if $\hterm_h(x)
                         \land
                           \hterm_h(t)$} \\
            &\text{$\quad\mathord{}
                         \land
                           \gfree_p(x)
                             \land
                               \gfree_p(t)$;} \\
        h \setdiff \sharesamevar_p(x, t),
            &\text{if $\hterm_h(x)
                         \land
	                   \hterm_h(t)$} \\
            &\text{$\quad\mathord{}
                         \land
                           \sharelin_p(x, t)$} \\
            &\text{$\quad\mathord{}
                          \land
                            \orlin_p(x,t)$;} \\
        h \setdiff \sharewith_p(x),
            &\text{if $\hterm_h(x)
                         \land
                           \lin_p(x)$;} \\
        h \setdiff \sharewith_p(t),
            &\text{if $\hterm_h(t)
                         \land
                           \lin_p(t)$;} \\
        h \setdiff \bigl(
                       \sharewith_p(x)
                         \union
                       \sharewith_p(t)
                     \bigr),
            &\text{otherwise.} \\
      \end{cases}
\end{align*}
The abstract unification function
$\fund{\amgu}{(H \times P) \times \Bind}{H \times P}$,
for any $\langle h, p \rangle \in H \times P$
and $(x \mapsto t) \in \Bind$, is given by
\[
  \amgu\bigl(
         \langle h, p \rangle,
         x \mapsto t
       \bigr)
    \defeq
      \Bigl\langle
        \amguH\bigl(
                \langle h, p \rangle,
                x \mapsto t
              \bigr),
        \amguP(p, x \mapsto t)
      \Bigr\rangle.
\]
\end{definition}

In the computation of $h'$ (the new finiteness component resulting
from the abstract evaluation of a binding) there are eight cases
based on properties holding for the concrete terms
described by $x$ and $t$.
\begin{enumerate}
\item
In the first case, the concrete term described by $x$ is both finite
and ground. Thus, after a successful execution of the binding,
any concrete term described by $t$ will be finite.
Note that $t$ could have contained variables
which may be possibly bound to cyclic terms
just before the execution of the binding.
\item
The second case is symmetric to the first one.
Note that these are the only cases when a ``positive'' propagation
of finiteness information is correct.
In contrast, in all the remaining cases, the goal is to limit
as much as possible the propagation of ``negative'' information,
i.e., the possible cyclicity of terms.
\item
The third case exploits the classical results proved in research
work on
occurs-check reduction~\cite{CrnogoracKS96,Sondergaard86}.
Accordingly, it is required that both $x$ and $t$ describe finite terms
that do not share.
The use of the implicitly disjunctive predicate $\orlin_p$
allows for the application of this case even when neither $x$ nor $t$
are known to be definitely linear.
For instance, as observed in~\cite{CrnogoracKS96}, this may happen
when the component $P$ embeds the domain $\Pos$
for groundness analysis.\footnote{%
  Let $t$ be $y$. Let also $P$ be $\Pos$.
  Then, given the $\Pos$ formula $\phi \defeq (x \lor y)$,
  both $\ind_\phi(x,y)$ and $\orlin_\phi(x,y)$ satisfy the conditions
  in Definition 4.
  Note that from $\phi$ we cannot infer that $x$ is definitely linear
  and neither that $y$ is definitely linear.}
\item
The fourth case
exploits the observation that cyclic terms cannot be
created when unifying two finite terms that are either ground or free.
Ground-or-freeness~\cite{BagnaraZH00,BagnaraZH05TPLP}
is a safe, more precise and inexpensive replacement
for the classical freeness property
when combining sharing analysis domains.
\item
The fifth case applies when unifying a linear and finite term
with another finite term possibly sharing with it,
provided they can only share linearly
(namely, all the shared variables occur linearly in the considered terms).
In such a context, only the shared variables can introduce cycles.
\item
In the sixth case, we drop the assumption about the finiteness
of the term described by $t$. As a consequence, all variables
sharing with $x$ become possibly cyclic.
However, provided $x$ describes a finite and linear term,
all finite variables independent from $x$ preserve their finiteness.
\item
The seventh case is symmetric to the sixth one.
\item
The last case states that term finiteness is preserved
for all variables that are independent from both $x$ and $t$.
\end{enumerate}
The following result, together with the assumption on $\amguP$
as specified in Definition~\ref{def:predicates-and-functs-P},
ensures that abstract unification on the combined domain
$H \times P$ is correct.
\begin{theorem}
\label{thm:soundness-of-amguH}
Let $\langle h, p \rangle \in H \times P$
and $(x \mapsto t) \in \Bind$, where $\{x\} \union \vars(t) \sseq \VI$.
Let also $\sigma \in \concrH(h) \inters \concrP(p)$ and
\(
  h' = \amguH\bigl(
               \langle h, p \rangle,
               x \mapsto t
             \bigr)
\).
Then
\[
  \tau \in \mgs\bigl(\sigma \union \{ x = t \}\bigr)
    \implies
      \tau \in \concrH(h').
\]
\end{theorem}

Abstract projection on the composite domain $H \times P$
is much simpler than abstract unification,
because in this case there is no interaction between
the two components of the abstract domain.

\begin{definition} \summary{(Abstract projection on $H \times P$.)}
\label{def:projH}
The function $\fund{\projH}{H \times \VI}{H}$ captures the effects,
on the $H$ component, of projecting away a variable.
For each $h \in H$ and $x \in \VI$,
\[
  \projH(h, x)
    \defeq
      h \union \{ x \}.
\]
The abstract variable projection function
$\fund{\proj}{(H \times P) \times \VI}{H \times P}$,
for any $\langle h, p \rangle \in H \times P$
and $x \in \VI$, is given by
\[
  \proj\bigl(
         \langle h, p \rangle,
         x
       \bigr)
    \defeq
      \bigl\langle
        \projH(h, x),
        \projP(p, x)
      \bigr\rangle.
\]
\end{definition}
As a consequence, as far as the $H$ component is concerned,
the correctness of the projection function does not depend on
the assumption on $\projP$ as specified
in Definition~\ref{def:predicates-and-functs-P}.
\begin{theorem}
\label{thm:soundness-of-projH}
Let $x \in \VI$, $h \in H$ and $\sigma \in \concrH(h)$.
Then
\[
  \tau \in \projectD x \st \{ \sigma \}
    \implies
      \tau \in \concrH\bigl( \projH(h, x) \bigr).
\]
\end{theorem}

We do not consider the disjunction and conjunction operations here.
The implementation (and therefore proof of correctness) for
disjunction is straightforward and omitted.
The implementation of \emph{independent conjunction}
where the descriptions are renamed apart is also straightforward.
On the other hand, full conjunction, which is only needed
for a top-down analysis framework, can be approximated by combining
unification and independent conjunction, obtaining a correct
(although possibly less precise) analysis.

Several abstract domains for sharing analysis can be used
to implement the parameter component $P$.
As a basic implementation, one could consider the well-known
set-sharing domain of Jacobs and Langen~\cite{JacobsL89}.
In such a case, most of the required correctness results
have already been established in~\cite{HillBZ02TPLP}.
Note however that, since no freeness and linearity information
is recorded in the plain set-sharing domain, some of the predicates
of Definition~\ref{def:predicates-and-functs-P}
need to be grossly approximated.
For instance, the predicate $\gfree_p$ will provide useful
information only when applied to an argument that is known to be
definitely ground.
Another possibility would be to use the domain based on pair-sharing,
definite groundness and definite linearity described in \cite{King00}.
A more precise choice is constituted by the $\SFL$ domain
(an acronym standing from Set-sharing plus Freeness plus Linearity)
introduced in \cite{HillZB01TR,Zaffanella01th}.
Even in this case, all the non-trivial correctness results
have already been proved.
In particular, in~\cite{HillZB04TPLP,Zaffanella01th} it is shown
that the abstraction function satisfies the requirement
of Definition~\ref{def:P} and that the abstract unification operator
is correct with respect to rational-tree unification.
In order to better highlight the generality of our specification
of the sharing component $P$, the instantiation of $P$ to $\SFL$
is presented in Appendix~\ref{sec:implementing-P}.
Notice that the quest for more precision does not end with $\SFL$:
a number of possible precision improvements are presented and discussed in
\cite{BagnaraZH00,BagnaraZH05TPLP}.

\section{Finite-Tree Dependencies}
\label{sec:finite-tree-dependencies}

The precision of the finite-tree
analysis based on $H \times P$ is highly dependent
on the precision of the generic component $P$. As  explained before,
the information provided by $P$ on groundness, freeness, linearity,
and sharing of variables is exploited, in the combination $H \times P$,
to circumscribe as much as possible
the creation and propagation of cyclic terms.
However, finite-tree analysis can also benefit from other kinds
of relational information.
In particular, we now show how \emph{finite-tree dependencies}
allow a positive propagation of finiteness information.

Let us consider the finite terms $t_1 = f(x)$, $t_2 = g(y)$,
and $t_3 = h(x,y)$:
it is clear that,
for each assignment of rational terms to $x$ and $y$,
$t_3$ is finite if and only if $t_1$ and $t_2$ are so.
We can capture this by the Boolean formula
$t_3 \piff (t_1 \land t_2)$.\footnote{The introduction of such Boolean
formulas, called \emph{dependency formulas}, is originally due to
P.~W.~Dart~\cite{Dart91}.}
The reasoning is based on the following facts:
\begin{enumerate}
\item
$t_1$, $t_2$, and $t_3$ are finite terms, so that the finiteness
of their instances depends only on the finiteness of the terms
that take the place of $x$ and $y$.
\item
$\vars(t_3) \Sseq \vars(t_1) \union \vars(t_2)$,
that is, $t_3$ \emph{covers} both $t_1$ and $t_2$;
this means that, if an assignment to the variables of $t_3$
produces a finite instance of $t_3$, that very assignment
will necessarily result in finite instances of $t_1$ and $t_2$.
Conversely, an assignment producing non-finite instances of
$t_1$ or $t_2$ will forcibly result in a non-finite instance of $t_3$.
\item
Similarly, $t_1$ and $t_2$, taken together, cover $t_3$.
\end{enumerate}
The important point to notice is that this dependency
will keep holding for any
further simultaneous instantiation of $t_1$, $t_2$, and $t_3$.
In other words, such dependencies are preserved by forward computations
(which proceed by consistently instantiating program variables).

Consider $x \mapsto t \in Bind$ where $t \in \HTerms$
and $\vars(t) = \{ y_1, \ldots, y_n \}$.
After this binding has been successfully applied,
the destinies of $x$ and $t$ concerning term-finiteness are
tied together: forever.
This tie can be described by the dependency formula
\begin{equation}
\label{eq:generic-bind-dependency-formula}
  x \piff (y_1 \land \cdots \land y_n),
\end{equation}
meaning that $x$ will be bound to a finite term
if and only if $y_i$ is bound to a finite term,
for each $i = 1$, \dots,~$n$.
While the dependency expressed by (\ref{eq:generic-bind-dependency-formula})
is a correct description of any computation state following the
application of the binding $x \mapsto t$, it is not as precise as
it could be.  Suppose that $x$ and $y_k$ are indeed the same variable.
Then (\ref{eq:generic-bind-dependency-formula}) is logically equivalent
to
\begin{equation}
\label{eq:imprecise-dependency-formula}
  x
    \pimplies
      (y_1 \land \cdots \land y_{k-1} \land y_{k+1} \land \cdots \land y_n).
\end{equation}
Although this is correct
---whenever $x$ is bound to a finite term, all the other variables
will be bound to finite terms--- it misses the point
that $x$ has just been bound, irrevocably, to a non-finite term:
no forward computation can change this.
Thus, the implication (\ref{eq:imprecise-dependency-formula}) holds vacuously.
A more precise and correct description for the state of affairs
caused by the cyclic binding is, instead,
the negated atom $\neg x$,
whose intuitive reading is ``$x$ is not (and never will be) finite.''

We are building an abstract domain for finite-tree dependencies
where we are making the deliberate choice of including only information
that cannot be withdrawn by forward computations.
The reason for this choice is that we want the concrete constraint
accumulation process to be paralleled, at the abstract level,
by another constraint accumulation process:
logical conjunction of Boolean formulas.
For this reason, it is important to distinguish between
\emph{permanent} and \emph{contingent} information.
Permanent information, once established for a program point $p$,
maintains its validity in all points that follow $p$
in any forward computation.
Contingent information, instead, does not carry its validity
beyond the point where it is established.
An example of contingent information is given by the $h$ component
of $H \times P$:
having $x \in h$ in the description of some program
point means that $x$ is definitely bound to a finite term
\emph{at that point}; nothing is claimed about the finiteness
of $x$ at later program points and, in fact, unless $x$ is ground,
$x$ can still be bound to a non-finite term.
However, if at some program point $x$ is finite and ground,
then $x$ will remain finite.
In this case we will ensure our Boolean dependency formula
entails the positive atom $x$.

At this stage, we already know something about the abstract domain
we are designing.
In particular, we have positive and negated atoms,
the requirement of describing program predicates of any arity
implies that arbitrary conjunctions of these atomic formulas
must be allowed and, finally, it is not difficult to observe
that the merge-over-all-paths operation \cite{CousotC77}
will be logical disjunction, so that the domain will have to
be closed under this operation.
This means that the carrier of our domain must be able to express
any Boolean function over the finite set $\VI$ of the variables of
interest: $\Bfun$ is the carrier.

\begin{definition} \summary{($\fund{\concrFD}{\Bfun}{\wp(\RSubst)}$.)}
\label{def:concrFD}
The function $\fund{\hval}{\RSubst}{\Bval}$ is defined,
for each $\sigma \in \RSubst$ and each $x \in \VI$,
by
\[
  \hval(\sigma)(x) = 1
    \quad\defiff\quad
      x \in \hvars(\sigma).
\]
The concretization function $\fund{\concrFD}{\Bfun}{\wp(\RSubst)}$
is defined, for $\phi \in \Bfun$, by
\[
  \concrFD(\phi)
    \defeq
      \Bigl\{\,
        \sigma \in \RSubst
      \Bigm|
        \forall \tau \in \down\sigma \itc
          \phi\bigl(\hval(\tau)\bigr) = 1
      \,\Bigr\}.
\]
\end{definition}

The domain of positive Boolean functions $\Pos$ used, among other things,
for groundness analysis is so popular that our use of the domain
$\Bfun$ deserves some further comments.
For the representation of finite-tree dependencies, the presence
in the domain of negative functions such as $\neg x$,
meaning that $x$ is bound to an infinite term,
is an important feature.
One reason why it is so is that knowing about definite non-finiteness
can improve the information on definite finiteness.
The easiest example goes as follows: if we know that either $x$ or $y$
is finite (i.e., $x \lor y$) and we know that $x$ is \emph{not} finite
(i.e., $\neg x$), then we can deduce that $y$ must be finite
(i.e.,~$y$).
It is important to observe that this reasoning can be applied,
verbatim, to groundness: a knowledge of non-groundness may improve
groundness information.
The big difference is that non-finiteness is information of the
permanent kind while non-groundness is only contingent.
As a consequence, a knowledge of finiteness and non-finiteness can
be monotonically accumulated along computation paths
by computing the logical conjunction of Boolean formulae.
An approach where groundness and non-groundness information
is represented by elements of $\Bfun$ would need to use a much
more complex operation and significant extra information
to correctly model the constraint accumulation process.

The other reason why the presence of negative functions
in the domain is beneficial is efficiency.
The most efficient implementations of $\Pos$ and $\Bfun$,
such as the ones described in \cite{ArmstrongMSS98,BagnaraS99},
are based on Reduced Ordered Binary Decision Diagrams
(ROBDD) \cite{Bryant92}.
While an ROBDD representing the imprecise information given
by the formula (\ref{eq:imprecise-dependency-formula})
has a worst case complexity that is exponential in $n$,
the more precise formula $\neg x$ has constant complexity.

The following theorem shows how most of the operators
needed to compute the concrete semantics of a logic program
can be correctly approximated on the abstract domain $\Bfun$.
Notice how the addition of equations is modeled by logical conjunction
and projection of a variable is modeled by existential quantification.
\begin{theorem}
\label{thm:concrFD-properties}
Let $\Sigma, \Sigma_1, \Sigma_2 \in \wp(\RSubst)$ and
$\phi, \phi_1, \phi_2 \in \Bfun$
be such that
$\concrFD(\phi) \Sseq \Sigma$,
$\concrFD(\phi_1) \Sseq \Sigma_1$, and
$\concrFD(\phi_2) \Sseq \Sigma_2$.
Let also $(x \mapsto t) \in \Bind$,
where $\{x\} \union \vars(t) \sseq \VI$.
Then the following hold:
\begin{subequations}
\renewcommand{\theequation}{\ref{thm:concrFD-properties}\alph{equation}}
\begin{align}
\label{eq:concrFD-properties:binding}
  \concrFD\Bigl(x \piff \bigland \vars(t)\Bigr)
    &\Sseq
      \bigl\{ \{ x \mapsto t \} \bigr\}; \\
\label{eq:concrFD-properties:neg-x}
  \concrFD(\neg x)
    &\Sseq
      \bigl\{ \{ x \mapsto t \} \bigr\},
        \text{ if $x \in \vars(t)$;} \\
\label{eq:concrFD-properties:x}
  \concrFD(x)
    &\Sseq
      \bigl\{\,
        \sigma \in \RSubst
      \bigm|
        x \in \gvars(\sigma) \inters \hvars(\sigma)
      \,\bigr\}; \\
\label{eq:concrFD-properties:and}
  \concrFD(\phi_1 \land \phi_2)
    &\Sseq
      \bigl\{\,
        \mgs(\sigma_1 \union \sigma_2)
      \bigm|
        \sigma_1 \in \Sigma_1, \sigma_2 \in \Sigma_2
      \,\bigr\}; \\
\label{eq:concrFD-properties:or}
  \concrFD(\phi_1 \lor \phi_2)
    &\Sseq
      \Sigma_1 \union \Sigma_2; \\
\label{eq:concrFD-properties:exists}
  \concrFD(\exists x \st \phi)
    &\Sseq
      \projectD x \st \Sigma.
\end{align}
\end{subequations}
\end{theorem}
Cases~(\ref{eq:concrFD-properties:binding}),
(\ref{eq:concrFD-properties:neg-x}),
and~(\ref{eq:concrFD-properties:and})
of Theorem~\ref{thm:concrFD-properties}
ensure that the following  definition of $\amguFD$
provides a correct approximation on $\Bfun$
of the concrete unification of rational trees.
\begin{definition}
\label{def:amguFD}
The function
\(
  \fund{\amguFD}{\Bfun \times \Bind}{\Bfun}
\)
captures the effects of a binding on a finite-tree dependency formula.
Let $\phi \in \Bfun$ and $(x \mapsto t) \in \Bind$
be such that $\{x\} \union \vars(t) \sseq \VI$.
Then
\[
  \amguFD(\phi, x \mapsto t)
    \defeq
      \begin{cases}
        \phi \land \bigl(x \piff \bigland \vars(t) \bigr),
            &\text{if $x \notin \vars(t)$;} \\
        \phi \land \neg x,
            &\text{otherwise.}
      \end{cases}
\]
\end{definition}
Other semantic operators, such as the consistent renaming of variables,
are very simple and omitted for the sake of brevity.

The next result shows how finite-tree dependencies
may improve the finiteness information
encoded in the $h$ component of the domain $H \times P$.
\begin{theorem}
\label{thm:concrH-concrFD}
Let $h \in H$ and $\phi \in \Bfun$.
Let also
\(
  h' \defeq \truev\bigl( \phi \land \bigland h \bigr)
\).
Then
\[
  \concrH(h) \inters \concrFD(\phi)
    = \concrH(h') \inters \concrFD(\phi).
\]
\end{theorem}

\begin{example}
Consider the following program,
where it is assumed that the only ``external'' query is
`\verb+?- r(X, Y)+':
\begin{centeredprogram}
p(X, Y) :- X = f(Y, \_). \\
q(X, Y) :- X = f(\_, Y). \\
r(X, Y) :- p(X, Y), q(X, Y), acyclic\_term(X).
\end{centeredprogram}
Then the predicate \verb+p/2+ in the
clause defining \verb+r/2+ will be called
with \verb+X+ and \verb+Y+ both unbound.
Computing on the abstract domain $H \times P$ gives us
the finiteness description $h_p = \{x,y\}$,
expressing the fact that both \verb+X+ and \verb+Y+
are  bound to finite terms.
Computing on the finite-tree dependencies domain $\Bfun$,
gives us the Boolean formula $\phi_p = x \pimplies y$
(\verb+Y+ is finite if \verb+X+ is so).

Considering now the call to the predicate \verb+q/2+, we note that,
since variable \verb+X+ is already bound to a non-variable term
sharing with \verb+Y+, all the finiteness information encoded
by $H$ will be lost (i.e., $h_q = \emptyset$).
So, both \verb+X+ and \verb+Y+ are detected as possibly cyclic.
However, the finite-tree dependency information is preserved,
since we have
$\phi_q = (x \pimplies y) \land (x \pimplies y) = x \pimplies y$.

Finally, consider the effect
of the abstract evaluation of \verb+acyclic_term(X)+.
On the $H \times P$ domain we can only infer that
variable \verb+X+ cannot be bound to an infinite term,
while \verb+Y+ will be still considered as possibly cyclic,
so that $h_r = \{x\}$.
On the domain $\Bfun$ we can just confirm
that the finite-tree dependency computed so far still holds,
so that $\phi_r = x \pimplies y$
(no stronger finite-tree dependency can be inferred,
since the finiteness of \verb+X+ is only contingent).
Thus, by applying the result of \textup{Theorem~\ref{thm:concrH-concrFD}},
we can recover the finiteness of \verb+Y+:
\[
  h'_r = \truev\Bigl(\phi_r \land \bigland h_r\Bigr)
       = \truev\bigl((x \pimplies y) \land x\bigr)
       = \truev(x \land y)
       = \{x,y\}.
\]
\end{example}

Information encoded in $H \times P$ and $\Bfun$
is not completely orthogonal and
the following result provides a kind of consistency check.

\begin{theorem}
\label{thm:concrH-concrFD-consistency}
Let $h \in H$ and $\phi \in \Bfun$.
Then
\[
  \concrH(h) \inters \concrFD(\phi) \neq \emptyset
    \quad \implies \quad
      h \inters \falsev\Bigl( \phi \land \bigland h \Bigr) = \emptyset.
\]
\end{theorem}
Note however that, provided the abstract operators
%computing on the two components
are correct, the computed descriptions
will always be mutually consistent, unless $\phi = \bot$.

\section{Groundness Dependencies}
\label{sec:groundness-dependencies}

Since information about the groundness of variables
is crucial for many applications,
it is natural to consider a static analysis domain
including both a finite-tree and a groundness component.
In fact, any reasonably precise implementation
of the parameter component $P$ of the abstract domain
specified in Section~\ref{sec:finite-tree-domain} will include some kind
of groundness information.\footnote{One could define $P$
so that it explicitly contains
the abstract domain $\Pos$. Even when this is not the case,
it should be noted that, as soon as the parameter $P$ includes
the set-sharing domain of Jacobs and Langen~\cite{JacobsL92},
then it will subsume the groundness information captured
by the domain $\Def$~\cite{CodishSS99,CortesiFW98}.}
We highlight similarities, differences and connections relating
the domain $\Bfun$ for finite-tree dependencies
to the abstract domain $\Pos$ for groundness dependencies.
Note that these results also hold
when considering a combination of $\Bfun$
with the groundness domain $\Def$~\cite{ArmstrongMSS98}.

We first define how elements of $\Pos$ represent sets of substitutions
in rational solved form.
\begin{definition} \summary{($\fund{\concrGD}{\Pos}{\wp(\RSubst)}$.)}
\label{def:concrGD}
The function $\fund{\gval}{\RSubst}{\Bval}$ is defined as follows,
for each $\sigma \in \RSubst$ and each $x \in \VI$:
\[
  \gval(\sigma)(x) = 1
    \quad\defiff\quad
      x \in \gvars(\sigma).
\]
The concretization function $\fund{\concrGD}{\Pos}{\wp(\RSubst)}$
is defined, for each $\psi \in \Pos$,
\[
  \concrGD(\psi)
    \defeq
      \Bigl\{\,
        \sigma \in \RSubst
      \Bigm|
        \forall \tau \in \down\sigma \itc
          \psi\bigl(\gval(\tau)\bigr) = 1
      \,\Bigr\}.
\]
\end{definition}

The following is a simple variant of the standard abstract unification
operator for groundness analysis over finite-tree domains:
the only difference concerns the case of cyclic bindings
\cite{Bagnara97th}.
\begin{definition}
\label{def:amguGD}
The function
\(
  \fund{\amguGD}{\Pos \times \Bind}{\Pos}
\)
captures the effects of a binding on a groundness dependency formula.
Let $\psi \in \Pos$ and $(x \mapsto t) \in \Bind$
be such that $\{x\} \union \vars(t) \sseq \VI$.
Then
\[
  \amguGD(\psi, x \mapsto t)
    \defeq
      \psi \land \Bigl(
                   x \piff \bigland \bigl(
                                      \vars(t) \setdiff \{x\}
                                    \bigr)
                 \Bigr).
\]
\end{definition}

The next result shows how, by exploiting the finiteness component $H$,
the finite-tree dependencies ($\Bfun$) component
and the groundness dependencies ($\Pos$) component
can improve each other.
\begin{theorem}
\label{thm:concrH-concrFD-concrGD}
Let $h \in H$, $\phi \in \Bfun$ and $\psi \in \Pos$.
Let also $\phi' \in \Bfun$ and $\psi' \in \Pos$ be defined as
$\phi' = \exists \VI \setdiff h \st \psi$ and
$\psi' = \pos(\exists \VI \setdiff h \st \phi)$.
Then
\begin{subequations}
\renewcommand{\theequation}{\ref{thm:concrH-concrFD-concrGD}\alph{equation}}
\begin{align}
\label{eq:concrH-concrFD-->>concrGD}
  \concrH(h) \inters \concrFD(\phi) \inters \concrGD(\psi)
    &= \concrH(h)
        \inters \concrFD(\phi)
          \inters \concrGD(\psi \land \psi');\\
\label{eq:concrH-concrGD-->>concrFD}
  \concrH(h) \inters \concrFD(\phi) \inters \concrGD(\psi)
    &= \concrH(h)
        \inters \concrFD(\phi \land \phi')
          \inters \concrGD(\psi).
\end{align}
\end{subequations}
\end{theorem}
Moreover,
even without any knowledge
of the $H$ component,  combining Theorem~\ref{thm:concrH-concrFD}
and Eq.~(\ref{eq:concrH-concrFD-->>concrGD}), the
groundness dependencies component can be improved.

\begin{theorem}
\label{thm:concrFD-->>concrGD}
Let $\phi \in \Bfun$ and $\psi \in \Pos$.
Then
\[
  \concrFD(\phi) \inters \concrGD(\psi)
    = \concrFD(\phi)
          \inters \concrGD\Bigl(\psi \land \bigland \truev(\phi)\Bigr).
\]
\end{theorem}

The following example shows that,
when computing on rational trees,
finite-tree dependencies may provide groundness information
that is not captured by the usual approaches.

\begin{example}
Consider the program:
\begin{centeredprogram}
p(a, Y). \\
p(X, a). \\
q(X, Y) :- p(X, Y), X = f(X, Z).
\end{centeredprogram}
The abstract semantics of \verb+p/2+,
for both finite-tree and groundness dependencies, is
$\phi_p = \psi_p = x \lor y$.
The finite-tree dependency for \verb+q/2+ is
$\phi_q = (x \lor y) \land \neg x = \neg x \land y$.
Using \textup{Definition~\ref{def:amguGD}},
the groundness dependency for \verb+q/2+ is
\begin{align*}
  \psi_q
    &= \exists z \st \bigl((x \lor y) \land (x \piff z)\bigr)
    = x \lor y.
\intertext{%
This can be improved, using \textup{Theorem~\ref{thm:concrFD-->>concrGD}},
to
}
  \psi'_q &= \psi_q \land \bigland \truev(\phi_q) = y.
\end{align*}
\end{example}

It is worth noticing that the groundness information can be improved
regardless of whether, like $\Pos$, the groundness domain captures
disjunctive
information:  groundness information represented
by the less expressive domain $\Def$ \cite{ArmstrongMSS98} can
be improved as well.
The next example illustrates this point.
\begin{example}
Consider the following program:
\begin{centeredprogram}
p(a, a). \\
p(X, Y) :- X = f(X, \_). \\
q(X, Y) :- p(X, Y), X = a.
\end{centeredprogram}
Consider the predicate \verb+p/2+.
Concerning finite-tree dependencies,
the abstract semantics of \verb+p/2+ is expressed by the Boolean formula
\(
  \phi_p = (x \land y) \lor \neg x = x \pimplies y
\)
(\verb+Y+ is finite if \verb+X+ is so).
In contrast, the $\Pos$-groundness abstract semantics of \verb+p/2+
is a plain ``don't know'': the Boolean formula
\(
  \psi_p = (x \land y) \lor \top = \top
\).
In fact, the groundness of \verb+X+ and \verb+Y+
can be completely decided by the call-pattern of \verb+p/2+.

Consider now the predicate \verb+q/2+.
The finiteness semantics of \verb+q/2+ is given by
$\phi_q = (x \pimplies y) \land x = x \land y$,
whereas the $\Pos$ formula expressing groundness dependencies is
$\psi_q = \top \land x = x$.
By \textup{Theorem~\ref{thm:concrFD-->>concrGD}}, we obtain
\[
  \psi'_q = \psi_q \land \bigland \truev(\phi_q) = x \land y,
\]
therefore recovering the groundness of variable $y$.
\end{example}

Since better groundness information, besides being useful in itself,
may also improve the precision of many other analyses
such as sharing~\cite{BagnaraZH00,BagnaraZH05TPLP,CodishSS99},
the reduction steps given by
Theorems~\ref{thm:concrH-concrFD-concrGD}
and~\ref{thm:concrFD-->>concrGD} can trigger
improvements to the precision of other components.
Theorem~\ref{thm:concrH-concrFD-concrGD} can also be exploited
to recover precision after the application of a widening operator
on either the groundness dependencies or the finite-tree dependencies
component.

\section{Experimental Results}
\label{sec:exp-results}

The work described here has been experimentally evaluated in the
framework provided by \china{}~\cite{Bagnara97th},
a data-flow analyzer for constraint logic languages
(i.e., ISO Prolog, \CLPR{}, \texttt{clp(FD)} and so forth).
\china{} performs bottom-up analysis deriving information on both
call-patterns and success-patterns by means of program transformations
and optimized fixpoint computation techniques.\footnote{More
precisely, \china{} uses a variation of the \emph{Magic Templates}
algorithm \cite{Ramakrishnan88}, in order to obtain goal-dependent
information, and a sophisticated chaotic iteration strategy proposed
in \cite{Bourdoncle93,Bourdoncle93th} (recursive fixpoint iteration on
the weak topological ordering defined by partitioning of the call
graph into strongly-connected subcomponents).}
An abstract description is computed for the call- and success-patterns
for each predicate defined in the program.

We implemented and compared the three domains
$\Pattern(P)$,
$\Pattern(H \times P)$ and
$\Pattern(\Bfun \times H \times P)$,%
\footnote{For ease of notation, the domain names are shortened
to P, H and B, respectively.}
where the parameter component $P$ has been instantiated to
the domain $\Pos \times \PSDFL$~\cite{BagnaraZH00,HillZB04TPLP,Zaffanella01th}
for tracking groundness, freeness, linearity
and (non-redundant) set-sharing information.
The $\Pattern(\cdot)$ operator~\cite{BagnaraHZ00} further upgrades
the precision of its argument by adding explicit structural information.
Note that the analyzer tracks the finiteness of the terms that can
be bound to those abstract variables occurring as leaves in the
acyclic term structure computed by the $\Pattern(\cdot)$ component;
therefore, in order to show that an abstract variable is definitely
bound to a finite term, the basic domain $\Pattern(P)$ has to prove
that this variable is definitely free.%
\footnote{Put in other words, by considering just the variables
occurring inside the pattern structure, we systematically disregard
those cases when the basic domain is able to prove that a particular
argument position is definitely bound to a finite and ground term
such as $f(a)$. Clearly, the same approach is consistently adopted
when considering the more accurate analysis domains.}

Concerning the $\Bfun$ component, the implementation
was straightforward,
since all the techniques described in \cite{BagnaraS99}
(and almost all the code, including the widenings)
was reused unchanged, obtaining comparable efficiency.
As a consequence, most of the implementation effort was in
the coding of the abstract operators on the $H$ component and
in the reduction processes between the different components.
A key choice, in this sense, is \emph{when} the reduction steps given in
Theorems~\ref{thm:concrH-concrFD}
and~\ref{thm:concrH-concrFD-concrGD} should be applied.
When striving for maximum precision, a trivial strategy is
to perform reductions immediately
after any application of any abstract operator.
This is how predicates like \verb+acyclic_term/1+
should be handled: after adding the variables of the argument
to the $H$ component, the reduction process is applied to
propagate the new information to all domain components.
However, such an approach turns out to be unnecessarily inefficient.
In fact, the next result shows that
Theorems~\ref{thm:concrH-concrFD}
and~\ref{thm:concrH-concrFD-concrGD}
cannot lead to a precision improvement
if applied just after the abstract evaluation of
the merge-over-all-paths or the existential quantification operations
(provided the initial descriptions are already reduced).

\begin{theorem}
\label{thm:concrH-concrFD-concrGD-subsumed}
Let $x \in \VI$, $h, h' \in H$ $\phi,\phi' \in \Bfun$
and $\psi,\psi' \in \Pos$ and suppose that
$\concrH(h) \inters \concrFD(\phi) \neq \emptyset$.
Let
\begin{align*}
  h_1    &\defeq h \inters h',
& \phi_1 &\defeq \phi \lor \phi',
& \psi_1 &\defeq \psi \lor \psi', \\
  h_2 &\defeq \projH(h, x),
& \phi_2 &\defeq \exists x \st \phi,
& \psi_2 &\defeq \exists x \st \psi.
\end{align*}
Let also
\begin{align*}
  h     &\Sseq \truev\Bigl( \phi \land \bigland h \Bigr),
& \phi  &\models (\exists \VI \setdiff h \st \psi),
& \psi  &\models \pos( \exists \VI \setdiff h \st \phi), \\
\displaybreak[1]
  h'    &\Sseq \truev\Bigl( \phi' \land \bigland h' \Bigr),
& \phi' &\models (\exists \VI \setdiff h' \st \psi'),
& \psi' &\models \pos(\exists \VI \setdiff h' \st \phi'). \\
\intertext{%
Then, for $i = 1$,~$2$,
}
  h_i    &\Sseq \truev\Bigl( \phi_i \land \bigland h_i \Bigr),
& \phi_i &\models (\exists \VI \setdiff h_i \st \psi_i),
& \psi_i &\models \pos(\exists \VI \setdiff h_i \st \phi_i).
\end{align*}
\end{theorem}

\begin{table*}[t]
\centering

\begin{tabular}{||c||r|r|r||}
\hhline{|t:=:t:=t:=t:=:t|}
 \multicolumn{1}{||c||}{Prec. class}
 & \multicolumn{1}{c|}{P}
 & \multicolumn{1}{c|}{H}
 & \multicolumn{1}{c||}{B} \\
\hhline{|:=::=:=:=:|}
$\phantom{10 \leq \mathord{}}p = 100$ & 2 & 84 & 86 \\
\hhline{||-||-|-|-||}
$80 \leq p < 100$ & 1 & 31 & 36 \\
\hhline{||-||-|-|-||}
$60 \leq p < 80\phantom{1}$ & 7 & 26 & 23\\
\hhline{||-||-|-|-||}
$40 \leq p < 60\phantom{1}$ & 6 & 41 & 40\\
\hhline{||-||-|-|-||}
$20 \leq p < 40\phantom{1}$ & 47 & 47 & 46 \\
\hhline{||-||-|-|-||}
$\phantom{1}0 \leq p < 20\phantom{1}$ & 185 & 19 & 17 \\
\hhline{|b:=:b:=b:=b:=:b|}
\end{tabular}

\bigskip
\bigskip

\begin{tabular}{||c||r|r||}
\hhline{|t:=:t:=t=:t|}
 \multicolumn{1}{||c||}{Prec. improvement}
 & \multicolumn{1}{c|}{$\textrm{P} \rightarrow \textrm{H}$}
 & \multicolumn{1}{c||}{$\textrm{H} \rightarrow \textrm{B}$} \\
\hhline{|:=::=:=:|}
$\phantom{-20 < \mathord{}} i > 20\phantom{-}$ & 185 & 4 \\
\hhline{||-||-|-||}
$\phantom{-}10 < i \leq 20\phantom{-}$ & 31 & 3 \\
\hhline{||-||-|-||}
$\phantom{-1}5 < i \leq 10\phantom{-}$ & 11 & 6 \\
\hhline{||-||-|-||}
$\phantom{-1}2 < i \leq 5\phantom{-0}$ & 4 & 10 \\
\hhline{||-||-|-||}
$\phantom{-1}0 < i \leq 2\phantom{-0}$ & 2 & 24 \\
\hhline{||-||-|-||}
no improvement & 15 & 201 \\
\hhline{|b:=:b:=b:=:b|}
\end{tabular}

\caption{The precision on finite variables when using P, H and B.}
\label{tab: P vs H vs B}
\end{table*}

A goal-dependent analysis was run for all the programs
in our benchmark suite.\footnote{The suite comprises all the logic
programs we have access to (including everything we could find by
systematically dredging the Internet):
364 programs, 24 MB of code, 800 K lines.
Besides classical benchmarks,
several real programs of respectable size are included,
the largest one containing 10063 clauses in 45658 lines of code.
The suite also comprises a few synthetic benchmarks,
which are artificial programs explicitly constructed to stress
the capabilities of the analyzer and of its abstract domains
with respect to precision and/or efficiency.
The interested reader can find more information
at the URI \texttt{http://www.cs.unipr.it/China/}.
}
For 116 of them, the analyzer detects that the program in not amenable
to goal-dependent analysis, either because the entry points are unknown
or because the program uses builtins in a way that every predicate can
be called with any call-pattern, so that the analysis provides results
that are so imprecise to be irrelevant.
The precision results for the remaining 248 programs are summarized in
Table~\ref{tab: P vs H vs B}.
Here, the precision is measured as the percentage of the total number of
variables that the analyser can show to be finite.
Two alternative views are provided.

In the first view, each column is labeled by an analysis domain
and each row is labeled by a precision interval.
For instance,
the value `31' at the intersection of column `H'
and row `$80 \leq p < 100$'
is to be read as
``\emph{for 31 benchmarks, the percentage $p$ of the total number of
variables that the analyzer can show to be finite using the
domain \textrm{H} is between 80\% and 100\%.}''

The second view provides a better picture
of the precision \emph{improvements} obtained
when moving from P to H (in the column `$\textrm{P} \rightarrow \textrm{H}$')
and from H to B (in the column `$\textrm{H} \rightarrow \textrm{B}$').
For instance,
the value `10' at the intersection of
column `$\textrm{H} \rightarrow \textrm{B}$' and row `$2 < i \leq 5$'
is to be read as
``\emph{when moving from \textrm{H} to \textrm{B},
for 10 benchmarks the improvement $i$ in the percentage of the
total number of variables shown to be finite
was between 2\% and 5\%.}''

It can be seen from Table~\ref{tab: P vs H vs B} that,
even though the H domain is remarkably precise,
the inclusion of the $\Bfun$ component allows for
a further, and sometimes significant, precision improvement
for a number of benchmarks.
It is worth noting that the current implementation of \china{}
does not yet fully exploit the finite-tree dependencies arising
when evaluating many of the built-in predicates,
therefore incurring an avoidable precision loss.
We are working on this issue
and we expect that the specialized implementation
of the abstract evaluation of some built-ins
will result in more and better precision improvements.
The experimentation has also shown that, in practice,
the $\Bfun$ component does not improve the groundness information.

Concerning efficiency, our experimentation shown that the techniques
we propose are really practical.  The total analysis
time for the 248 programs for which we give precision results
in Table~\ref{tab: P vs H vs B} is 596 seconds for P, 602 seconds for H,
and 1211 seconds for B.\footnote{On a PC system equipped with
an Athlon XP 2800 CPU, 1 GB of RAM memory and running GNU/Linux.}
It should be stressed that, as mentioned before,
the implementation of $\Bfun$ was derived in a straightforward way from the
one of $\Pos$ described in \cite{BagnaraS99}.  We believe that a different
tuning of the widenings we employ in that component could reduce the gap
between the efficiency of H and the one of B.

\section{Conclusion}
\label{sec:conclusion}

Several modern logic-based languages offer a computation domain based
on rational trees.
On the one hand, the use of such trees is encouraged by the
possibility of using efficient and correct unification algorithms and
by an increase in expressivity.
On the other hand, these gains are countered by the extra problems
rational trees bring with themselves and that can be summarized as
follows: several built-ins, library predicates, program analysis and
manipulation techniques are only well-defined for program fragments
working with finite trees.

As a consequence, those applications that exploit rational trees
tend to do so in a very controlled way, that is,
most program variables can only be bound to finite terms.
By detecting the program variables that may be bound to infinite terms
with a good degree of accuracy, we can significantly reduce
the disadvantages of using rational trees.

In this paper we have proposed an abstract-interpretation based solution to
this problem, where the composite abstract domain $H \times P$
allows tracking of the creation and propagation of infinite terms.
Even though this information is crucial to any finite-tree analysis,
propagating the guarantees of finiteness that come from several
built-ins (including those that are explicitly provided to test
term-finiteness) is also important.
Therefore, we have introduced a domain of Boolean functions
$\Bfun$ for finite-tree dependencies which,
when coupled to the domain $H \times P$, can enhance its expressive power.
Since $\Bfun$ has many similarities with the domain $\Pos$
used for groundness analysis, we have investigated how these two
domains relate to each other and, in particular,
the synergy arising from their combination
in the ``global'' domain of analysis.

\section*{Acknowledgment}

We would like to express our gratitude
to the Journal referees for their useful comments that have
helped improve the final versions of the paper.

\hyphenation{ Ba-gna-ra Bie-li-ko-va Bruy-noo-ghe Common-Loops DeMich-iel
  Dober-kat Er-vier Fa-la-schi Fell-eisen Gam-ma Gem-Stone Glan-ville Gold-in
  Goos-sens Graph-Trace Grim-shaw Her-men-e-gil-do Hoeks-ma Hor-o-witz Kam-i-ko
  Kenn-e-dy Kess-ler Lisp-edit Lu-ba-chev-sky Nich-o-las Obern-dorf Ohsen-doth
  Par-log Para-sight Pega-Sys Pren-tice Pu-ru-sho-tha-man Ra-guid-eau Rich-ard
  Roe-ver Ros-en-krantz Ru-dolph SIG-OA SIG-PLAN SIG-SOFT SMALL-TALK Schee-vel
  Schlotz-hauer Schwartz-bach Sieg-fried Small-talk Spring-er Stroh-meier
  Thing-Lab Zhong-xiu Zac-ca-gni-ni Zaf-fa-nel-la Zo-lo }

\appendix
\newpage
\section{An Instance of the Parameter Domain $P$}
\label{sec:implementing-P}

As discussed in Section~\ref{sec:finite-tree-domain},
several abstract domains for sharing analysis can be used
to implement the parameter component $P$.
We here consider the abstract domain
$\SFL$~\cite{HillZB04TPLP,Zaffanella01th},
integrating the set-sharing domain of Jacobs and Langen
with definite freeness and linearity information.

\begin{definition} \summary{(The set-sharing domain $\SH$.)}
The set $\SH$ is defined by
$\SH \defeq \wp(\SG)$,
where
\(
  \SG \defeq \wp(\VI) \setdiff \{ \emptyset \}
\)
is the set of \emph{sharing groups}.
$\SH$ is ordered by subset inclusion.
\end{definition}

The information about definite freeness and linearity is encoded
by two sets of variables, one for each property.

\begin{definition} \summary{(The domain $\SFL$.)}
Let $F \defeq \wp(\VI)$ and $L \defeq \wp(\VI)$
be partially ordered by reverse subset inclusion.
The domain $\SFL$ is defined by the Cartesian product
\(
  \SFL \defeq \SH \times F \times L
\)
ordered by `$\mathord{\leqSFL}$',
the component-wise extension of the orderings defined on the sub-domains;
the bottom element is
$\botSFL \defeq \langle \emptyset, \VI, \VI \rangle$.
\end{definition}

In the next definition we introduce a few well-known operations
on the set-sharing domain $\SH$.
These will be used to define the operations on the domain $\SFL$.
\begin{definition} \summary{(Abstract operators on $\SH$.)}
\label{def:abs-funcs-SH}
For each $\sh \in \SH$ and each $V \sseq \VI$,
the extraction of the
\emph{relevant component of $\sh$ with respect to $V$}
is given by the function
$\fund{\rel}{\wp(\VI)\times\SH}{\SH}$
defined as
\[
  \rel(V, \sh)
    \defeq
      \{\, S \in \sh \mid S \inters V \neq \emptyset \,\}.
\]

For each $\sh \in \SH$ and each $V \sseq \VI$,
the function $\fund{\irel}{\wp(\VI)\times\SH}{\SH}$
gives the \emph{irrelevant component of $\sh$ with respect to $V$}.
It is defined as
\[
  \irel(V, \sh)
    \defeq \sh \setdiff \rel(V,\sh).
\]

The function
$\fund{(\cdot)^\star}{\SH}{\SH}$,
called \emph{star-union},
is given, for each $\sh \in \SH$, by
\[
  \sh^\star
    \defeq
      \biggl\{\,
        S \in \SG
      \biggm|
        \exists n \geq 1
          \st
            \exists T_1, \ldots, T_n \in \sh
              \st S = \bigunion_{i=1}^n T_i
      \,\biggr\}.
\]

For each $\sh_1, \sh_2 \in \SH$,
the function
$\fund{\bin}{\SH\times\SH}{\SH}$,
called \emph{binary union},
is given by
\[
  \bin(\sh_1, \sh_2)
    \defeq
      \{\,
        S_1 \union S_2
      \mid
        S_1 \in \sh_1,
        S_2 \in \sh_2
      \,\}.
\]

For each $\sh \in \SH$ and each $(x \mapsto t) \in \Bind$,
the function $\fund{\cyclicreduce_x^t}{\SH}{\SH}$
strengthens the sharing set $\sh$ by forcing the coupling
of $x$ with $t$:
\[
  \cyclicreduce_x^t(\sh)
    \defeq
      \irel\bigl(\{x\} \union \vars(t), \sh\bigr)
        \union
          \rel\bigl(\vars(t) \setdiff \{x\}, \sh\bigr).
\]

For each $\sh \in \SH$ and each $x \in \VI$,
the function $\fund{\projSH}{\SH \times \VI}{\SH}$
projects away variable $x$ from $\sh$:
\[
  \projSH(\sh, x)
    \defeq
      \bigl\{ \{x\} \bigr\}
	\union
	  \bigl\{\,
            S \setdiff \{x\}
          \bigm|
            S \in \sh, S \neq \{x\}
          \,\bigr\}.
\]
\end{definition}

It is now possible to define the implementation, on the domain $\SFL$,
of all the predicates and functions specified
in Definition~\ref{def:predicates-and-functs-P}.

\begin{definition} \summary{(Abstract operators on $\SFL$.)}
\label{def:abs-funcs-SFL}
For each $\sfl \in \SFL$ and $s,t \in \HTerms$,
where $\sfl = \langle \sh, f, l \rangle$ and
$\vars(s) \union \vars(t) \sseq \VI$,
let $\sh_s = \rel\bigl(\vars(s), \sh\bigr)$
and $\sh_t = \rel\bigl(\vars(t), \sh\bigr)$.
Then
\begin{align*}
%%% definite independence
  \ind_\sfl(s, t)
    &\defeq
      \bigl( \sh_s \inters \sh_t = \emptyset \bigr); \\
%%% definite groundness
  \ground_\sfl(t)
    &\defeq
      \bigl(\vars(t) \sseq \VI \setdiff \vars(\sh) \bigr); \\
%%% occur linearly
  \occlin_\sfl(y,t)
    &\defeq
      \ground_\sfl(y)
        \lor
          \Bigl(
            \occlin(y, t)
              \land
            (y \in l) \\
    &\qquad \qquad
	      \land
            \forall z \in \vars(t)
              \itc
                \bigl(
                  y \neq z \implies \ind_\sfl(y,z)
                \bigr)
           \Bigr); \\
%%% can only share linearly
   \sharelin_\sfl(s,t)
     &\defeq
       \forall y \in \vars(\sh_s \inters \sh_t)
         \itc \\
     &\qquad \quad
           y \in \vars(s) \implies \occlin_\sfl(y, s)\\
           &\qquad \quad
		   \land
           y \in \vars(t) \implies \occlin_\sfl(y, t); \\
%%% definite freeness
  \free_\sfl(t)
    &\defeq
      \exists y \in \VI \st (y = t) \land (y \in f); \\
%%% ground-or-free
  \gfree_\sfl(t)
    &\defeq
      \ground_\sfl(t) \lor \free_\sfl(t); \\
%%% definite linearity
  \lin_\sfl(t)
    &\defeq
      \forall y \in \vars(t)
        \itc
          \occlin_\sfl(y, t); \\
%% or-linearity
  \orlin_\sfl(s,t)
    &\defeq
      \lin_\sfl(s) \lor \lin_\sfl(t); \\
%%% may share same var
  \sharesamevar_\sfl(s,t)
    &\defeq
      \vars(\sh_s \inters \sh_t); \\
%%% may share with
  \sharewith_\sfl(t)
    &\defeq
      \vars(\sh_t).
\end{align*}

The function $\fund{\amguSFL}{\SFL\times\Bind}{\SFL}$
captures the effects of a binding on an element of $\SFL$.
Let $\sfl = \langle \sh, f, l \rangle \in \SFL$ and
$(x \mapsto t) \in \Bind$, where $\{x\} \union \vars(t) \sseq \VI$.
Let also
\[
  \sh'
    \defeq
      \cyclicreduce_x^t(\sh_{-} \union \sh''),
\]
where
\begin{align*}
  \sh_x
    &\defeq
      \rel\bigl(\{x\}, \sh\bigr),
& \sh_t
    &\defeq
      \rel\bigl(\vars(t), \sh\bigr), \\
  \sh_{xt}
    &\defeq
      \sh_x \inters \sh_t,
& \sh_{-}
    &\defeq
      \irel\bigl(\{x\} \union \vars(t), \sh\bigr),
\end{align*}
%%%%%%%%%
\begin{equation*}
  \sh''
    \defeq
      \begin{cases}
        \bin(\sh_x, \sh_t),
          &\text{if $\free_{\sfl}(x)
                       \lor
                     \free_{\sfl}(t)$;} \\
        \bin\bigl(
              \sh_x \union \bin(\sh_x, \sh_{xt}^\star),
          & \\
        \qquad\qquad
              \sh_t \union \bin(\sh_t, \sh_{xt}^\star)
            \bigr),
          &\text{if $\lin_{\sfl}(x) \land \lin_{\sfl}(t)$;} \\
        \bin(\sh_x^\star, \sh_t),
          &\text{if $\lin_{\sfl}(x)$;} \\
        \bin(\sh_x, \sh_t^\star),
          &\text{if $\lin_{\sfl}(t)$;} \\
        \bin(\sh_x^\star, \sh_t^\star),
          &\text{otherwise.}
      \end{cases}
\end{equation*}

Letting $S_x \defeq \sharewith_\sfl(x)$
and $S_t \defeq \sharewith_\sfl(t)$,
we also define
\begin{align*}
  f'
    &\defeq
      \begin{cases}
        f,
          &\text{if $\free_{\sfl}(x) \land \free_{\sfl}(t)$;} \\
        f \setdiff S_x,
          &\text{if $\free_{\sfl}(x)$;} \\
        f \setdiff S_t,
          &\text{if $\free_{\sfl}(t)$;} \\
        f \setdiff (S_x \union S_t),
          &\text{otherwise;} \\
      \end{cases} \\
\displaybreak[0]
  l'
    &\defeq
      \bigl( \VI \setdiff \vars(\sh') \bigr) \union f' \union l'', \\
\intertext{%
where
}
  l''
    &\defeq
      \begin{cases}
        l \setdiff (S_x \inters S_t),
          &\text{if $\lin_{\sfl}(x)
                       \land
                     \lin_{\sfl}(t)$;} \\
        l \setdiff S_x,
          &\text{if $\lin_{\sfl}(x)$;} \\
        l \setdiff S_t,
          &\text{if $\lin_{\sfl}(t)$;} \\
        l \setdiff (S_x \union S_t),
          &\text{otherwise.} \\
      \end{cases}
\end{align*}
Then
\[
  \amguSFL\bigl(\sfl, x \mapsto t\bigr)
    \defeq
      \langle \sh', f', l' \rangle.
\]

The function $\fund{\projSFL}{\SFL \times \VI}{\SFL}$
correctly captures the operation of projecting away
a variable from an element of $\SFL$.
For each $\sfl \in \SFL$ and $x \in \VI$,
\[
  \projSFL(\sfl, x)
    \defeq
      \begin{cases}
        \botSFL,
          &\text{if $\sfl = \botSFL$;} \\
        \bigl\langle
	  \projSH(\sh, x),
          f \union \{ x \},
          l \union \{ x \}
        \bigr\rangle,
          &\text{if $\sfl = \langle \sh, f, l \rangle \neq \botSFL$.}
      \end{cases}
\]
\end{definition}

Observe that a set-sharing domain such as $\SFL$
is strictly more precise for term finiteness
information than a pair-sharing domain such as
$\PSDFL$~\cite{HillZB04TPLP,Zaffanella01th}
(where the set-sharing component $\SH$ in $\SFL$ is replaced by
the domain $\PSD$ as defined in~\cite{BagnaraHZ02TCS,ZaffanellaHB02TPLP}).
To see this,
consider the abstract evaluation of the binding $x \mapsto y$
and the description $\langle h, \sfl \rangle \in H \times \SFL$,
where $h = \{x,y,z\}$ and $\sfl = \langle \sh, f, l \rangle$ is
such that
\(
  \sh = \bigl\{ \{x,y\}, \{x,z\}, \{y,z\} \bigr\}
\),
$f = \emptyset$ and $l = \{x,y,z\}$.
Then $z \notin \sharesamevar_\sfl(x, y)$ so that we have $h' = \{z\}$.
In contrast,
when using
a pair sharing domain such as $\PSDFL$
the element $\sfl$ is equivalent to
$\sfl' = \langle \sh', f, l \rangle$,
where
\(
  \sh' = \sh \union \bigl\{ \{x,y,z\} \bigr\}
\).
Hence we have $z \in \sharesamevar_{\sfl'}(x, y)$
and $h' = \emptyset$.
Thus, in $\sh$ the information provided by the sharing group $\{x,y,z\}$
is redundant for the pair-sharing and groundness properties,
but not redundant for term finiteness.
Note that the above observation holds regardless of the pair-sharing
variant considered, so that similar examples can be obtained
for $\ASub$~\cite{CodishDY91,Sondergaard86}
and $\ScozzariShPSh$~\cite{Scozzari00}.

Although the domain $\SFL$ described here is very precise
and used to implement the parameter component $P$
for computing our experimental results,
it is not intended as the target of
the generic specification given in
Definition~\ref{def:predicates-and-functs-P};
more powerful sharing domains can also satisfy this schema,
including all the enhanced combinations considered
in~\cite{BagnaraZH00,BagnaraZH05TPLP}.
For instance,
as the predicate $\gfree_\sfl$ defined on $\SFL$ does not fully exploit
the disjunctive nature of its generic specification $\gfree_p$,
the precision of the analysis may be improved by adding a domain component
explicitly tracking ground-or-freeness,
as proposed in~\cite{BagnaraZH00,BagnaraZH05TPLP}.
The same argument applies to the predicate $\orlin_\sfl$,
with respect to $\orlin_p$, when considering the
combination with the groundness domain $\Pos$.

\section{Proofs of the Stated Results}
\label{sec:proofs}

This appendix
provides the proofs of the results stated in the paper.
Section~\ref{sec:proofs-prelims} introduces the notations and
preliminary concepts that are subsequently used in the proofs. In
Section~\ref{sec:proofs-equality-theories} we recall a few
general results holding for (syntactic) equality theories and
provide the
proof of Proposition~\ref{prop:down-entailment-equivalence}.
The definition of (strongly) variable idempotent substitutions
is given in Section~\ref{sec:VSubst}, together with some properties
holding for them;
these are then used in
Section~\ref{sec:results-on-operators} to  prove some general results
on operators on substitutions in $\RSubst$,
Propositions~\ref{prop:rt-hvars-RSubst} and~\ref{prop:down-gvars-hvars}.
 Section~\ref{sec:results-on-operators} is  propaedeutic to
 Section~\ref{sec:proofs-abstrH},  where we prove
 Theorem~\ref{thm:equiv-RSubst-have-equiv-abstrH} and to
Section~\ref{sec:amguH-correctness}, where  we provide
the proofs of Theorems~\ref{thm:soundness-of-amguH}
and~\ref{thm:soundness-of-projH}.
Results in Section~\ref{sec:results-on-operators}  are then
used in
Section~\ref{sec:proofs-finite-tree-dependencies} to prove
Theorems~\ref{thm:concrFD-properties},
\ref{thm:concrH-concrFD}
and~\ref{thm:concrH-concrFD-consistency}, and in
Section~\ref{sec:proofs-groundness-dependencies} to prove
Theorems~\ref{thm:concrH-concrFD-concrGD}
and~\ref{thm:concrH-concrFD-concrGD-subsumed}.

\subsection{Notations and Preliminaries for the Proofs}
\label{sec:proofs-prelims}

To simplify the expressions in the paper,
any variable in a formula that is not in the scope of
an explicit quantifier is assumed to be universally quantified.

A path $p \in \bigl(\Nset \setdiff \{0\}\bigr)^\star$
is any finite sequence of non-zero natural numbers.
The empty path is denoted by $\epsilon$, whereas
$i \concat p$ denotes the path obtained by concatenating
the sequence formed by the natural number $i \neq 0$ with the
sequence  of the path $p$.
Given a path $p$ and a (possibly infinite) term $t \in \Terms$,
we denote by $t[p]$ the subterm of $t$ found by following path $p$.
Formally,
\begin{equation*}
  t[p]
    = \begin{cases}
        t
          &\text{if $p = \epsilon$;} \\
        t_i[q]
          &\text{if \(p = i \concat q
                        \land
                      (1 \leq i \leq n)
                        \land
                      t = f(t_1, \ldots, t_n)
                    \).}
      \end{cases}
\end{equation*}
Note that $t[p]$ is only defined for those paths $p$ actually
corresponding to subterms of $t$.

The function $\fund{\size}{\HTerms}{\Nset}$ is defined,
for each $t \in \HTerms$, by
\[
  \size(t)
    \defeq
      \begin{cases}
        1,  &\text{if $t \in \Vars$;} \\
        1 + \sum_{i=1}^n \size(t_i),
            &\text{if $t = f(t_1, \ldots, t_n)$, where $n \geq 0$.}
      \end{cases}
\]

A substitution $\sigma$ is \emph{idempotent} if, for all $t \in \HTerms$,
we have $t\sigma\sigma = t\sigma$. The set of all idempotent substitutions
is denoted by $\ISubst$ and $\ISubst \sslt \RSubst$.

If $t \in \HTerms$,
we denote the set of variables that occur more than once in $t$ by:
\[
  \nonlinvars(t)
    \defeq
      \bigl\{\, y \in \vars(t) \bigm| \neg\occlin(y,t) \,\bigr\}.
\]

If $\bar{s} = (s_1, \ldots,s_n) \in \HTerms^n$
and $\bar{t} = (t_1,\ldots,t_n) \in \HTerms^n$
are two tuples of finite terms, then we let $\bar{s} = \bar{t}$
denote the set of equations between corresponding components of
$\bar{s}$ and $\bar{t}$. Namely,
\[
  (\bar{s} = \bar{t})
    \defeq
      \{\, s_i = t_i \mid 1 \leq i \leq n \,\}.
\]
Moreover, we overload the functions $\mvars$, $\occlin$ and $\nonlinvars$
to work on tuples of terms;
thus, we will say that $\bar{s}$ is linear
if and only if $\nonlinvars(\bar{s}) = \emptyset$.

\subsubsection{Equality Theories}
\label{subsec:equality-theories}

Let $\{ s, t, s_1, \ldots, s_n, t_1, \ldots, t_m \} \sseq \HTerms$.
We assume that any equality theory $T$ over $\Terms$
includes the \emph{congruence axioms}
denoted by the following schemata:
\begin{align}
\label{eq-ax:id}
s=s &,  \\
\label{eq-ax:sym}
s=t &\piff t=s, \\
\label{eq-ax:trans}
r=s \land s=t &\pimplies r=t, \\
\label{eq-ax:congr}
s_1=t_1 \land \cdots \land s_n=t_n
  &\pimplies
    f(s_1, \ldots, s_n) = f(t_1, \ldots, t_n).
\end{align}

In logic programming and most implementations of Prolog
it is usual to assume an equality theory based on syntactic identity.
This consists of the congruence axioms
together with the \emph{identity axioms} denoted by the following schemata,
where $f$ and $g$ are distinct function symbols or $n \neq m$:
\begin{gather}
\label{eq-ax:injective-functions}
f(s_1, \ldots, s_n) = f(t_1, \ldots, t_n)
  \pimplies
    s_1=t_1 \land \cdots \land s_n=t_n, \\
\label{eq-ax:diff-funct}
\neg \bigl(f(s_1, \ldots, s_n) = g(t_1, \ldots, t_m)\bigr).
\end{gather}
The axioms characterized by
schemata~(\ref{eq-ax:injective-functions}) and~(\ref{eq-ax:diff-funct})
ensure the equality theory depends only on the syntax.
The equality theory for a non-syntactic domain replaces these axioms
by ones that depend instead on the semantics of the domain and,
in particular, on the interpretation given to functor symbols.

The equality theory of Clark~\cite{Clark78}
on which pure logic programming is based,
usually called the \emph{Herbrand} equality theory
and denoted $\HT$, is given by the congruence axioms,
the identity axioms, and the axiom schema
\begin{equation}
\label{eq-ax:occ-check}
  \forall z \in \Vars
    \itc
      \forall t \in (\HTerms \setdiff \Vars)
        \itc
          z \in \vars(t) \pimplies \neg (z = t).
\end{equation}
Axioms characterized by the schema~(\ref{eq-ax:occ-check})
are called the \emph{occurs-check axioms} and are an essential
part of the standard unification procedure in SLD-resolution.

An alternative approach used in some implementations
of Prolog,
does not require the occurs-check axioms.
This approach is based on the theory of
rational trees $\RT$~\cite{Colmerauer82,Colmerauer84}.
It assumes the congruence axioms and the identity axioms together with a
\emph{uniqueness axiom} for each substitution in rational solved form.
Informally speaking these state that,
after assigning a ground rational tree to each parameter variable,
the substitution uniquely defines a ground rational tree
for each of its domain variables.

In the sequel we will use the expression ``equality theory''
to denote any consistent, decidable theory $T$ satisfying the congruence
axioms.
We will also use the expression ``syntactic equality theory'' to denote
any equality theory $T$ also satisfying
the identity axioms.\footnote{Note that, as a consequence of
axiom~(\ref{eq-ax:diff-funct})
and the assumption that there are at least two
distinct function symbols in the language, one of which is a constant,
there exist two terms $a_1, a_2 \in \GTerms \inters \HTerms$ such that,
for any syntactic equality theory $T$, we have $T \entails a_1 \neq a_2$.}
Note that both $\HT$ and $\RT$ are syntactic equality theories.
When the equality theory $T$ is clear from the context,
it is convenient to adopt the notations $\sigma \implies \tau$ and
$\sigma \iff \tau$, where $\sigma,\tau$ are sets of equations, to denote
$T\entails \forall(\sigma \pimplies \tau)$ and
$T\entails \forall(\sigma \piff \tau)$, respectively.

Given an equality theory $T$, and a set of equations in rational solved form
$\sigma$, we say that $\sigma$ is \emph{satisfiable} in $T$ if
\(
  T
    \entails
      \forall \Vars \setdiff \dom(\sigma)
        \itc \exists \dom(\sigma) \st \sigma
\).
Observe that, given an arbitrary equality theory $T$,
a substitution in $\RSubst$ may not be satisfiable in $T$.
For example, $\exists x \st \bigl\{ x = f(x) \bigr\}$ is false
in the Clark equality theory.
However, as every element of $\RSubst$
satisfies the identity axioms as well as the
axioms~(\ref{eq-ax:injective-functions}) and~(\ref{eq-ax:diff-funct})
and, as the uniqueness axioms do not affect satisfiability,
every element of $\RSubst$ is satisfiable in $\RT$.

\subsection{Properties of Equality Theories}
\label{sec:proofs-equality-theories}

\begin{pf*}{Proof of Proposition~\vref{prop:down-entailment-equivalence}.}
Suppose $\tau \in \down \sigma$. Then,
by Definition~\ref{def:downarrow},
for some $\upsilon \in \RSubst$,
$\tau \in \mgs(\sigma \union \upsilon)$.
and hence
\(
    \RT \entails \forall\bigl(
                           \tau \piff (\sigma \union \upsilon)
                        \bigr)
\).
Therefore
$\RT \entails \forall(\tau \pimplies \sigma)$.

Conversely, suppose 
\(
  \RT \entails \forall(\tau \pimplies \sigma).
\)
Then
\(
    \RT \entails \forall\bigl(
                           \tau \pimplies (\sigma \union \tau)
                        \bigr)
\)
so that, as
\(
     \entails \forall\bigl(
                           (\sigma \union \tau) \pimplies \tau
                        \bigr),
\) 
we have
\(
    \RT \entails \forall\bigl(
                          \tau \piff \sigma \union \tau) 
                        \bigr)
\).
Therefore $\tau \in \mgs(\sigma \union \tau)$ so that,
by Definition~\ref{def:downarrow},
$\tau \in \down \sigma$.
\qed\end{pf*}

We now prove a number of results about substitutions in $\RSubst$,
assuming suitable equality theories, that will be used in the proofs
of our main results.

\begin{lemma}
\label{lem:add-binding-new}
Let $\sigma \in \RSubst$ and
$\{x \mapsto t\} \in \RSubst$
be both satisfiable in the equality theory $T$,
where $x \notin \dom(\sigma)$ and
$\vars(t) \inters \dom(\sigma) = \emptyset$.
Define also $\sigma' \defeq \sigma \union \{x \mapsto t\}$.
Then $\sigma' \in \RSubst$ and $\sigma'$ is satisfiable in $T$.
\end{lemma}

\begin{pf}%[of Lemma~\vref{lem:add-binding-new}]
Note that $\sigma'$ is a substitution,
since $\sigma \in \RSubst$ and $x \notin \dom(\sigma)$.
Moreover, as $\vars(t) \inters \dom(\sigma) = \emptyset$,
$\sigma'$  cannot contain circular subsets.
Hence, $\sigma' \in \RSubst$.

Since both $\sigma$ and $\{ x \mapsto t \}$ are satisfiable in $T$,
we have
\begin{align}
\notag
  T &\entails
      \forall \Vars \setdiff \dom(\sigma)
        \itc
          \exists \dom(\sigma) \st \sigma, \\
\notag
  T &\entails
      \forall \Vars \setdiff \{x\}
        \itc
          \exists x \st \{x = t\}. \\
\intertext{%
Letting $V = \Vars \setdiff \bigl(\dom(\sigma) \union \{x\}\bigr)$,
we can rewrite these as
}
\label{lem:add-binding-new:sigma}
  T &\entails
      \forall V
        \itc
          \forall x \itc \exists  \dom(\sigma) \st \sigma, \\
\label{lem:add-binding-new:x}
  T &\entails
      \forall V
        \itc
          \forall \dom(\sigma) \itc \exists  x \st \{x = t\}. \\
\intertext{%
Then, as
$\vars(x = t) \inters \dom(\sigma) = \emptyset$, it follows
from~(\ref{lem:add-binding-new:x}) that
}
\notag
  T &\entails
      \forall V
        \itc
          \exists  x \st \{x = t\}. \\
\intertext{%
Combining this with~(\ref{lem:add-binding-new:sigma}) gives
}
\notag
  T &\entails
      \forall V
        \itc
          \Bigl(
            \bigl(\forall x \itc \exists  \dom(\sigma) \st \sigma\bigr)
            \land
            \bigl(\exists  x \st \{x = t\}\bigr)
          \Bigr). \\
\intertext{%
Thus we have
}
\notag
  T &\entails
      \forall V
        \itc
          \exists x \st \bigl(\exists  \dom(\sigma) \st \sigma
          \land \{x = t\}\bigr), \\
\intertext{%
and hence, as $\vars(x = t) \inters \dom(\sigma) = \emptyset$,
}
\notag
  T &\entails
      \forall V
        \itc
          \exists x \st \exists  \dom(\sigma) \st
            \bigl(\sigma \land \{x = t\}\bigr). \\
\intertext{%
Therefore,
}
\notag
  T &\entails
      \forall V
        \itc
          \exists \bigl(\dom(\sigma) \union \{x\} \bigr)
            \st
              \sigma \union \{x = t\}.
\end{align}
Thus $\sigma'$ is satisfiable in $T$.
\qed\end{pf}

\begin{lemma}
\label{lem:application}
Assume $T$ is an equality theory and $\sigma \in \RSubst$.
Then, for each $t \in \HTerms$,
\[
  T
    \entails
      \forall \bigl(\sigma \pimplies (t=t\sigma)\bigr).
\]
\end{lemma}
\begin{pf}
Proved in~\cite[Lemma~2]{HillBZ02TPLP}.
\qed\end{pf}

\begin{lemma}
\label{lem:iff-application}
Assume $T$ is an equality theory and $\sigma \in \RSubst$.
Then, for each $s,t \in \HTerms$,
\[
  T
    \entails
      \forall
        \bigl(
           \sigma \union \{s=t\}
             \piff
               \sigma \union \{s=t\sigma\}
        \bigr).
\]
\end{lemma}
\begin{pf}
First, note, using the congruence axioms
(\ref{eq-ax:sym}) and (\ref{eq-ax:trans}), that,
for any terms $p,q,r \in \HTerms$,
\begin{equation}
\label{eq:lem:iff-application:iff-trans}
T \entails \forall (p=q \land q=r)
\piff \forall (p=r \land q=r).
\end{equation}

Secondly note that, using Lemma \ref{lem:application},
for any substitution $\tau\in \RSubst$ and term $r \in \HTerms$,
$T
    \entails
      \forall \bigl(\tau \pimplies (r=r\tau)\bigr)$.
Thus
\begin{equation}
\label{eq:lem:iff-application:iff-appl}
T
    \entails
      \forall \bigl(\tau \piff \tau \union \{r=r\tau\}\bigr).
\end{equation}

Using these results, we obtain
\begin{align*}
  T
    &\entails
       \forall \bigl(\sigma  \union \{s=t\}
                   \piff
                         \sigma \union
                    \{s=t, t=t\sigma\}
              \bigr),
 &&\text{[by (\ref{eq:lem:iff-application:iff-appl})]}\\
  T
    &\entails
       \forall \bigl(\sigma  \union \{s=t\}
                   \piff
                         \sigma \union
                    \{s=t\sigma, t=t\sigma\}
              \bigr),
 &&\text{[by (\ref{eq:lem:iff-application:iff-trans})]}\\
  T
    &\entails
       \forall \bigl(\sigma  \union \{s=t\}
                   \piff
                         \sigma \union
                    \{s=t\sigma\}
              \bigr).
 &&\text{[by (\ref{eq:lem:iff-application:iff-appl})]}
\end{align*}
\qed\end{pf}

\begin{lemma}
\label{lem:rt-equal-terms}
Let $\sigma \in \RSubst$ be satisfiable in a syntactic equality theory
$T$ and $s,t \in \HTerms$, where
$T \entails \forall\bigl(\sigma \pimplies (s = t)\bigr)$.
Then $\rt(s, \sigma) = \rt(t, \sigma)$.
\end{lemma}
\begin{pf}%[of Lemma~\vref{lem:rt-equal-terms}]
We suppose, towards a contradiction,
that
$\rt(s, \sigma) \neq \rt(t, \sigma)$.
Then there exists a finite path $p$ such that:
\renewcommand{\labelenumi}{\alph{enumi}.}
\begin{enumerate}
\item
$x = \rt(s, \sigma)[p] \in \Vars \setdiff \dom(\sigma)$,
$y = \rt(t, \sigma)[p] \in \Vars \setdiff \dom(\sigma)$
and $x \neq y$;
or
\item
$x = \rt(s, \sigma)[p] \in \Vars \setdiff \dom(\sigma)$ and
$r = \rt(t, \sigma)[p] \notin \Vars$ or, symmetrically,
$r = \rt(s, \sigma)[p] \notin \Vars$ and
$x = \rt(t, \sigma)[p] \in \Vars \setdiff \dom(\sigma)$;
or
\item
$r_1 = \rt(s, \sigma)[p] \notin \Vars$,
$r_2 = \rt(t, \sigma)[p] \notin \Vars$ and
$r_1$ and $r_2$ have different principal functors.
\end{enumerate}
\renewcommand{\labelenumi}{\arabic{enumi}.}
Then, by definition of `$\rt$',
there exists an index $i \in \Nset$
such that one of these holds:
\begin{enumerate}
\item
\label{case:rt-equal-terms:both-vars}
$x = s\sigma^i[p] \in \Vars \setdiff \dom(\sigma)$,
$y = t\sigma^i[p] \in \Vars \setdiff \dom(\sigma)$ and
$x \neq y$;
or
\item
\label{case:rt-equal-terms:one-vars}
$x = s\sigma^i[p] \in \Vars \setdiff \dom(\sigma)$ and
$r = t\sigma^i[p] \notin \Vars$ or, in a symmetrical way,
$r = s\sigma^i[p] \notin \Vars$ and
$x = t\sigma^i[p] \in \Vars \setdiff \dom(\sigma)$;
or
\item
\label{case:rt-equal-terms:both-compound}
$r_1 = s\sigma^i[p] \notin \Vars$ and
$r_2 = t\sigma^i[p] \notin \Vars$
have different principal functors.
\end{enumerate}

By Lemma~\ref{lem:application}, we have
\(
  T
    \entails
      \forall\bigl(
               \sigma \pimplies (s\sigma^i = t\sigma^i)
             \bigr)
\);
from this, since $T$ is a syntactic equality theory,
we obtain that
\begin{equation}
\label{eq:rt-equal-terms:application}
  T
    \entails
      \forall\bigl(
               \sigma \pimplies (s\sigma^i[p] = t\sigma^i[p])
             \bigr).
\end{equation}

We now prove that each case leads to a contradiction.

Consider case~\ref{case:rt-equal-terms:both-vars}.
Let $r_1,r_2 \in \GTerms \inters \HTerms$
be two terms having different principal functors,
so that $T \entails \forall(r_1 \neq r_2)$.
Then, as $\sigma$ is satisfiable in $T$,
by Lemma~\ref{lem:add-binding-new}, we have that
$\sigma' = \sigma \union \{ x \mapsto r_1, y \mapsto r_2 \} \in \RSubst$
is satisfiable in $T$ and also
$T \entails \forall(\sigma' \pimplies \sigma)$,
$T \entails \forall\bigl(\sigma' \pimplies (x = r_1)\bigr)$,
$T \entails \forall\bigl(\sigma' \pimplies (y = r_2)\bigr)$.
This is a contradiction,
since, by~(\ref{eq:rt-equal-terms:application}), we have
$T \entails \forall\bigl(\sigma \pimplies (x=y)\bigr)$.

Consider case~\ref{case:rt-equal-terms:one-vars}.
Without loss of generality, consider the first subcase,
where $x = s\sigma^i[p] \in \Vars \setdiff \dom(\sigma)$
and $r = t\sigma^i[p] \notin \Vars$.
Let $r' \in \GTerms \inters \HTerms$ be such that $r$ and $r'$
have different principal functors,
so that $T \entails \forall(r \neq r')$.
By Lemma~\ref{lem:add-binding-new}, as $\sigma$ is satisfiable in $T$,
$\sigma' = \sigma \union \{ x \mapsto r' \} \in \RSubst$
is satisfiable in $T$;
we also have that
$T \entails \forall(\sigma' \pimplies \sigma)$ and
$T \entails \forall\bigl(\sigma' \pimplies (x = r')\bigr)$.
This is a contradiction as,
by~(\ref{eq:rt-equal-terms:application}),
$T \entails \forall\bigl(\sigma \pimplies (x=r)\bigr)$.

Finally, consider case~\ref{case:rt-equal-terms:both-compound}.
In this case
$T \entails \forall(r_1 \neq r_2)$.
This immediately leads to a contradiction,
since, by~(\ref{eq:rt-equal-terms:application}),
$T \entails \forall\bigl(\sigma \pimplies (r_1 = r_2)\bigr)$.
\qed\end{pf}

\begin{lemma}
\label{lem:ground-and-finite-unifies-with-same-size}
Let $T$ be a syntactic equality theory.
Let $s \in \HTerms \inters \GTerms$ and $t \in \Terms$ be
such that $\size(t) > \size(s)$.
Then $T \entails \forall(s \neq t)$.
\end{lemma}
\begin{pf}
By induction on $m = \size(s)$.
For the base case, when $m = 1$, we have that $s$ is
a term functor of arity $0$. Since $\size(t) > 1$,
then $t = f(t_1, \ldots, t_n)$, where $n > 0$.
Then, by the identity axioms, we have $T \entails \forall(s \neq t)$.

For the inductive case, when $m > 1$, assume that
the result holds for all $m' < m$ and
let $s = f(s_1, \ldots, s_n)$, where $n > 0$.
Since $\size(t) > m$, we have
$t = f'(t_1, \ldots, t_{n'})$, where $n' > 0$.
If $f \neq f'$ or $n \neq n'$ then, by the identity axioms,
we have $T \entails \forall(s \neq t)$.
Otherwise, let $f = f'$ and $n = n'$.
Note that, for all $i \in \{1, \ldots, n\}$, we have
$\size(s_i) < m$. Also, there exists an index $j \in \{1, \ldots, n\}$
such that $\size(t_j) > \size(s_j)$.
By the inductive hypothesis, $T \entails \forall(s_j \neq t_j)$
so that, by the identity axioms,
$T \entails \forall(s \neq t)$.
\qed\end{pf}

The next two propositions establish useful properties of
 the function $\mathord{\rt}$.
\begin{proposition}
\label{prop:rt-vars-HTerms}
Let $\sigma \in \RSubst$ and $t \in \HTerms$.
Then
\begin{subequations}
\renewcommand{\theequation}{\ref{prop:rt-vars-HTerms}\alph{equation}}
\begin{gather}
\label{case:rt-vars-HTerms:rt-vars}
  \vars\bigl(\rt(t, \sigma)\bigl) \inters \dom(\sigma) = \emptyset, \\
\label{case:rt-vars-HTerms:rt-HTerms}
  \rt(t, \sigma) \in \HTerms
    \iff
      \exists i \in \Nset \st \rt(t, \sigma) = t\sigma^i.
\end{gather}
\end{subequations}
\end{proposition}
\begin{pf}
\par\indent
(\ref{case:rt-vars-HTerms:rt-vars})
Let $x \in \dom(\sigma)$ and, towards a contradiction,
suppose $x \in \vars\bigl(\rt(t, \sigma)\bigr)$.
Thus, there exists a finite path $p$
such that $x = \rt(t, \sigma)[p]$.
Thus, by definition of `$\rt$',
there exists an index $i \in \Nset$
such that $x = \sigma^i(t)[p]$.
Since $x \in \dom(\sigma)$, then $x \neq x\sigma$,
so that $x \neq \sigma^{i+1}(t)[p]$.
Also note that, since $\sigma \in \RSubst$,
$\sigma$ contains no circular subsets,
so that we have $x \neq \sigma^j(t)[p]$, for each index $j > i$.
This implies $x \neq \rt(t, \sigma)[p]$,
which is a contradiction.
Since no such finite path $p$ can exist, we can conclude
$x \notin \vars\bigl(\rt(t, \sigma)\bigr)$.

(\ref{case:rt-vars-HTerms:rt-HTerms})
Since substitutions map finite terms into finite terms,
a finite number of applications cannot produce an infinite term,
so that the left implication holds.
Proving the right implication by contraposition,
suppose that $\rt(t, \sigma) \neq t\sigma^i$, for all $i \in \Nset$.
Then, by definition of `$\rt$', we have $t\sigma^i \neq t\sigma^{i+1}$,
for all $i \in \Nset$.
Letting $n \in \Nset$ be the number of bindings in $\sigma \in \RSubst$,
for all $i \in \Nset$ we have that $\size(t\sigma^i) < \size(t\sigma^{i+n})$,
because $\sigma$ has no circular subsets.
Thus $\rt(t, \sigma) \notin \HTerms$,
because there is no finite upper bound
to the number of function symbols occurring in $\rt(t, \sigma)$.
\qed\end{pf}

\begin{proposition}
\label{prop:rt-and-hterms}
Let $\sigma,\tau \in \RSubst$ be satisfiable in
a syntactic equality theory $T$ and $W \sseq \Vars$,
where
\(
  T \entails
     \forall(\exists W \st \tau \pimplies \exists W \st \sigma),
\)
and $x \in \Vars \setdiff W$.
Then
\[
  \rt(x, \tau) \in \HTerms
    \implies 
       \rt(x,\sigma) \in \HTerms.
\]
\end{proposition}
\begin{pf}
We assume that $\rt(x, \tau) \in \HTerms$ but
$\rt(x,\sigma) \notin \HTerms$ and derive a contradiction.
By hypothesis $\rt(x, \tau) \in \HTerms$, so that
by Proposition~\ref{prop:rt-vars-HTerms},
there exists $i \in \Nset$ such that
$\rt(x, \tau) = x\tau^i$ and also
$\vars(x\tau^i) \inters \dom(\tau) = \emptyset$.
Let $t \in \GTerms \inters \HTerms$ and
\[
  \upsilon
    \defeq
      \bigl\{\,
        y \mapsto t
      \bigm|
        y \in \vars(x\tau^i)
      \,\bigr\}.
\]
Then, as $\tau$ is satisfiable in $T$, by Lemma~\ref{lem:add-binding-new},
$\tau' \defeq \tau \union \upsilon \in \RSubst$
is also satisfiable in $T$.
Moreover, we have that $x\tau^i\tau' \in \GTerms \inters \HTerms$.
Define now $n \defeq \size(x\tau^i\tau')$.
As $\rt(x, \sigma) \notin \HTerms$,
there exists $j \in \Nset$ such that $\size(x\sigma^j) > n$.
Therefore, by Lemma~\ref{lem:ground-and-finite-unifies-with-same-size},
\begin{equation}
\label{eq:rt-and-hterms:not-unifies}
  T \entails \forall(x\tau^i\tau' \neq x\sigma^j).
\end{equation}
By Lemma~\ref{lem:application},
\begin{equation}
\label{eq:rt-and-hterms:xtautauprime}
  T \entails
        \forall\bigl(
                 \tau \pimplies (x = x\tau^i)
               \bigr).
\end{equation}
Also, by Lemma~\ref{lem:application},
\(
  T \entails
        \forall\bigl(
                 \sigma \pimplies (x = x\sigma^j)
               \bigr)
\)
so that, as $T$ is a first-order theory, 
\begin{equation}
\label{eq:rt-and-hterms:existsW-xsigmaj}
  T \entails
        \forall\bigl(
                 \exists W \st \sigma \pimplies \exists W \st (x = x\sigma^j)
               \bigr).
\end{equation}
By definition of $\tau'$,
\(
  \forall(\tau' \pimplies \tau).
\)
Hence, by hypothesis and
the logically true statement
\(
  \forall(\tau \pimplies \exists W \st \tau)
\),
we obtain
\(
  T \entails
  \forall(\tau' \pimplies \exists W \st \sigma).
\)
Observe that $\vars(x = x\tau^i\tau') = \{x\}$ and,
as a consequence,
$\vars(x = x\tau^i\tau') \inters W = \emptyset$.
Therefore,
by~(\ref{eq:rt-and-hterms:xtautauprime})
and~(\ref{eq:rt-and-hterms:existsW-xsigmaj}), we obtain
\begin{align*}
  & T
      \entails
        \forall\bigl(
                 \tau'
                   \pimplies
                     (x = x\tau^i\tau' \land \exists W \st x = x\sigma^j)
               \bigr)\\
  &\qquad\iff
    T
      \entails
        \forall\bigl(
                 \tau'
                   \pimplies
                     \exists W \st
                       (x = x\tau^i\tau' \land x = x\sigma^j)
               \bigr) \\
  &\qquad\iff
    T
      \entails
        \forall\bigl(
                 \tau'
                   \pimplies
                     \exists W \st (x\tau^i\tau' = x\sigma^j)
               \bigr)
\end{align*}
which contradicts~(\ref{eq:rt-and-hterms:not-unifies}).
\qed\end{pf}

\subsection{Variable-Idempotence}
\label{sec:VSubst}

In~\cite{HillBZ02TPLP}, (weak) variable-idempotent substitutions
were introduced as a subclass of substitutions in rational solved form
in order to allow a more convenient reasoning about
the sharing of variables for possibly non-idempotent substitutions.
In~\cite{HillBZ98b} a stronger definition was used, taking into
consideration also the variables in the domain of the substitution.
\emph{Strong} variable-idempotence is a useful concept when dealing
with the finiteness of a rational term and the multiplicity of
variables occurring in it (e.g., when linearity is a property of
interest).
In the following we consider this stronger definition,
also adopted in~\cite{HillZB04TPLP,Zaffanella01th}.

\begin{definition} \summary{(Variable-idempotence.)}
A substitution $\sigma \in \RSubst$ is said to be
(strongly) \emph{variable-idempotent}
if and only if for all $t \in \HTerms$ we have
\[
  \vars(t\sigma\sigma) = \vars(t\sigma).
\]
The set of variable-idempotent substitutions is denoted $\VSubst$.
\end{definition}

Note that we have
$\ISubst \sslt \VSubst \sslt \RSubst$.

\begin{definition}
\summary{(\cS-transformation.)}
\label{def:Sstep}
The relation $\reld{\Sstep}{\RSubst}{\RSubst}$,
called \emph{\cS-step}, is defined by
\[
\genfrac{}{}{}{}
  {
    (x \mapsto t) \in \sigma \qquad (y \mapsto s) \in \sigma \qquad x \neq y
  }
  {
    \sigma
      \Sstep
        \bigl( \sigma \setdiff \{y \mapsto s\} \bigr)
          \union \bigl\{y \mapsto s\{x \mapsto t\} \bigr\}
  }.
\]
If we have a finite sequence of \cS-steps
$\sigma_1 \Sstep \cdots \Sstep \sigma_n$ mapping $\sigma_1$ to $\sigma_n$,
then we write $\sigma_1 \Sstepstar \sigma_n$ and say that
$\sigma_1$ can be rewritten, by \cS-transformation, to $\sigma_n$.
\end{definition}

The following theorems show that considering substitutions
in $\VSubst$ is not a restrictive hypothesis.

\begin{theorem}
\label{thm:Sstepstar-equiv}
Suppose $\sigma \in \RSubst$ and $\sigma \Sstepstar \sigma'$.
Then we have $\sigma' \in \RSubst$, $\dom(\sigma) = \dom(\sigma')$,
and $\vars(\sigma) = \vars(\sigma')$.
Moreover, if $T$ is any equality theory,
we have $T \entails \forall(\sigma \piff \sigma')$.
\end{theorem}
\begin{pf}
Proved in~\cite[Theorem 1]{HillBZ02TPLP}.
\qed\end{pf}

\begin{theorem}
\label{thm:strong-var-idem}
Suppose $\sigma \in \RSubst$.
Then there exists  $\sigma' \in \VSubst$ such that
$\sigma \Sstepstar \sigma'$ and, for all $\tau \sseq \sigma'$,
$\tau \in \VSubst$.
\end{theorem}
\begin{pf}
The proof is the same given for~\cite[Theorem 2]{HillBZ02TPLP},
where a weaker result, using \emph{weak} variable-idempotence, was stated.
\qed\end{pf}

\begin{theorem}
\label{thm:VSubst-generality}
Let $T$ be an equality theory and $\sigma \in \RSubst$.
Then there exists  $\sigma' \in \VSubst$ such that
$\dom(\sigma) = \dom(\sigma')$,
$\vars(\sigma) = \vars(\sigma')$,
$T \entails \forall(\sigma \piff \sigma')$ and
for all $\tau \sseq \sigma'$, $\tau \in \VSubst$.
\end{theorem}
\begin{pf}
The result easily follows from
Theorems~\ref{thm:Sstepstar-equiv} and~\ref{thm:strong-var-idem}.
\qed\end{pf}

The next result concerning a useful property of
variable idempotent substitutions will be needed
in Subsection \ref{sec:amguH-correctness} for proving
Theorem~\ref{thm:soundness-of-amguH}.

\begin{lemma}
\label{lem:groundness-mgs-properties}
Let $\sigma \in \VSubst$ be satisfiable
in a syntactic equality theory $T$.
Let $s \in \HTerms \inters \GTerms$ and $t \in \HTerms$ and
suppose that $T \entails \forall(\sigma \pimplies  s=t)$.
Then $s = t\sigma$.
\end{lemma}
\begin{pf}
By hypothesis, $T \entails \forall(\sigma \pimplies  s=t)$
and $s,t \in \HTerms$ so that we can apply
Lemma~\ref{lem:rt-equal-terms} to obtain
\begin{equation}
\label{lem:groundness-mgs-properties:rt}
\rt(s, \sigma) = \rt(t, \sigma).
\end{equation}
By Proposition~\ref{prop:rt-vars-HTerms},
there exists $i,j \in \Nset$ such that
$\rt(s, \sigma) = s\sigma^i$ and $\rt(t, \sigma) = t\sigma^j$
and
\(
\dom(\sigma) \inters \vars(t\sigma^j) = \emptyset.
\)
Thus, if $j=0$, we have
$t\sigma^j = t = t\sigma$.
On the other hand, if $j > 0$, 
as $\sigma \in \VSubst$, $\vars(t\sigma^j) = \vars(t\sigma)$
so that
$\dom(\sigma) \inters \vars(t\sigma) = \emptyset$ and hence
$t\sigma = t\sigma^j$.
As $s \in \GTerms$, $\vars(s) = \emptyset$ so that 
 $s = s\sigma^i$.
Thus, by~(\ref{lem:groundness-mgs-properties:rt}) we have $s = t\sigma$.
\qed\end{pf}	

\subsection{Some Results on the Groundness and Finiteness Operators}
\label{sec:results-on-operators}

The following proposition is proved in~\cite{HillBZ02TPLP}, and shows
that the function `$\gvars$' precisely captures the intended property.

\begin{proposition}
\label{prop:RSubst-gvars}
Let $\sigma \in \RSubst$ and $x \in \Vars$.
Then
\[
  y \in \gvars(\sigma)
    \iff
      \rt(y, \sigma) \in \GTerms.
\]
\end{proposition}

When computing $\hvars(\sigma)$ by means of the fixpoint computation
given in Definition~\vref{def:finiteness-functs}, the fixpoint
is reached after a single iteration if $\sigma \in \VSubst$.
\begin{lemma}
\label{lem:hvars-eq-hvars1-VSubst}
For each $\sigma \in \VSubst$ we have
\(
  \hvars(\sigma) = \hvars_1(\sigma)
\).
\end{lemma}
\begin{pf}
We show that $\hvars_2(\sigma) \sseq \hvars_1(\sigma)$.
Let $y \in \hvars_2(\sigma)$.
By Definition~\ref{def:finiteness-functs}, we have two cases:
\begin{enumerate}
\item
if $y \in \hvars_1(\sigma)$ then there is nothing to prove;
\item
assume now $y \in \dom(\sigma)$
and $\vars(y\sigma) \sseq \hvars_1(\sigma)$.
By Definition~\ref{def:finiteness-functs}, we have two subcases:
\begin{enumerate}
\item
$\vars(y\sigma) \sseq \Vars \setdiff \dom(\sigma)$.

Then $\vars(y\sigma) \sseq \hvars_0(\sigma)$,
so that $y \in \hvars_1(\sigma)$;
\item
$V = \vars(y\sigma) \inters \dom(\sigma) \neq \emptyset$
and, for all $z \in V$,
$\vars(z\sigma) \inters  \dom(\sigma)= \emptyset$.

Let $z \in V$ so that $z \in \vars(y\sigma)$.
By hypothesis, we have $\sigma \in \VSubst$
so that $z \in \vars(y\sigma\sigma)$.
As $z \in \dom(\sigma)$ and
 $\vars(z\sigma) \inters \dom(\sigma) = \emptyset$,
$z \notin \vars(z\sigma)$.
This means that $z \notin \vars(y\sigma\sigma)$,
which is a contradiction since $\sigma \in \VSubst$.
\end{enumerate}
\end{enumerate}
\qed\end{pf}

\begin{proposition}
\label{prop:hvars-alt-charact-VSubst}
For each $\sigma \in \VSubst$, we have
\[
  \hvars(\sigma)
    = \bigl\{\,
         y \in \Vars
       \bigm|
         \vars(y\sigma) \inters \dom(\sigma) = \emptyset
       \,\bigr\}.
\]
\end{proposition}
\begin{pf}
The result is obtained by applying
Lemma~\ref{lem:hvars-eq-hvars1-VSubst}
and then unfolding Definition~\ref{def:finiteness-functs}.
\qed\end{pf}

\begin{proposition}
\label{prop:hvars-rt-is-rsigma-VSubst}
Let $\sigma \in \VSubst$ and $r \in \HTerms$,
where $\vars(r) \sseq \hvars(\sigma)$.
Then
\begin{align*}
  \rt(r,\sigma) &= r\sigma, \\
  \vars(r\sigma) \inters \dom(\sigma) &= \emptyset.
\end{align*}
\end{proposition}
\begin{pf}
Suppose $y \in \vars(r)$.
Then,
by Proposition~\ref{prop:hvars-alt-charact-VSubst},
$\vars(y\sigma) \inters \dom(\sigma) = \emptyset$.
Thus, for any $i > 0$, we have $y\sigma^i = y\sigma \in \HTerms$.
Thus $\rt(y,\sigma) = y\sigma$.
As this holds for all $y \in \vars(r)$,
it follows that $\rt(r,\sigma) = r\sigma$ and
$\vars(r\sigma) \inters \dom(\sigma) = \emptyset$.
\qed\end{pf}

The following result shows that,
for a variable-idempotent substitution,
the finiteness operator precisely captures
the intended property.

\begin{lemma}
\label{lem:rt-hvars-VSubst}
Let $\sigma \in \VSubst$ and $y \in \Vars$.
Then
\[
  \rt(y, \sigma) \in \HTerms
    \quad \iff \quad
      y \in \hvars(\sigma).
\]
\end{lemma}
\begin{pf}
Since $\sigma \in \VSubst$,
by Proposition~\ref{prop:hvars-alt-charact-VSubst}
we have $y \in \hvars(\sigma)$ if and only if
$\vars(y\sigma) \inters \dom(\sigma) = \emptyset$.

Let $\vars(y\sigma) \inters \dom(\sigma) = \emptyset$.
Then, for any $i > 0$, we have $y\sigma^i = y\sigma \in \HTerms$.
Hence $\rt(y, \sigma) = y\sigma \in \HTerms$.

In order to prove the other inclusion,
let now $\rt(y, \sigma) \in \HTerms$.
By Proposition~\ref{prop:rt-vars-HTerms},
there exists an $i \in \Nset$ such that $\rt(y, \sigma) = y\sigma^i$
and $\vars(y\sigma^i) \inters \dom(\sigma) = \emptyset$.
Since $\sigma \in VSubst$, we have $\vars(y\sigma^i) = \vars(y\sigma)$,
so that $\vars(y\sigma) \inters \dom(\sigma) = \emptyset$.
\qed\end{pf}

In order to prove Proposition~\ref{prop:rt-hvars-RSubst},
i.e., to show that the finiteness operator
precisely captures the intended property even for arbitrary
substitutions in $\RSubst$, we now prove that this operator
is invariant under the application of $S$-steps.

\begin{lemma}
\label{lem:Sstep-equiv-hvars}
Let $\sigma,\sigma' \in \RSubst$ where $\sigma \Sstep \sigma'$.
Then $\hvars(\sigma) = \hvars(\sigma')$.
\end{lemma}
\begin{pf}
Let $(x \mapsto t), (y \mapsto s) \in \sigma$,
where $x \neq y$, such that
\begin{equation*}
\sigma' = \bigl(\sigma \setdiff \{y \mapsto s\}\bigr)
            \union
              \bigl\{y \mapsto s\{x \mapsto t\}\bigr\}.
\end{equation*}
If $x \notin \vars(s)$ then we have $\sigma = \sigma'$
and the result trivially holds. Thus, we assume $x \in \vars(s)$.
We prove the two inclusions separately.

In order to prove $\hvars(\sigma) \sseq \hvars(\sigma')$
we show, by induction on $m \geq 0$, that we have
\begin{align*}
  \hvars_m(\sigma)
    &\sseq
      \hvars_m(\sigma').\\
\intertext{%
For the base case, when $m = 0$,
by Theorem~\ref{thm:Sstepstar-equiv}
we have $\dom(\sigma) = \dom(\sigma')$ so that
}
  \hvars_0(\sigma)
    &= \Vars \setdiff \dom(\sigma) \\
    &= \Vars \setdiff \dom(\sigma') \\
    &= \hvars_0(\sigma').
\end{align*}
For the inductive step, when $m > 0$,
assume $\hvars_{m-1}(\sigma) \sseq \hvars_{m-1}(\sigma')$
and let $z \in \hvars_m(\sigma)$.
By Definition~\ref{def:finiteness-functs}, we have two cases:
if $z \in \hvars_{m-1}(\sigma)$ then the result follows by a
straight application of the inductive hypothesis;
otherwise, we have
\begin{align*}
  z \in \dom(\sigma)
    &\land
      \vars(z\sigma) \sseq \hvars_{m-1}(\sigma). \\
\intertext{%
Now, if $z\neq y$ we have $z\sigma = z\sigma'$,
so that, by Theorem~\ref{thm:Sstepstar-equiv}
and the inductive hypothesis we have
}
  z \in \dom(\sigma')
    &\land
      \vars(z\sigma') \sseq \hvars_{m-1}(\sigma'), \\
\end{align*}
so that, by Definition~\ref{def:finiteness-functs},
$z \in \hvars_m(\sigma')$.
Otherwise, if $z=y$, then
\begin{align*}
  \vars(z\sigma)
    &= \vars(s) \\
    &\sseq \hvars_{m-1}(\sigma).\\
\intertext{%
Since, by hypothesis, $x \in \vars(s)$,
}
  \vars(z\sigma')
    &=
      \vars\bigl(s\{x \mapsto t\}\bigr) \\
    &=
      \bigl(\vars(s) \setdiff \{x\}\bigr) \union \vars(t),\\
\intertext{%
and we need to show $\vars(z\sigma') \sseq \hvars_{m-1}(\sigma')$.
By the inductive hypothesis we have
}
  \vars(s)
    &\sseq
      \hvars_{m-1}(\sigma');\\
\intertext{%
Note that, since $x \in \vars(s)$, it follows
$x \in \hvars_{m-1}(\sigma')$ so that,
by Definition~\ref{def:finiteness-functs},
}
  \vars(t)
    &\sseq \hvars_{m-2}(\sigma') \\
    &\sseq \hvars_{m-1}(\sigma').
\end{align*}

In order to prove $\hvars(\sigma) \Sseq \hvars(\sigma')$
we show, by induction on $m \geq 0$, that we have
\begin{align*}
  \hvars_{m+1}(\sigma)
    &\Sseq
      \hvars_m(\sigma').\\
\intertext{%
For the base case, when $m = 0$,
by Definition~\ref{def:finiteness-functs}
and Theorem~\ref{thm:Sstepstar-equiv} we have
}
  \hvars_1(\sigma)
    &\Sseq \hvars_0(\sigma) \\
    &= \Vars \setdiff \dom(\sigma) \\
    &= \Vars \setdiff \dom(\sigma') \\
    &= \hvars_0(\sigma').
\end{align*}
For the inductive step, when $m > 0$,
assume $\hvars_m(\sigma) \Sseq \hvars_{m-1}(\sigma')$
and let $z \in \hvars_m(\sigma')$.
By Definition~\ref{def:finiteness-functs}, we have two cases:
if $z \in \hvars_{m-1}(\sigma')$ then the result follows
by the inductive hypothesis and by Definition~\ref{def:finiteness-functs};
otherwise, we have
\begin{align*}
  z \in\dom(\sigma')
    &\land
      \vars(z\sigma') \sseq \hvars_{m-1}(\sigma'). \\
\intertext{%
Now, if $z\neq y$ we have $z\sigma = z\sigma'$,
so that, by Theorem~\ref{thm:Sstepstar-equiv}
and the inductive hypothesis we have
}
  z \in \dom(\sigma)
    &\land
      \vars(z\sigma) \sseq \hvars_{m}(\sigma), \\
\end{align*}
so that, by Definition~\ref{def:finiteness-functs},
$z \in \hvars_{m+1}(\sigma)$.
Otherwise, if $z=y$,
by definition of $\sigma'$, the inductive hypothesis
and Definition~\ref{def:finiteness-functs}, we have
\begin{align*}
  \vars(z\sigma')
    &=
      \vars\bigl(s\{x \mapsto t\}\bigr) \\
    &=
      \bigl(\vars(s) \setdiff \{x\}\bigr) \union \vars(t) \\
    &\sseq \hvars_{m-1}(\sigma') \\
    &\sseq \hvars_{m}(\sigma) \\
    &\sseq \hvars_{m+1}(\sigma).
\intertext{%
Also note that we have
}
  \vars(x\sigma)
    &= \vars(t) \\
    &\sseq \hvars_{m}(\sigma) \\
\end{align*}
so that, by Definition~\ref{def:finiteness-functs} we have
\[
  x \in \hvars_{m+1}(\sigma).
\]
The result follows by observing that
\[
  \vars(z\sigma)
     = \vars(s)
     = \bigl(\vars(s) \setdiff \{x\}\bigr) \union \{x\}.
\]
\qed\end{pf}

\begin{lemma}
\label{lem:Sstepstar-equiv-ft}
Let $\sigma,\sigma' \in \RSubst$,
where $\sigma \Sstepstar \sigma'$.
Then
\(
  \hvars(\sigma) = \hvars(\sigma')
\).
\end{lemma}
\begin{pf}
By induction on the length $n \geq 0$ of the derivation.
For the base case, when $n = 0$, there is nothing to prove.
Suppose now that
\[
  \sigma = \sigma_0
    \Sstep \cdots \Sstep \sigma_{n-1} \Sstep \sigma_n = \sigma',
\]
where $n > 0$.
By the inductive hypothesis, since the derivation
$\sigma \Sstepstar \sigma_{n-1}$ has length $n-1$,
we have $\hvars(\sigma) = \hvars(\sigma_{n-1})$.
Then the thesis follows by Lemma~\ref{lem:Sstep-equiv-hvars}.
\qed\end{pf}

\begin{pf*}{Proof of Proposition~\vref{prop:rt-hvars-RSubst}.}
By Theorem~\ref{thm:VSubst-generality},
there exists  $\sigma' \in \VSubst$ such that
$\sigma \Sstepstar \sigma'$ and, for all equality theories $T$,
$T \entails \forall (\sigma \piff \sigma')$.
By Lemma~\ref{lem:rt-hvars-VSubst},
for all $x \in \Vars$, $\rt(x, \sigma') \in \HTerms$ if and only if
$x \in \hvars(\sigma')$.
By Lemma~\ref{lem:Sstepstar-equiv-ft},
we have $\hvars(\sigma) = \hvars(\sigma')$ and,
by Proposition~\ref{prop:rt-and-hterms},
for all $x \in \Vars$, $\rt(x, \sigma') \in \HTerms$ if and only if
$\rt(x, \sigma) \in \HTerms$.
Therefore,
for any $x \in \Vars$, $\rt(x, \sigma) \in \HTerms$ if and only if
$x \in \hvars(\sigma)$
\qed\end{pf*}

\begin{pf*}{Proof of Proposition~\vref{prop:down-gvars-hvars}.}
We prove the two statements (\ref{eq:down-gvars-hvars:hvars})
and (\ref{eq:down-gvars-hvars:gvars-inters-hvars}) separately.

(\ref{eq:down-gvars-hvars:hvars}).
By hypothesis, $\tau \in \down \sigma$,
Thus, by Proposition~\ref{prop:down-entailment-equivalence},
\(
  \RT \entails
        \forall(\tau \pimplies \sigma)
\).
Suppose $x \in \hvars(\tau)$.
Then, by Proposition~\ref{prop:rt-hvars-RSubst},
we have  $\rt(x, \tau) \in \HTerms$.
Therefore we can apply Proposition~\ref{prop:rt-and-hterms} to obtain
$\rt(x, \sigma) \in \HTerms$ and hence,
by Proposition~\ref{prop:rt-hvars-RSubst},
$x \in \hvars(\sigma)$.

(\ref{eq:down-gvars-hvars:gvars-inters-hvars}).
Suppose $x \in \hvars(\sigma) \inters \gvars(\sigma)$.
Then, by Propositions~\ref{prop:rt-hvars-RSubst} and
\ref{prop:RSubst-gvars},
$\rt(x, \sigma) \in \GTerms \inters \HTerms$.
Thus, by case~(\ref{case:rt-vars-HTerms:rt-HTerms})
of Proposition~\ref{prop:rt-vars-HTerms},
there exists $i \in \Nset$ such that
$\rt(x, \sigma) = x\sigma^i$ and also
$\vars(x\sigma^i) = \emptyset$.
Thus $\rt(x\sigma^i, \tau) = x\sigma^i$.
Since by hypothesis we have $\tau \in \down \sigma$,
by Lemma~\ref{lem:application} and transitivity we obtain that
\(
  \RT \entails
        \forall\bigl(
                 \tau \pimplies (x = x\sigma^i)
               \bigr)
\).
Thus, by Lemma~\ref{lem:rt-equal-terms},
$\rt(x, \tau) = \rt(x\sigma^i, \tau) = x\sigma^i$.
Therefore, by Propositions~\ref{prop:rt-hvars-RSubst} and
\ref{prop:RSubst-gvars},
$x \in \gvars(\tau) \inters \hvars(\tau)$.
\qed\end{pf*}

\begin{proposition}
\label{prop:project-hvars}
Let $\sigma,\tau \in \RSubst$ be satisfiable in a syntactic equality theory $T$
and $W \sseq \Vars$,
where
\(
  T \entails
     \forall(\exists W \st \sigma \piff \exists W \st \tau).
\)
Then
\[
  \hvars(\sigma) \setdiff W
    = \hvars(\tau) \setdiff W.
\]
\end{proposition}
\begin{pf}%%[of Proposition~\vref{prop:project-hvars}]
Suppose $z \in \hvars(\sigma) \setdiff W$.
By Proposition~\ref{prop:rt-hvars-RSubst},
$\rt(z, \sigma) \in \HTerms$ and hence,
by Proposition~\ref{prop:rt-and-hterms},
$\rt(z, \tau) \in \HTerms$.
Therefore, by Proposition~\ref{prop:rt-hvars-RSubst}, $z \in \hvars(\tau)$.

The reverse inclusion follows by symmetry.
\qed\end{pf}

\begin{corollary}
\label{cor:all-mgs-have-same-hvars}
Let $e \sseq \Eqs$ be satisfiable in the syntactic equality theory $T$.
If $\sigma,\tau \in \mgs(e)$, then $\hvars(\sigma) = \hvars(\tau)$.
\end{corollary}
\begin{pf}
By definition of $\mgs$, we have $\sigma,\tau \in \RSubst$,
$T \entails \forall(\sigma \piff e)$
and $T \entails \forall(\tau \piff e)$ so that
$T \entails \forall(\sigma \piff \tau)$.
Thus the result follows
by Proposition~\ref{prop:project-hvars}.
\qed\end{pf}

\subsection{Abstracting Finiteness}
\label{sec:proofs-abstrH}
				   
\begin{pf*}{Proof of Theorem~\vref{thm:equiv-RSubst-have-equiv-abstrH}.}
By Definition~\ref{def:abstr-concr-H},
$\abstrH(\sigma) = \hvars(\sigma) \inters \VI$ and
$\abstrH(\tau) = \hvars(\tau) \inters \VI$.
The result is a simple consequence of Proposition~\ref{prop:project-hvars},
since $\RT$ is a syntactic equality theory,
$\sigma, \tau \in \RSubst$ are satisfiable in $\RT$
and, by hypothesis, $\RT \entails \forall(\sigma \piff \tau)$.
\qed\end{pf*}

\subsection{Correctness of Abstract Unification on $H \times P$}
\label{sec:amguH-correctness}

For the rest of the appendix it is assumed that the equality theory
$\RT$ holds.
Note that this means that the congruence and identity axioms hold
and also that
every substitution in $\RSubst$ is satisfiable in $\RT$.

\begin{lemma}
\label{lem:share-linearly-mgs-properties}
Let $\bar{s} = (s_1, \ldots, s_n) \in \HTerms^n$ be linear,
and suppose the tuple of terms $\bar{t} = (t_1, \ldots, t_n) \in \HTerms^n$
is such that
$\vars(\bar{s}) \inters \nonlinvars(\bar{t}) = \emptyset$
and $\mgs(\bar{s} = \bar{t}) \neq \emptyset$.
Then there exists
$\mu \in \mgs(\bar{s} = \bar{t})$
such that,
for each variable
\(
  z \in
   \dom(\mu)
     \setdiff
   \bigl(\vars(\bar{s}) \inters \vars(\bar{t})\bigr)
\),
we have
$\vars(z\mu) \inters \dom(\mu) = \emptyset$.
\end{lemma}

\begin{pf}
The proof is by induction on the number of variables in
$\vars(\bar{s}) \union \vars(\bar{t})$.

Suppose first that,
for some $i=1$, \dots, $n$,
we have $s_i = f(r_1, \ldots, r_m)$
and $t_i = f(u_1, \ldots, u_m)$ (with $m \geq 0$).
Let
\begin{align*}
  \bar{s}' &\defeq (s_1, \ldots, s_{i-1},
                         r_1, \ldots, r_m,
                         s_{i+1}, \ldots, s_n),\\
  \bar{t}' &\defeq (t_1, \ldots, t_{i-1},
                         u_1, \ldots, u_m,
                         t_{i+1}, \ldots, t_n).
\end{align*}
Then
\(
   \mvars(\bar{s}')
   =
   \mvars(\bar{s})
\)
and
\(
   \mvars(\bar{t}')
   =
   \mvars(\bar{t})
\)
so that, as $\bar{s}$ is linear,
$\bar{s}'$ is linear,
$\vars(\bar{s}') \inters \nonlinvars(\bar{t}') = \emptyset$
and
\(
  \vars(\bar{s}') \inters \vars(\bar{t}')
    =
      \vars(\bar{s}) \inters \vars(\bar{t})
\).
Moreover, by the congruence axiom (\ref{eq-ax:congr}),
\(
  \mgs(\bar{s}' = \bar{t}') = \mgs(\bar{s} = \bar{t})
\).
We repeat this process until all terms in
$\bar{s}'$ and $\bar{t}'$ can not be decomposed any further.
(Note that in the case that $s_i$ and $t_i$ are identical constants,
we can remove them from $\bar{s}'$ and $\bar{t}'$, since
the  corresponding equation $s_i = t_i$ holds vacuously.)
Thus, as $\bar{s}$ and $\bar{t}$
are finite sequences of finite terms,
we can assume that, for all $i=1$, \dots, $n$,
either $s_i \in \Vars$ or $t_i \in \Vars$.

Secondly, suppose that
for some $i=1$, \dots,~$n$, $s_i = t_i$.
By the previous paragraph, we can assume that $s_i \in \Vars$.
Let
\begin{align*}
  \bar{s}_i &\defeq
     (s_1, \ldots, s_{i-1}, s_{i+1}, \ldots, s_n), \\
  \bar{t}_i &\defeq
     (t_1, \ldots, t_{i-1}, t_{i+1}, \ldots, t_n).
\end{align*}
Then
\(
  \mvars(\bar{s}_i) \union \{s_i\}
    = \mvars(\bar{s})
\)
and
\(
  \mvars(\bar{t}_i) \union \{s_i\}
    = \mvars(\bar{t})
\)
so that, as $\bar{s}$ is linear,
$\bar{s}_i$ is linear,
$\vars(\bar{s}_i) \inters \nonlinvars(\bar{t}_i) = \emptyset$
and
\[
  \bigl(\vars(\bar{s}_i) \inters \vars(\bar{t}_i)\bigr) \union \{s_i\}
    =
      \vars(\bar{s}) \inters \vars(\bar{t}).
\]
As $\bar{s}$ is linear and
$\vars(\bar{s}) \inters \nonlinvars(\bar{t}) = \emptyset$,
$s_i \notin \vars(\bar{s}_i) \union \vars(\bar{t}_i)$
and hence, for all $\mu \in \mgs(\bar{s} = \bar{t})$,
we have $s_i \notin \dom(\mu)$.
Therefore
\[
   \dom(\mu)
       \setdiff
     \bigl(\vars(\bar{s}) \inters \vars(\bar{t})\bigr)
  =
   \dom(\mu)
       \setdiff
     \bigl(\vars(\bar{s}_i) \inters \vars(\bar{t}_i)\bigr).
\]
Furthermore, by the congruence axiom (\ref{eq-ax:id}),
\(
  \mgs(\bar{s}_i = \bar{t}_i)
    = \mgs(\bar{s} = \bar{t}).
\)
Thus, as $\bar{s}$ and $\bar{t}$
are sequences of finite length $n$,
we can assume that $s_i \neq t_i$, for all $i=1$, \dots,~$n$.

Therefore, for the rest of the proof, we will assume that
for each $i=1$, \dots,~$n$,
$s_i \neq t_i$
and
either $s_i \in \Vars$ or $t_i \in \Vars$.

For the base case, we have
$\vars(\bar{s}) \union \vars(\bar{t}) = \emptyset$
and the result holds.

For the inductive step,
$\vars(\bar{s}) \union \vars(\bar{t}) \neq \emptyset$
so that $n>0$.
As the order of the equations in $\bar{s} = \bar{t}$
is not relevant to the hypothesis,
we assume, without loss of generality that if, for some $i=1$, \dots,~$n$,
$\vars(s_i) \inters \vars(t_i) = \emptyset$,
then $\vars(s_1) \inters \vars(t_1) = \emptyset$.
There are three cases we consider separately:
\begin{enumerate}
\item[a.]
for all $i=1$, \dots,~$n$,
$\vars(s_i) \inters \vars(t_i) \neq \emptyset$;
\item[b.]
$s_1 \in \Vars \setdiff \vars(t_1)$;
\item[c.]
$t_1 \in \Vars \setdiff \vars(s_1)$.
\end{enumerate}

\textbf{Case a.} For all $i=1$, \dots,~$n$,
$\vars(s_i) \inters \vars(t_i) \neq \emptyset$.

For each $i=1$, \dots,~$n$, we are assuming that
either $s_i \in \Vars$ or $t_i \in \Vars$,
Therefore, for each $i=1$, \dots,~$n$,
$s_i \in \vars(t_i)$ or $t_i \in \vars(s_i)$
so that, without loss of generality,
we can assume, for some $k$, where $0 \leq k \leq n$,
$s_i \in \Vars$ if $1 \leq i \leq k$
and $t_i \in \Vars$ if $k+1 \leq i \leq n$.

Let
\[
  \mu \defeq \{ s_1 = t_1, \ldots, s_k = t_k\}
               \union
             \{ t_{k+1} = s_{k+1}, \ldots, t_n = s_n\}.
\]
We now show that $\mu \sseq \Eqs$ is in rational solved form.
As $\bar{s}$ is linear,
$(s_1, \ldots, s_k)$ is linear.
As $\bar{s}$ is linear and
$t_i \in \vars(s_i)$ if $k+1 \leq i \leq n$,
then $(t_{k+1}, \ldots, t_n)$ is linear and
$\{s_1, \ldots, s_k\} \inters \{t_{k+1}, \ldots, t_n\} = \emptyset$.
As we are assuming that, for all $i=1$, \dots,~$n$,
$s_i \neq t_i$ and
$\vars(s_i) \inters \vars(t_i) \neq \emptyset$,
it follows that
$t_i \notin \Vars$ when $1 \leq i \leq k$ and
$s_i \notin \Vars$ when $k+1 \leq i \leq n$,
so that each equation in $\mu$ is a binding
and $\mu$ has no circular subsets.
Thus $\mu \in \RSubst$
and hence, by the congruence axiom (\ref{eq-ax:sym}),
$\mu \in \mgs(\bar{s} = \bar{t})$.

As $s_i \in \vars(t_i)$ when $1 \leq i \leq k$
and $t_i \in \vars(s_i)$ when $k+1 \leq i \leq n$,
\(
  \dom(\mu) \setdiff \bigl(\vars(\bar{s}) \inters \vars(\bar{t})\bigr)
    = \emptyset
\).
Therefore the required result holds.

\textbf{Case b.}
$s_1 \in \Vars \setdiff \vars(t_1)$.

Let
\begin{equation}
\label{eq:lem:share-linearly-mgs-properties:case-b:s1-t1}
\begin{split}
  \bar{s}_1 &\defeq
     (s_2, \ldots, s_n), \\
  \bar{t}_1 &\defeq
     \bigl(t_2\{s_1 \mapsto t_1\}, \ldots, t_n\{s_1 \mapsto t_1\}\bigr).
\end{split}
\end{equation}
As $\bar{s}$ is linear, $s_1 \notin \vars(\bar{s}_1)$.
Also, all occurrences of $s_1$ in $\bar{t}$
are replaced in $\bar{t}_1$ by $t_1$ so that,
as $s_1  \notin \vars(t_1)$,
$s_1 \notin \vars(\bar{t}_1)$.
Thus
\begin{equation}
\label{eq:lem:share-linearly-mgs-properties:case-b:s1-notin-s1-andt1}
s_1 \notin \vars(\bar{s}_1) \union \vars(\bar{t}_1).
\end{equation}
Therefore
\(
  \vars(\bar{s}_1) \union \vars(\bar{t}_1)
    \sslt \vars(\bar{s}) \union \vars(\bar{t})
\).
Now since $\bar{s}$ is linear, $\bar{s}_1$ is linear.
Thus, to apply the inductive hypothesis to
$\bar{s}_1$ and $\bar{t}_1$, we have to show that
\begin{equation}
\label{eq:lem:share-linearly-mgs-properties:case-b:mvars-pty-for-s1-t1}
\vars(\bar{s}_1) \inters \nonlinvars(\bar{t}_1) = \emptyset.
\end{equation}
Suppose that $u \in \vars(\bar{s}_1)$
so that $u \in \vars(\bar{s})$.
Now, by hypothesis, we have
$\vars(\bar{s}) \inters \nonlinvars(\bar{t}) = \emptyset$.
Thus $s_1,u \notin \nonlinvars(\bar{t})$.
If $u \in \vars\bigl((t_2, \ldots, t_n)\bigr)$
so that $u \notin \vars(t_1)$,
then $u \notin \nonlinvars(\bar{t}_1)$.
On the other hand, if $u \notin \vars\bigl((t_2, \ldots, t_n)\bigr)$,
then, as $s_1 \notin \nonlinvars\bigl((t_2, \ldots, t_n)\bigr)$
and $u  \notin \nonlinvars(t_1)$, $u \notin \nonlinvars(\bar{t}_1)$.
Thus,
for all $u \in \vars(\bar{s}_1)$,
$u \notin \nonlinvars(\bar{t}_1)$.
Hence
(\ref{eq:lem:share-linearly-mgs-properties:case-b:mvars-pty-for-s1-t1})
holds.
It follows that the inductive hypothesis for
$\bar{s}_1$ and $\bar{t}_1$ holds.
Therefore there exists $\mu_1 \in \RSubst$ where
\[
  \mu_1 \in \mgs(\bar{s}_1 = \bar{t}_1)
\]
such that,
for each
\(
  z
    \in
      \dom(\mu_1)
        \setdiff \bigl(\vars(\bar{s}_1) \inters \vars(\bar{t}_1)\bigr)
\),
$\vars(z\mu_1) \inters \dom(\mu_1) = \emptyset$.

Let
\begin{equation}
\label{eq:lem:share-linearly-mgs-properties:case-b:mu}
  \mu \defeq \{s_1 = t_1\mu_1\} \union \mu_1.
\end{equation}
We now show that $\mu \sseq \Eqs$ is in
$\mgs(\bar{s} = \bar{t})$.
First we show that $\mu$ is in rational solved form.
By (\ref{eq:lem:share-linearly-mgs-properties:case-b:s1-notin-s1-andt1}),
\begin{align}
\label{eq:lem:share-linearly-mgs-properties:case-b:s1-notin-mu1}
s_1 &\notin \vars(\mu_1),\\
\intertext{%
and, as $s_1 \notin \vars(t_1)$, we have
}
\label{eq:lem:share-linearly-mgs-properties:case-b:s1-notin-t1mu1}
s_1 &\notin \vars(t_1\mu_1).
\end{align}
Thus, as $\mu_1 \in \RSubst$,
 $\mu$ has no identities or circular subsets
so that $\mu \in \RSubst$.
By Lemma \ref{lem:iff-application},
$\mu \in \mgs(\bar{s} = \bar{t})$.

Let
\begin{equation}
\label{eq:lem:share-linearly-mgs-properties:case-b:z-in-dommu-diff-st}
  z \in
   \dom(\mu)
     \setdiff
   \bigl(\vars(\bar{s}) \inters \vars(\bar{t})\bigr).
\end{equation}
Then we have to show that
\begin{equation}
\label{eq:lem:share-linearly-mgs-properties:case-b:zmu-inters-dommu-empty}
  \vars(z\mu) \inters \dom(\mu) = \emptyset.
\end{equation}
It follows from
(\ref{eq:lem:share-linearly-mgs-properties:case-b:mu}) and
(\ref{eq:lem:share-linearly-mgs-properties:case-b:z-in-dommu-diff-st}) that
either $z \in \dom(\mu_1)$ so that $z\mu = z\mu_1$
or $z = s_1$ and $z\mu = t_1\mu_1$.
We consider these two cases separately.

Suppose first that $z \in \dom(\mu_1)$.
By (\ref{eq:lem:share-linearly-mgs-properties:case-b:s1-t1}),
we have both
$\vars(\bar{s}_1) \sseq  \vars(\bar{s})$
and
$\vars(\bar{t}_1) \sseq  \vars(\bar{t})$,
so that
\(
  \vars(\bar{s}_1) \inters \vars(\bar{t}_1)
    \sseq \vars(\bar{s}) \inters \vars(\bar{t})
\).
Hence we have
\(
  z
    \in
      \dom(\mu_1)
        \setdiff \bigl(\vars(\bar{s}_1) \inters \vars(\bar{t}_1)\bigr)
\).
Thus we obtain, by the inductive hypothesis,
$\vars(z\mu_1) \inters \dom(\mu_1) = \emptyset$.
Now, as $z \in \dom(\mu_1)$
and (\ref{eq:lem:share-linearly-mgs-properties:case-b:s1-notin-mu1})
holds,
 $s_1 \notin \vars(z\mu_1)$.
Thus, as $\dom(\mu) = \dom(\mu_1) \union \{s_1\}$,
$\vars(z\mu_1) \inters \dom(\mu) = \emptyset$.
Hence, as $z\mu = z\mu_1$,
(\ref{eq:lem:share-linearly-mgs-properties:case-b:zmu-inters-dommu-empty})
holds.

Secondly suppose that $z = s_1$.
Then we have that
$s_1 \notin \vars(\bar{s}) \inters \vars(\bar{t})$.
Hence $\bar{t}_1 = (t_2, \ldots, t_n)$.
Let $u$ be any variable in $\vars(t_1)$.
Then
we have that $u \notin \vars(\bar{s}_1) \inters \vars(\bar{t}_1)$,
since $\vars(\bar{s}) \inters \nonlinvars(\bar{t}) = \emptyset$.
If $u \in \dom(\mu_1)$, then we can apply the inductive hypothesis
to obtain
$\vars(u\mu_1) \inters \dom(\mu_1) = \emptyset$.
On the other hand, if $u \notin \dom(\mu_1)$, we have $u = u\mu_1$ and
$\vars(u\mu_1) \inters \dom(\mu_1) = \emptyset$.
Hence $\vars(t_1\mu_1) \inters \dom(\mu_1) = \emptyset$.
Thus, as $\dom(\mu) = \dom(\mu_1) \union \{s_1\}$,
by (\ref{eq:lem:share-linearly-mgs-properties:case-b:s1-notin-t1mu1}),
$\vars(t_1\mu_1) \inters \dom(\mu) = \emptyset$.
Therefore, as $z\mu = t_1\mu_1$,
(\ref{eq:lem:share-linearly-mgs-properties:case-b:zmu-inters-dommu-empty})
holds.

\textbf{Case c.}
$t_1 \in \Vars \setdiff \vars(s_1)$.

Let
\begin{equation}
\label{eq:lem:share-linearly-mgs-properties:case-c:s1-t1}
\begin{split}
  \bar{s}_1 &\defeq
     \bigl(s_2\{t_1 \mapsto s_1\}, \ldots, s_n\{t_1 \mapsto s_1\}\bigr), \\
  \bar{t}_1 &\defeq
     \bigl(t_2\{t_1 \mapsto s_1\}, \ldots, t_n\{t_1 \mapsto s_1\}\bigr).
\end{split}
\end{equation}
All occurrences of $t_1$ in $\bar{s}$ and $\bar{t}$
are replaced in $\bar{s}_1$ and $\bar{t}_1$ by $s_1$ so that,
since $t_1  \notin \vars(s_1)$,
\begin{equation}
\label{eq:lem:share-linearly-mgs-properties:case-c:t1-notin-s1-andt1}
t_1 \notin \vars(\bar{s}_1) \union \vars(\bar{t}_1).
\end{equation}
Therefore
\(
  \vars(\bar{s}_1) \union \vars(\bar{t}_1)
    \sslt \vars(\bar{s}) \union \vars(\bar{t})
\).
Now, $\bar{s}_1$ is linear since $\bar{s}$ is linear.
Thus, to apply the inductive hypothesis to
$\bar{s}_1$ and $\bar{t}_1$, we have to show that
\begin{equation}
\label{eq:lem:share-linearly-mgs-properties:case-c:mvars-pty-for-s1-t1}
\vars(\bar{s}_1) \inters \nonlinvars(\bar{t}_1) = \emptyset.
\end{equation}
Suppose $u$ is any variable in $\vars(\bar{s}_1)$.
Then either $u \in \vars\bigl((s_2, \ldots, s_n)\bigr)$
or we have $u \in \vars(s_1)$
and $t_1 \in \vars\bigl((s_2, \ldots, s_n)\bigr)$.
By hypothesis,
$\vars(\bar{s}) \inters \nonlinvars(\bar{t}) = \emptyset$,
so that $u \notin \nonlinvars(\bar{t})$.
If $u \in \vars\bigl((s_2, \ldots, s_n)\bigr)$,
then, as $\bar{s}$ is linear, $u \notin \vars(s_1)$.
Thus, it follows from
(\ref{eq:lem:share-linearly-mgs-properties:case-c:s1-t1}) that
 $u \notin \nonlinvars(\bar{t}_1)$.
If $t_1 \in \vars\bigl((s_2, \ldots, s_n)\bigr)$,
then we have $t_1 \notin \vars\bigl((t_2, \ldots, t_n)\bigr)$ so that,
again by (\ref{eq:lem:share-linearly-mgs-properties:case-c:s1-t1}),
$\bar{t}_1 = (t_2, \ldots, t_n)$.
Thus,
for all $u \in \vars(\bar{s}_1)$,
$u \notin \nonlinvars(\bar{t}_1)$.
Hence
(\ref{eq:lem:share-linearly-mgs-properties:case-c:mvars-pty-for-s1-t1})
holds.
It follows that the inductive hypothesis for
$\bar{s}_1$ and $\bar{t}_1$ holds.
Therefore there exists $\mu_1 \in \RSubst$ where
\[
  \mu_1 \in \mgs(\bar{s}_1 = \bar{t}_1)
\]
such that,
for each
\(
  z
    \in
      \dom(\mu_1)
        \setdiff \bigl(\vars(\bar{s}_1) \inters \vars(\bar{t}_1)\bigr)
\),
we have $\vars(z\mu_1) \inters \dom(\mu_1) = \emptyset$.

Let
\begin{equation}
\label{eq:lem:share-linearly-mgs-properties:case-c:mu}
  \mu \defeq \{t_1 = s_1\mu_1\} \union \mu_1.
\end{equation}
We now show that $\mu \sseq \Eqs$ is in
$\mgs(\bar{s} = \bar{t})$.
First we show that $\mu$ is in rational solved form.
By (\ref{eq:lem:share-linearly-mgs-properties:case-c:t1-notin-s1-andt1}),
\begin{align}
\label{eq:lem:share-linearly-mgs-properties:case-c:t1-notin-mu1}
t_1 &\notin \vars(\mu_1),\\
\intertext{%
and, as $t_1 \notin \vars(s_1)$, we have
}
\label{eq:lem:share-linearly-mgs-properties:case-c:t1-notin-s1mu1}
t_1 &\notin \vars(s_1\mu_1).
\end{align}
Thus, as $\mu_1 \in \RSubst$,
$\mu$ has no identities or circular subsets
so that $\mu \in \RSubst$.
By Lemma \ref{lem:iff-application},
$\mu \in \mgs(\bar{s} = \bar{t})$.

Let
\begin{equation}
\label{eq:lem:share-linearly-mgs-properties:case-c:z-in-dommu-diff-st}
  z \in
   \dom(\mu)
     \setdiff
   \bigl(\vars(\bar{s}) \inters \vars(\bar{t})\bigr).
\end{equation}
Then we have to show that
\begin{equation}
\label{eq:lem:share-linearly-mgs-properties:case-c:zmu-inters-dommu-empty}
  \vars(z\mu) \inters \dom(\mu) = \emptyset.
\end{equation}
It follows from
(\ref{eq:lem:share-linearly-mgs-properties:case-c:mu}) and
(\ref{eq:lem:share-linearly-mgs-properties:case-c:z-in-dommu-diff-st}) that
either $z \in \dom(\mu_1)$ so that $z\mu = z\mu_1$
or $z = t_1$ and $z\mu = s_1\mu_1$.
We consider these two cases separately.

Suppose first that $z \in \dom(\mu_1)$.
To apply the inductive hypothesis to $z$, we need to show that,
\[
  \vars(\bar{s}_1) \inters \vars(\bar{t}_1)
   \sseq \vars(\bar{s}) \inters \vars(\bar{t}).
\]
To see this, let us suppose $u \in \vars(\bar{s}_1) \inters \vars(\bar{t}_1)$.
Then,
by~(\ref{eq:lem:share-linearly-mgs-properties:case-c:s1-t1}),
either we have $u \in \vars\bigl((s_2, \ldots, s_n)\bigr)$ or
$u \in \vars(s_1)$ and $t_1 \in \vars\bigl((s_2, \ldots, s_n)\bigr)$.
If $u \in \vars\bigl((s_2, \ldots, s_n)\bigr)$,
then $u \in \vars(\bar{s})$ so that, as $\bar{s}$ is linear,
we have also $u \notin \vars(s_1)$
and hence $u \in \vars\bigl((t_2, \ldots, t_n)\bigr)$.
Alternatively, if
$u \in \vars(s_1)$ and $t_1 \in \vars\bigl((s_2, \ldots, s_n)\bigr)$,
then $u,t_1 \in \vars(\bar{s})$.
Moreover, by hypothesis,
$\vars(\bar{s}) \inters \nonlinvars(\bar{t}) = \emptyset$,
so that $t_1 \notin \vars\bigl((t_2, \ldots, t_n)\bigr)$.
Thus $\bar{t}_1 = (t_2, \ldots, t_n)$
and hence $u \in \vars(\bar{t})$.
Therefore, in both cases,
$u \in \vars(\bar{s}) \inters \vars(\bar{t})$.
It follows that
\(
  z \in
   \dom(\mu_1)
     \setdiff
   \bigl(\vars(\bar{s}_1) \inters \vars(\bar{t}_1)\bigr)
\).
Thus, by the inductive hypothesis,
we have $\vars(z\mu_1) \inters \dom(\mu_1) = \emptyset$.
Now, as $z \in \dom(\mu_1)$
and (\ref{eq:lem:share-linearly-mgs-properties:case-c:t1-notin-mu1})
holds, $t_1 \notin \vars(z\mu_1)$.
Thus, as $\dom(\mu) = \dom(\mu_1) \union \{t_1\}$,
$\vars(z\mu_1) \inters \dom(\mu) = \emptyset$.
Hence, as $z\mu = z\mu_1$,
(\ref{eq:lem:share-linearly-mgs-properties:case-c:zmu-inters-dommu-empty})
holds.

Secondly, suppose that $z = t_1$.
Then
$t_1 \notin \vars(\bar{s}) \inters \vars(\bar{t})$
and, consequently, $\bar{s}_1 = (s_2, \ldots, s_n)$.
Let $u$ be any variable in $\vars(s_1)$.
Then, as $\bar{s}$ is linear,
we have $u \notin \vars(\bar{s}_1)$ so that
$u \notin \vars(\bar{s}_1) \inters \vars(\bar{t}_1)$.
Thus, if $u \in \dom(\mu_1)$, we can apply the inductive hypothesis
to $u$ and obtain
$\vars(u\mu_1) \inters \dom(\mu_1) = \emptyset$.
On the other hand, if $u \notin \dom(\mu_1)$, $u = u\mu_1$ and
$\vars(u\mu_1) \inters \dom(\mu_1) = \emptyset$.
Hence $\vars(s_1\mu_1) \inters \dom(\mu_1) = \emptyset$.
Thus, as $\dom(\mu) = \dom(\mu_1) \union \{t_1\}$,
by (\ref{eq:lem:share-linearly-mgs-properties:case-c:t1-notin-s1mu1}),
$\vars(s_1\mu_1) \inters \dom(\mu) = \emptyset$.
Therefore, as $z\mu = s_1\mu_1$,
(\ref{eq:lem:share-linearly-mgs-properties:case-c:zmu-inters-dommu-empty})
holds.
\qed\end{pf}

\begin{lemma}
\label{lem:semi-linearity-mgs-properties}
Suppose that the tuple of terms
$\bar{s} \defeq (s_1,\ldots,s_n) \in \HTerms^n$ is linear,
$\bar{t} \defeq (t_1, \ldots, t_n) \in \HTerms^n$ and
$\mgs(\bar{s} = \bar{t}) \neq \emptyset$.
Then there exists
$\mu \in \mgs(\bar{s} = \bar{t})$
and,
for each $z \in \dom(\mu) \setdiff \vars(\bar{s})$,
the following properties hold:
\begin{enumerate}
\item
\label{pty:lem:semi-linearity-mgs-properties:is-in-s}
$\vars(z\mu) \sseq \vars(\bar{s})$;
\item
\label{pty:lem:semi-linearity-mgs-properties:is-free}
$\vars(z\mu) \inters \dom(\mu) = \emptyset$.
\end{enumerate}
\end{lemma}

\begin{pf}
The proof is by induction on the number of variables in
$\vars(\bar{s}) \union \vars(\bar{t})$.

Suppose first that,
for some $i=1$, \dots,~$n$,
we have $s_i = f(r_1, \ldots, r_m)$
and $t_i = f(u_1, \ldots, u_m)$ ($m \geq 0$).
Let
\begin{align*}
  \bar{s}' &\defeq (s_1, \ldots, s_{i-1},
                         r_1, \ldots, r_m,
                         s_{i+1}, \ldots, s_n),\\
  \bar{t}' &\defeq (t_1, \ldots, t_{i-1},
                         u_1, \ldots, u_m,
                         t_{i+1}, \ldots, t_n).
\end{align*}
Then
\(
  \mvars(\bar{s}') = \mvars(\bar{s})
\)
and
\(
  \mvars(\bar{t}') = \mvars(\bar{t})
\)
so that, as $\bar{s}$ is linear,
  $\bar{s}'$ is linear.
Moreover, by the congruence axiom (\ref{eq-ax:congr}),
\(
  \mgs(\bar{s}' = \bar{t}') = \mgs(\bar{s} = \bar{t})
\).
We repeat this process until all terms in
$\bar{s}'$ and $\bar{t}'$ can not be decomposed any further.
(Note that in the case that $s_i$ and $t_i$ are identical constants,
we can remove them from $\bar{s}'$ and $\bar{t}'$, since
the  corresponding equation $s_i = t_i$ holds vacuously.)
Thus, as $\bar{s}$ and $\bar{t}$
are finite sequences of finite terms,
we can assume that, for all $i=1$, \dots,~$n$,
either $s_i \in \Vars$ or $t_i \in \Vars$.

Secondly, suppose that
for some $i=1$, \dots,~$n$, $s_i = t_i$.
By the previous paragraph, we can assume that $s_i \in \Vars$.
Let
\begin{align*}
  \bar{s}_i &\defeq
     (s_1, \ldots, s_{i-1}, s_{i+1}, \ldots, s_n), \\
  \bar{t}_i &\defeq
     (t_1, \ldots, t_{i-1}, t_{i+1}, \ldots, t_n).
\end{align*}
Then
\(
  \mvars(\bar{s}_i) \union \{s_i\} = \mvars(\bar{s})
\)
and
\(
  \mvars(\bar{t}_i) \union \{s_i\} = \mvars(\bar{t})
\)
so that, as $\bar{s}$ is linear,
$\bar{s}_i$ is linear.
Therefore
\[
  \dom(\mu) \setdiff \vars(\bar{s})
     \sseq
       \dom(\mu) \setdiff \vars(\bar{s}_i).
\]
Furthermore, by the congruence axiom (\ref{eq-ax:id}),
\(
  \mgs(\bar{s}_i = \bar{t}_i) = \mgs(\bar{s} = \bar{t})
\).
Thus, as $\bar{s}$ and $\bar{t}$
are sequences of finite length $n$,
we can assume that $s_i \neq t_i$, for all $i=1$, \dots,~$n$.

Therefore, for the rest of the proof, we will assume that
$s_i \neq t_i$ and
either $s_i \in \Vars$ or $t_i \in \Vars$,
for all $i=1$, \dots,~$n$.

For the base case, we have
$\vars(\bar{s}) \union \vars(\bar{t}) = \emptyset$
and the result holds.

For the inductive step,
$\vars(\bar{s}) \union \vars(\bar{t}) \neq \emptyset$
so that $n>0$.
As the order of the equations in $\bar{s} = \bar{t}$
is not relevant to the hypothesis,
we assume, without loss of generality that if, for some $i=1$, \dots,~$n$,
$\vars(s_i) \inters \vars(t_i) = \emptyset$
then, we have $\vars(s_1) \inters \vars(t_1) = \emptyset$.
There are four cases we consider separately:
\begin{enumerate}
\item[a.]
for all $i=1$, \dots,~$n$,
$\vars(s_i) \inters \vars(t_i) \neq \emptyset$;
\item[b.]
$s_1 \in \Vars \setdiff \vars(t_1)$;
\item[c.]
$t_1 \in \Vars \setdiff \vars(\bar{s})$ and $s_1 \notin \Vars$;
\item[d.]
$t_1 \in \vars(\bar{s}) \setdiff \vars(s_1)$ and $s_1 \notin \Vars$.
\end{enumerate}

\textbf{Case a.} For all $i=1$, \dots,~$n$,
$\vars(s_i) \inters \vars(t_i) \neq \emptyset$.

For each $i=1$, \dots,~$n$, we are assuming that
either $s_i \in \Vars$ or $t_i \in \Vars$,
Therefore, for each $i=1$, \dots,~$n$,
$s_i \in \vars(t_i)$ or $t_i \in \vars(s_i)$
so that, without loss of generality,
we can assume, for some $k$, where $0 \leq k \leq n$,
$s_i \in \Vars$ if $1 \leq i \leq k$
and $t_i \in \Vars$ if $k+1 \leq i \leq n$.

Let
\[
  \mu \defeq \{ s_1 = t_1, \ldots ,s_k = t_k\}
               \union
             \{ t_{k+1} = s_{k+1}, \ldots ,t_n = s_n\}.
\]
We show that $\mu \sseq \Eqs$ is in $\mgs(\bar{s} = \bar{t})$.
First we must show that $\mu \in \RSubst$.
As $\bar{s}$ is linear,
$(s_1, \ldots, s_k)$ is linear.
As $\bar{s}$ is linear and
$t_i \in \vars(s_i)$ if $k+1 \leq i \leq n$,
then $(t_{k+1}, \ldots, t_n)$ is linear and
$\{s_1, \ldots, s_k\} \inters \{t_{k+1}, \ldots, t_n\} = \emptyset$.
As we are assuming that, for all $i=1$, \dots,~$n$,
$s_i \neq t_i$ and
$\vars(s_i) \inters \vars(t_i) \neq \emptyset$,
it follows that
$t_i \notin \Vars$ when $1 \leq i \leq k$ and
$s_i \notin \Vars$ when $k+1 \leq i \leq n$,
so that each equation in $\mu$ is a binding
and $\mu$ has no circular subsets.
Thus $\mu \in \RSubst$
and hence, by the congruence axiom (\ref{eq-ax:sym}),
$\mu \in \mgs(\bar{s} = \bar{t})$.

As $\{t_{k+1}, \ldots, t_n\} \sseq \vars\bigl((s_{k+1}, \ldots, s_n)\bigr)$,
we have
$\dom(\mu) \setdiff \vars(\bar{s}) = \emptyset$.
Therefore the required result holds.

\textbf{Case b.} $s_1 \in \Vars \setdiff \vars(t_1)$.

Let
\begin{align*}
  \bar{s}_1 &\defeq
     (s_2, \ldots, s_n), \\
  \bar{t}_1 &\defeq
     \bigl(t_2\{s_1 \mapsto t_1\}, \ldots, t_n\{s_1 \mapsto t_1\}\bigr).
\end{align*}
As $\bar{s}$ is linear, $\bar{s}_1$ is linear
and $s_1 \notin \vars(\bar{s}_1)$.
Also, all occurrences of $s_1$ in $\bar{t}$
are replaced in $\bar{t}_1$ by $t_1$ so that,
as $s_1  \notin \vars(t_1)$ (by the assumption for this case),
 $s_1 \notin \vars(\bar{t}_1)$.
Thus
\begin{equation}
\label{eq:lem:semi-linearity-mgs-properties:case-b:w-notin-s1-andt1}
s_1 \notin \vars(\bar{s}_1) \union \vars(\bar{t}_1).
\end{equation}
It follows that
\(
  \vars(\bar{s}_1) \union \vars(\bar{t}_1)
    \sslt \vars(\bar{s}) \union \vars(\bar{t})
\)
so that the inductive hypothesis applies to
$\bar{s}_1$ and $\bar{t}_1$.
Thus there exists $\mu_1 \in \RSubst$ where
\[
  \mu_1 \in \mgs(\bar{s}_1 = \bar{t}_1)
\]
such that, for each $z \in \dom(\mu_1) \setdiff \vars(\bar{s}_1)$,
properties
\ref{pty:lem:semi-linearity-mgs-properties:is-in-s}
and
\ref{pty:lem:semi-linearity-mgs-properties:is-free}
hold using $\mu_1$ and $\bar{s}_1$.

Let
\[
  \mu \defeq \{s_1 = t_1\mu_1\} \union \mu_1.
\]
We show that $\mu \sseq \Eqs$ is in $\mgs(\bar{s} = \bar{t})$.
By (\ref{eq:lem:semi-linearity-mgs-properties:case-b:w-notin-s1-andt1}),
we have $s_1 \notin \vars(\mu_1)$ so that $s_1 \notin \dom(\mu_1)$.
Also, since $\mu_1 \in \RSubst$, $\mu$ has no identities or circular subsets.
Thus we have $\mu \in \RSubst$.
By Lemma \ref{lem:iff-application},
$\mu \in \mgs(\bar{s} = \bar{t})$.

Suppose that $z \in \dom(\mu) \setdiff \vars(\bar{s})$.
As
\begin{align}
\notag
  \vars(\bar{s}_1) \union \{s_1\}
    &=
      \vars(\bar{s}) \\
\intertext{%
  and
}
\notag
  \dom(\mu_1) \union \{s_1\}
    &=
      \dom(\mu),
\intertext{%
we have
}
\label{eq:lem:semi-linearity-mgs-properties:case-b:dom-less-s-invariant}
  \dom(\mu_1) \setdiff \vars(\bar{s}_1)
    &=
      \dom(\mu) \setdiff \vars(\bar{s}).
\end{align}
Therefore $z \in \dom(\mu_1) \setdiff \vars(\bar{s}_1)$ and
$z\mu_1 = z\mu$.
Thus the inductive
properties~\ref{pty:lem:semi-linearity-mgs-properties:is-in-s}
and~\ref{pty:lem:semi-linearity-mgs-properties:is-free}
using $\mu_1$ and $\bar{s}_1$
can be applied to $z$.
We show that
properties~\ref{pty:lem:semi-linearity-mgs-properties:is-in-s}
and~\ref{pty:lem:semi-linearity-mgs-properties:is-free}
using $\mu$ and $\bar{s}$ can be applied to $z$.
\begin{enumerate}
\item
By property~\ref{pty:lem:semi-linearity-mgs-properties:is-in-s},
$\vars(z\mu) \sseq \vars(\bar{s}_1)$
and hence,
$\vars(z\mu) \sseq \vars(\bar{s})$.

\item
By property~\ref{pty:lem:semi-linearity-mgs-properties:is-free},
we have $\vars(z\mu) \inters \dom(\mu_1) = \emptyset$.
Now $s_1 \notin \vars(z\mu)$ because
$s_1 \notin \vars(\bar{s}_1)$ (since $\bar{s}$ is linear) and
$\vars(z\mu) \sseq \vars(\bar{s}_1)$
(by property~\ref{pty:lem:semi-linearity-mgs-properties:is-in-s}).
Thus, as $\dom(\mu) = \dom(\mu_1) \union \{s_1\}$,
we have $\vars(z\mu) \inters \dom(\mu) = \emptyset$.
\end{enumerate}

\textbf{Case c.}
Assume that $t_1 \in \Vars \setdiff \vars(\bar{s})$
and $s_1 \notin \Vars$.

Let
\begin{align*}
  \bar{s}_1 &\defeq
     (s_2, \ldots, s_n), \\
  \bar{t}_1 &\defeq
     \bigl(t_2\{t_1 \mapsto s_1\}, \ldots, t_n\{t_1 \mapsto s_1\}\bigr).
\end{align*}
As $\bar{s}$ is linear, $\bar{s}_1$ is linear.
By the assumption for this case, $t_1 \notin \vars(\bar{s}_1)$.
Also, all occurrences of $t_1$ in $\bar{t}$
are replaced in $\bar{t}_1$ by $s_1$ so that
 $t_1 \notin \vars(\bar{t}_1)$.
Thus
\begin{equation}
\label{eq:lem:semi-linearity-mgs-properties:case-c:t_1-notin-s1-andt1}
t_1 \notin \vars(\bar{s}_1) \union \vars(\bar{t}_1).
\end{equation}
It follows that
\(
  \vars(\bar{s}_1) \union \vars(\bar{t}_1)
    \sslt \vars(\bar{s}) \union \vars(\bar{t})
\)
so that we can apply the inductive hypothesis to
$\bar{s}_1$ and $\bar{t}_1$.
Thus there exists $\mu_1 \in \RSubst$ where
\[
  \mu_1 \in \mgs(\bar{s}_1 = \bar{t}_1)
\]
such that, for each $z \in \dom(\mu_1) \setdiff \vars(\bar{s}_1)$,
properties
\ref{pty:lem:semi-linearity-mgs-properties:is-in-s}
and
\ref{pty:lem:semi-linearity-mgs-properties:is-free}
hold using $\mu_1$ and $\bar{s}_1$.
Note that,
by (\ref{eq:lem:semi-linearity-mgs-properties:case-c:t_1-notin-s1-andt1}),
$t_1 \notin \vars(\mu_1)$ and, in particular, $t_1 \notin \dom(\mu_1)$.

Let
\begin{equation}
\label{eq:lem:semi-linearity-mgs-properties:subcase1-zmu}
  \mu \defeq \{t_1 = s_1\mu_1\} \union \mu_1.
\end{equation}
As $s_1 \notin \Vars$ and $\mu_1 \in \RSubst$,
$\mu \in \Eqs$ has no identities or circular subsets so that $\mu \in \RSubst$.
By Lemma \ref{lem:iff-application},
$\mu \in \mgs(\bar{s} = \bar{t})$.

As $t_1 \in \dom(\mu)$
(by (\ref{eq:lem:semi-linearity-mgs-properties:subcase1-zmu}))
and $t_1 \notin \vars(\bar{s})$
(by the assumption for this case),
we have
\[
  \dom(\mu_1) \setdiff \vars(\bar{s}_1) \union \{t_1\}
 =
  \dom(\mu) \setdiff \vars(\bar{s}).
\]
Suppose that $z \in \dom(\mu) \setdiff \vars(\bar{s})$.
Then either $z \neq t_1$ so that $z\mu = z\mu_1$ and
the inductive
properties~\ref{pty:lem:semi-linearity-mgs-properties:is-in-s}
and
\ref{pty:lem:semi-linearity-mgs-properties:is-free}
 using $\mu_1$ and $\bar{s}_1$
can be applied to $z$
or $z = t_1$ and $z\mu = s_1\mu_1$.
We show that properties~\ref{pty:lem:semi-linearity-mgs-properties:is-in-s}
and
\ref{pty:lem:semi-linearity-mgs-properties:is-free}
using $\mu$ and $\bar{s}$
can be applied to $z$.
\begin{enumerate}
\item
Suppose $z \neq t_1$ so that $z\mu = z\mu_1$.
Using
property~\ref{pty:lem:semi-linearity-mgs-properties:is-in-s},
$\vars(z\mu_1) \sseq \vars(\bar{s}_1)$.
As $\vars(\bar{s}_1) \sseq \vars(\bar{s})$,
it follows that $\vars(z\mu) \sseq \vars(\bar{s})$.

Suppose that $z = t_1$ so that $z\mu = s_1\mu_1$.
Let $u$ be any variable in $s_1$.
As $\bar{s}$ is linear, $u \notin \vars(\bar{s}_1)$.
Thus, if $u \in \dom(\mu_1)$,
we can use
property~\ref{pty:lem:semi-linearity-mgs-properties:is-in-s}
to derive that $\vars(u\mu_1) \sseq \vars(\bar{s}_1)$.
If $u \notin \dom(\mu_1)$, then
$u\mu_1 = u$ so that $\vars(u\mu_1) \sseq \vars(s_1)$.
Moreover
$\vars(s_1) \union  \vars(\bar{s}_1) = \vars(\bar{s})$
so that
\begin{equation}
\label{eq:lem:semi-linearity-mgs-properties:case-c:s1mu1-in-s}
\vars(s_1\mu_1) \sseq \vars(\bar{s}).
\end{equation}
Hence $\vars(z\mu) \sseq \vars(\bar{s})$.

\item
Suppose $z \neq t_1$ so that $z\mu = z\mu_1$.
Then, as
property~\ref{pty:lem:semi-linearity-mgs-properties:is-free} holds,
we have $\vars(z\mu) \inters \dom(\mu_1) = \emptyset$.
Now $t_1 \notin \vars(z\mu)$ because
$\vars(z\mu) \sseq \vars(\bar{s}_1)$
(by
property~\ref{pty:lem:semi-linearity-mgs-properties:is-in-s})
and $t_1 \notin \vars(\bar{s}_1)$
(by (\ref{eq:lem:semi-linearity-mgs-properties:case-c:t_1-notin-s1-andt1})).
Thus, as $\dom(\mu) = \dom(\mu_1) \union \{t_1\}$,
we have $\vars(z\mu) \inters \dom(\mu) = \emptyset$.

Suppose that $z = t_1$ so that $z\mu = s_1\mu_1$.
Let $u$ be any variable in $\vars(s_1)$.
Then, as $\bar{s}$ is linear, $u \notin \vars(\bar{s}_1)$.
Then either $u \in \dom(\mu_1)$,
and we can apply
property~\ref{pty:lem:semi-linearity-mgs-properties:is-free}
to $u$ to obtain $\vars(u\mu_1) \inters \dom(\mu_1) = \emptyset$,
or $u = u\mu_1$,
and $\vars(u\mu_1) \inters \dom(\mu_1) = \emptyset$.
Hence we have $\vars(s_1\mu_1) \inters \dom(\mu_1) = \emptyset$.
Now $t_1 \notin \vars(s_1\mu_1)$ because
$\vars(s_1\mu_1) \sseq \vars(\bar{s})$
(by (\ref{eq:lem:semi-linearity-mgs-properties:case-c:s1mu1-in-s}))
and $t_1 \notin \vars(\bar{s})$
(by the assumption for this case).
Thus, as $\dom(\mu) = \dom(\mu_1) \union \{t_1\}$,
we have $\vars(z\mu) \inters \dom(\mu) = \emptyset$.
\end{enumerate}

\textbf{Case d.}
Assume that
$t_1 \in \vars(\bar{s}) \setdiff \vars(s_1)$ and $s_1 \notin \Vars$.

Let
\begin{align*}
  \bar{s}_1 &\defeq
     \bigl(s_2\{t_1 \mapsto s_1\}, \ldots, s_n\{t_1 \mapsto s_1\}\bigr), \\
  \bar{t}_1 &\defeq
     \bigl(t_2\{t_1 \mapsto s_1\}, \ldots, t_n\{t_1 \mapsto s_1\}\bigr).
\end{align*}

As $\bar{s}$ is linear,
there is only one occurrence of $t_1$ in $\{s_2, \ldots, s_n\}$,
and, in $\bar{s}_1$, this is replaced by $s_1$ which is also linear.
Thus $\bar{s}_1$ is linear, $\bar{s}_1 \sseq \bar{s}$ and
$t_1 \notin \vars(\bar{s}_1)$.
Also, all occurrences of $t_1$ in $\bar{t}$
are replaced in $\bar{t}_1$ by $s_1$ so that
$t_1 \notin \vars(\bar{t}_1)$.
Thus
\begin{equation}
\label{eq:lem:semi-linearity-mgs-properties:case-d:t_1-notin-s1-andt1}
t_1 \notin \vars(\bar{s}_1) \union \vars(\bar{t}_1).
\end{equation}
It follows that
\(
  \vars(\bar{s}_1) \union \vars(\bar{t}_1)
    \sslt \vars(\bar{s}) \union \vars(\bar{t})
\)
so that we can apply the inductive hypothesis to
$\bar{s}_1$ and $\bar{t}_1$.
Thus, there exists $\mu_1 \in \RSubst$ where
\[
  \mu_1 \in \mgs(\bar{s}_1 = \bar{t}_1)
\]
such that, for each $z \in \dom(\mu_1) \setdiff \vars(\bar{s}_1)$,
properties
\ref{pty:lem:semi-linearity-mgs-properties:is-in-s}
and
\ref{pty:lem:semi-linearity-mgs-properties:is-free}
hold using $\mu_1$ and $\bar{s}_1$.

Let
\[
  \mu \defeq \{t_1 = s_1\mu_1\} \union \mu_1.
\]
By (\ref{eq:lem:semi-linearity-mgs-properties:case-d:t_1-notin-s1-andt1}),
$t_1 \notin \vars(\mu_1)$.
Moreover $\mu_1 \in \RSubst$ and $s_1 \notin \Vars$ so that $\mu \in \Eqs$
has no identities or circular subset.
Thus $\mu \in \RSubst$.
By Lemma \ref{lem:iff-application},
$\mu \in \mgs(\bar{s} = \bar{t})$.

As
$\vars(\bar{s}_1) \union \{t_1\} = \vars(\bar{s})$
and
$\dom(\mu_1) \union \{t_1\} = \dom(\mu)$, we have
\[
  \dom(\mu_1) \setdiff \vars(\bar{s}_1)
 =
  \dom(\mu) \setdiff \vars(\bar{s}).
\]
Suppose $z \in \dom(\mu) \setdiff \vars(\bar{s})$.
Then $z \neq t_1$, $z\mu = z\mu_1$
and the inductive
properties~\ref{pty:lem:semi-linearity-mgs-properties:is-in-s}
and~\ref{pty:lem:semi-linearity-mgs-properties:is-free}
using $\mu_1$ and $\bar{s}_1$ can be applied to $z$.
We show that the properties
\ref{pty:lem:semi-linearity-mgs-properties:is-in-s}
and
\ref{pty:lem:semi-linearity-mgs-properties:is-free}
using $\mu$ and $\bar{s}$
can be applied to $z$.
\begin{enumerate}
\item
By property~\ref{pty:lem:semi-linearity-mgs-properties:is-in-s},
$\vars(z\mu) \sseq \vars(\bar{s}_1)$
and hence, as $\bar{s}_1 \sseq \bar{s}$,
$\vars(z\mu) \sseq \vars(\bar{s})$.

\item
By property~\ref{pty:lem:semi-linearity-mgs-properties:is-free},
we have $\vars(z\mu) \inters \dom(\mu_1) = \emptyset$.
Now $t_1 \notin \vars(z\mu)$ because
$t_1 \notin \vars(\bar{s}_1)$
(by (\ref{eq:lem:semi-linearity-mgs-properties:case-d:t_1-notin-s1-andt1}))
and $\vars(z\mu) \sseq \vars(\bar{s}_1)$
(by property~\ref{pty:lem:semi-linearity-mgs-properties:is-in-s}).
It follows that
$\vars(z\mu) \inters \dom(\mu) = \emptyset$,
since $\dom(\mu_1) \union \{t_1\} = \dom(\mu)$.
\end{enumerate}
\qed\end{pf}

\begin{proposition}
\label{prop:finite-ground-propagates-finiteness}
Let $p \in P$ and $(x \mapsto t) \in \Bind$,
where $\{x\} \union \vars(t) \sseq \VI$.
Let also $\sigma \in \concrP(p) \inters \VSubst$ and
suppose that $\{r,r'\} = \{x, t\}$, $\vars(r) \sseq \hvars(\sigma)$ and
$\rt(r,\sigma) \in \GTerms$.
Then, for all $\tau \in \mgs\bigl(\sigma \union \{x = t\}\bigr)$,
we have
\begin{equation}
\label{eq:prop:finite-ground-propagates-finiteness:to-prove}
\hvars(\sigma) \union \vars(r') \sseq \hvars(\tau).
\end{equation}
\end{proposition}

\begin{pf}
If $\sigma \union \{x = t\}$ is not satisfiable,
the result is trivial.
We therefore assume,
for the rest of the proof, that $\sigma \union \{x = t\}$
is satisfiable in $\RT$.
It follows from Corollary~\ref{cor:all-mgs-have-same-hvars}
that we just have to show that
\begin{enumerate}
\item
\label{prop:finite-ground-propagates-finiteness:statement-1}
$\vars(r') \sseq \hvars(\tau)$,
for some $\tau \in \mgs\bigl(\sigma \union \{x = t\}\bigr)$;
\item
\label{prop:finite-ground-propagates-finiteness:statement-2}
$\hvars(\sigma) \sseq \hvars(\tau)$,
for some $\tau \in \mgs\bigl(\sigma \union \{x = t\}\bigr)$.
\end{enumerate}
From these, we can then conclude that, for all
$\tau \in \mgs\bigl(\sigma \union \{x = t\}\bigr)$,
(\ref{eq:prop:finite-ground-propagates-finiteness:to-prove}) holds.

Note that, in both cases,
since $\sigma \in \VSubst$ and $\vars(r) \sseq \hvars(\sigma)$,
by Proposition~\ref{prop:hvars-rt-is-rsigma-VSubst}
we have $\rt(r, \sigma) = r\sigma$,
so that $r\sigma \in \HTerms \inters \GTerms$.

We first prove
statement~\ref{prop:finite-ground-propagates-finiteness:statement-1}.
We must show that there exists
$\tau \in \mgs \bigl(\sigma \union \{ x = t \}\bigr)$
such that $\vars(r') \sseq \hvars(\tau)$.

As $\mgs \bigl(\sigma \union \{ x = t \}\bigr) \neq \emptyset$,
by Theorem~\ref{thm:VSubst-generality} and the definition of $\mgs$
we can assume that there exists
$\tau \in \VSubst \inters \mgs \bigl(\sigma \union \{ x = t \}\bigr)$.
Thus
\[
  \tau \implies \bigl(\sigma\union \{ r = r'\}\bigr).
\]
By Lemma~\ref{lem:application} and the congruence axioms,
we have $\tau \implies  \{r\sigma = r'\}$.
Since $\tau \in \VSubst$ and $r\sigma \in \HTerms \inters \GTerms$,
Lemma~\ref{lem:groundness-mgs-properties} applies (with $s = r\sigma$)
so that $r\sigma = r'\tau \in \HTerms \inters \GTerms$.
Thus, by Proposition~\ref{prop:hvars-alt-charact-VSubst},
$\vars(r') \sseq \hvars(\tau)$.

We now prove
statement~\ref{prop:finite-ground-propagates-finiteness:statement-2}.
In this case, we show that there exists
$\tau \in \mgs \bigl(\sigma \union \{ x = t \}\bigr)$
such that $\hvars(\sigma) \sseq \hvars(\tau)$.

Let
\begin{align*}
  \{u_1, \ldots, u_l\}
     &\defeq
  \dom(\sigma) \inters \vars(r'\sigma), \\
\bar{s} &\defeq (u_1, \ldots, u_l, r\sigma), \\
\bar{t} &\defeq (u_1\sigma, \ldots, u_l\sigma, r'\sigma).
\end{align*}
By Lemma~\ref{lem:iff-application} and the congruence axioms,
$\sigma \union \{x = t\} \implies \bar{s} = \bar{t}$.
Thus, as $\sigma \union \{x = t\}$ is satisfiable in $\RT$,
$\mgs(\bar{s} = \bar{t}) \neq \emptyset$.
Then, by Theorem~\ref{thm:VSubst-generality},
there exists $\mu \in \VSubst \inters \mgs(\bar{s} = \bar{t})$.
Therefore, since $r\sigma \in \HTerms \inters \GTerms$ and
$\mu \implies  \{r\sigma = r'\sigma\}$,
Lemma~\ref{lem:groundness-mgs-properties} applies (with $s=r\sigma$)
so that we can conclude $r\sigma = r'\sigma\mu \in \HTerms \inters \GTerms$.
Hence, for all $w \in \dom(\mu)$,
\begin{equation}
\label{eq:prop:finite-ground-propagates-finiteness:wmu-empty}
\vars(w\mu) = \emptyset.
\end{equation}

Let
\begin{align*}
  \nu
    &\defeq
      \bigl\{\,
        z = z\sigma\mu
      \bigm|
        z \in \dom(\sigma) \setdiff \vars(r'\sigma)
      \,\bigr\}, \\
  \tau
    &\defeq
      \nu \union \mu.
\end{align*}
Then,
as $\sigma, \mu \in \RSubst$, it follows from
(\ref{eq:prop:finite-ground-propagates-finiteness:wmu-empty}) that
$\nu, \tau \in \Eqs$ have no identities or circular subsets
so that $\nu, \tau \in \RSubst$.
By Lemma~\ref{lem:iff-application},
$\tau \in \mgs\bigl(\sigma \union \{x = t\}\bigr)$.

Suppose that $y \in \hvars(\sigma)$.
Then we show that $y \in \hvars(\tau)$.
Using Proposition~\ref{prop:hvars-rt-is-rsigma-VSubst},
$\rt(y,\sigma) = y\sigma$ and
\begin{equation}
\label{eq:prop:finite-ground-propagates-finiteness:y-sigma-domsigma}
\vars(y\sigma) \inters \dom(\sigma) = \emptyset.
\end{equation}

We show that $\vars(y\tau) \inters \dom(\tau) = \emptyset$.
Now, if $y \notin \dom(\tau)$, the result holds trivially.
Suppose that $y \in \dom(\nu)$, then $y\tau = y\sigma\mu$
and $y\in \dom(\sigma)$.
Let $w$ be any variable in $\vars(y\sigma)$
so that, by
(\ref{eq:prop:finite-ground-propagates-finiteness:y-sigma-domsigma}),
$w \notin \dom(\sigma)$.
If $w \notin \dom(\mu)$, then $w=w\mu \notin \dom(\tau)$.
If $w \in \dom(\mu)$, then, by
(\ref{eq:prop:finite-ground-propagates-finiteness:wmu-empty}),
$\vars(w\mu) = \emptyset$.
Therefore,
$\vars(w\mu) \inters \dom(\tau) = \emptyset$.
It follows that
$\vars(y\nu) \inters \dom(\tau) = \emptyset$.
Finally, suppose $y \in \dom(\mu)$.
Then,
by (\ref{eq:prop:finite-ground-propagates-finiteness:wmu-empty}),
$\vars(y\mu) = \emptyset$.
Therefore $\vars(y\mu) \inters \dom(\tau) = \emptyset$.

Therefore, using Definition~\ref{def:finiteness-op}, we have that
$y \in \hvars(\tau)$ as required.
\qed\end{pf}

\begin{proposition}
\label{prop:hvars-correctness-lin-ind}
Let $p \in P$ and $(x \mapsto t) \in \Bind$,
where $\{x\} \union \vars(t) \sseq \VI$.
Let also $\sigma \in \concrP(p) \inters \VSubst$
and suppose that
$x \in \hvars(\sigma)$ and $\vars(t) \sseq \hvars(\sigma)$.
Suppose also that
$\ind_p(x,t)$ and that $\orlin_p(x,t)$ hold.
Then, for all
substitutions $\tau \in \mgs\bigl(\sigma \union \{x = t\}\bigr)$,
\begin{equation}
\label{eq:lem:hvars-correctness-lin-ind:to-prove}
  \hvars(\sigma)
  \sseq
  \hvars(\tau).
\end{equation}
\end{proposition}

\begin{pf}
If $\sigma \union \{x = t\}$ is not satisfiable,
the result is trivial.
We therefore assume, for the rest of the proof,
that $\sigma \union \{x = t\}$ is satisfiable in $\RT$.
It follows from Corollary~\ref{cor:all-mgs-have-same-hvars}
that we just have to show that there exists
$\tau \in \mgs\bigl(\sigma \union \{x = t\}\bigr)$
such that
(\ref{eq:lem:hvars-correctness-lin-ind:to-prove}) holds.

As $x \in \hvars(\sigma)$ and $\vars(t) \sseq \hvars(\sigma)$,
by using Proposition~\ref{prop:hvars-rt-is-rsigma-VSubst} we obtain
$\rt(x,\sigma) = x\sigma$ and $\rt(t,\sigma) = t\sigma$.
Also
\begin{equation}
\label{eq:lem:hvars-correctness-lin-ind:x-t-sigma-domsigma}
\begin{split}
\vars(x\sigma) \inters \dom(\sigma) &= \emptyset, \\
\vars(t\sigma) \inters \dom(\sigma) &= \emptyset.
\end{split}
\end{equation}
As $\ind_{p}(x,t)$ holds,
\begin{equation}
\label{eq:lem:hvars-correctness-lin-ind:x-t-ind}
  \vars(x\sigma) \inters \vars(t\sigma)
   = \emptyset.
\end{equation}
By hypothesis, $\orlin(x,t)$ holds so that,
by Definition~\ref{def:predicates-and-functs-P},
for some $r\in \{x,t\}$, $r\sigma$ is linear.
Let $r' \defeq \{x,t\} \setdiff \{r\}$.

By Lemma~\ref{lem:iff-application} and the congruence axioms,
$\sigma \union \{x = t\} \implies \{r\sigma = r'\sigma\}$.
Thus, as $\sigma \union \{x = t\}$ is satisfiable in $\RT$,
$\mgs(r\sigma = r'\sigma) \neq \emptyset$.
Thus we can apply
Lemma~\ref{lem:share-linearly-mgs-properties}
(where $\bar{s} = r\sigma$ and  $\bar{t} = r'\sigma$)
so that,
using~(\ref{eq:lem:hvars-correctness-lin-ind:x-t-ind}),
there exists $\mu \in \mgs(x\sigma = t\sigma)$
such that, for all $w \in \dom(\mu)$,
\begin{equation}
\label{eq:lem:hvars-correctness-lin-ind:wmu-dommu-empty}
  \vars(w\mu) \inters \dom(\mu) = \emptyset.
\end{equation}
Note that,
by (\ref{eq:lem:hvars-correctness-lin-ind:x-t-sigma-domsigma}),
\begin{equation}
\label{eq:lem:hvars-correctness-lin-ind:domnu-varsmu-empty}
  \dom(\sigma) \inters \vars(\mu) = \emptyset.
\end{equation}

Let
\begin{align*}
  \nu  &\defeq \bigl\{\, z = z\sigma\mu \bigm| z \in \dom(\sigma) \,\bigr\}, \\
  \tau &\defeq \nu \union \mu.
\end{align*}
Then,
as $\sigma, \mu \in \RSubst$, it follows from
(\ref{eq:lem:hvars-correctness-lin-ind:domnu-varsmu-empty}) that
$\nu, \tau \in \Eqs$ have no identities or circular subsets
so that $\nu, \tau \in \RSubst$.
By Lemma~\ref{lem:iff-application},
$\tau \in \mgs\bigl(\sigma \union \{x = t\}\bigr)$.

Suppose $y \in \hvars(\sigma)$.
Then we show that $y \in \hvars(\tau)$.
As $y \in \HTerms$,
we have, using Proposition~\ref{prop:hvars-rt-is-rsigma-VSubst},
$\rt(y,\sigma) = y\sigma$
and
\begin{equation}
\label{eq:lem:hvars-correctness-lin-ind:y-sigma-domsigma}
\vars(y\sigma) \inters \dom(\sigma) = \emptyset.
\end{equation}

We show that $\vars(y\tau) \inters \dom(\tau) = \emptyset$.
If $y \notin \dom(\tau)$, the result holds trivially.
Suppose that $y \in \dom(\nu)$, then $y\tau = y\sigma\mu$.
Let $w$ be any variable in $\vars(y\sigma)$.
Then, by
(\ref{eq:lem:hvars-correctness-lin-ind:y-sigma-domsigma}),
$w \notin \dom(\sigma)$.
If $w \notin \dom(\mu)$, then
$w = w\mu \notin \dom(\tau)$.
If $w \in \dom(\mu)$, then
$\vars(w\mu) \sseq \vars(\mu)$ so that,
by (\ref{eq:lem:hvars-correctness-lin-ind:domnu-varsmu-empty}),
$\vars(w\mu) \inters \dom(\nu) = \emptyset$.
Moreover
(\ref{eq:lem:hvars-correctness-lin-ind:wmu-dommu-empty})
applies so that $\vars(w\mu) \inters \dom(\mu) = \emptyset$.
Therefore
we have $\vars(w\mu) \inters \dom(\tau) = \emptyset$.
It follows that
$\vars(y\nu) \inters \dom(\tau) = \emptyset$.
Finally, suppose $y \in \dom(\mu)$.
Then $y\tau = y\mu$ and,
by (\ref{eq:lem:hvars-correctness-lin-ind:domnu-varsmu-empty}),
we have $\vars(y\mu) \inters \dom(\nu) = \emptyset$.
Also
(\ref{eq:lem:hvars-correctness-lin-ind:wmu-dommu-empty}) applies
where $w$ is replaced by $y$
so that $\vars(y\mu) \inters \dom(\mu) = \emptyset$.
Thus $\vars(y\mu) \inters \dom(\tau) = \emptyset$.

Therefore, using Definition~\ref{def:finiteness-op}, we have that
$y \in \hvars(\tau)$ as required.
\qed\end{pf}

\begin{proposition}
\label{prop:both-finite-preserve-finiteness-bis}
Let $p \in P$ and $(x \mapsto t) \in \Bind$,
where $\{x\} \union \vars(t) \sseq \VI$.
Let also $\sigma \in \concrP(p) \inters \VSubst$ and
suppose that $x \in \hvars(\sigma)$ and $\vars(t) \sseq \hvars(\sigma)$.
Suppose also that
$\gfree_p(x)$ and $\gfree_p(t)$ hold.
Then, for all $\tau \in \mgs\bigl(\sigma \union \{x = t\}\bigr)$,
we have
\begin{equation}
\label{eq:lem:both-finite-preserve-finiteness-bis:to-prove}
\hvars(\sigma) \sseq \hvars(\tau).
\end{equation}
\end{proposition}

\begin{pf}
If $\sigma \union \{x = t\}$ is not satisfiable,
the result is trivial.
We therefore assume, for the rest of the proof,
that $\sigma \union \{x = t\}$ is satisfiable in $\RT$.
It follows from Corollary~\ref{cor:all-mgs-have-same-hvars}
that we just have to show that there exists
$\tau \in \mgs\bigl(\sigma \union \{x = t\}\bigr)$
such that
(\ref{eq:lem:both-finite-preserve-finiteness-bis:to-prove}) holds.

By Definition~\ref{def:predicates-and-functs-P},
$\gfree_p(x)$ and $\gfree_p(t)$ imply that either
$\rt(x, \sigma) \in \GTerms$ or $\rt(x, \sigma) \in \Vars$, and either
$\rt(t, \sigma) \in \GTerms$ or $\rt(t, \sigma) \in \Vars$.
Since we have
$\rt(x, \sigma),\rt(t, \sigma) \in \HTerms$ and $\sigma \in \VSubst$,
as a consequence of
Proposition~\ref{prop:hvars-rt-is-rsigma-VSubst}, we have
$\rt(x, \sigma)=x\sigma$,
$\rt(t, \sigma)=t\sigma$
and $x\sigma, t\sigma \notin \dom(\sigma)$.
There are three cases:
\begin{itemize}
\item
$\vars(x\sigma) = \emptyset \lor \vars(t\sigma) = \emptyset$.
Then the result follows from
Proposition~\ref{prop:finite-ground-propagates-finiteness}.
\item
$x\sigma = t\sigma \in \Vars$.
Then letting $\tau = \sigma$ gives the required result.
\item
$x\sigma,t\sigma \in \Vars$ are distinct variables.
Let $\tau = \sigma \union \{x\sigma = t\sigma\}$.
Then, as $x\sigma, t\sigma \notin \dom(\sigma)$,
$\tau \in \RSubst$.
Hence,
by Lemma~\ref{lem:iff-application},
$\tau \in \mgs \bigl(\sigma \union \{x = t\}\bigr)$.
Let $y$ be any variable in $\hvars(\sigma)$.
We show that $y \in \hvars(\tau)$.

Suppose first that $y \neq x\sigma$.
Then $y\tau = y\sigma$.
Thus using Proposition~\ref{prop:hvars-rt-is-rsigma-VSubst},
$\rt(y,\sigma) = y\tau$ and
$\vars(y\tau) \inters \dom(\sigma) = \emptyset$.
Thus
$\vars(y\tau) \inters \dom(\tau) \sseq \{x\sigma\}$.
However, $x\sigma\tau = t\sigma \notin \dom(\tau)$ so that,
by Definition~\ref{def:finiteness-functs},
$\vars(y\tau) \sseq \hvars_1(\tau)$
and hence $y \in \hvars_2(\tau)$.
Therefore, by  Definitions~\ref{def:finiteness-functs}
and~\ref{def:finiteness-op},
we have $y \in \hvars(\tau)$.

Secondly, suppose that $y = x\sigma$.
Then $y\tau = t\sigma$.
So that, as $t\sigma \in \Vars \setdiff \dom(\sigma)$
and $x\sigma \neq t\sigma$,
$\vars(y\tau) \inters \dom(\tau) = \emptyset$.
Therefore, using Definition~\ref{def:finiteness-op}, we have that
$y \in \hvars(\tau)$ as required.
\end{itemize}
\qed\end{pf}

\begin{proposition}
\label{prop:hvars-correctness-share-linearly}
Let $p \in P$ and $(x \mapsto t) \in \Bind$,
where $\{x\} \union \vars(t) \sseq \VI$.
Let $\sigma \in \concrP(p) \inters \VSubst$ and
suppose that $x \in \hvars(\sigma)$ and $\vars(t) \sseq \hvars(\sigma)$.
Furthermore, suppose that
$\orlin_p(x,t)$ and $\sharelin_p(x,t)$ hold.
Then,
for all
substitutions $\tau \in \mgs\bigl(\sigma \union \{x = t\}\bigr)$,
we have
\begin{equation}
\label{eq:lem:hvars-correctness-share-linearly:to-prove}
  \hvars(\sigma) \setdiff \sharesamevar_{p}(x,t)
  \sseq
  \hvars(\tau).
\end{equation}
\end{proposition}

\begin{pf}
If $\sigma \union \{x = t\}$ is not satisfiable,
the result is trivial.
We therefore assume, for the rest of the proof,
that $\sigma \union \{x = t\}$ is satisfiable in $\RT$.
It follows from Corollary~\ref{cor:all-mgs-have-same-hvars}
that we just have to show that there exists
$\tau \in \mgs\bigl(\sigma \union \{x = t\}\bigr)$
such that
(\ref{eq:lem:hvars-correctness-share-linearly:to-prove}) holds.

As $x \in \hvars(\sigma)$ and $\vars(t) \sseq \hvars(\sigma)$,
by using Proposition~\ref{prop:hvars-rt-is-rsigma-VSubst} we obtain
$\rt(x,\sigma) = x\sigma$ and $\rt(t,\sigma) = t\sigma$.
Also
\begin{equation}
\label{eq:lem:hvars-correctness-share-linearly:x-t-sigma-domsigma}
\vars(x\sigma) \inters \dom(\sigma) = \emptyset, \quad
\vars(t\sigma) \inters \dom(\sigma) = \emptyset.
\end{equation}
By hypothesis, $\orlin_p(x,t)$ holds so that,
by Definition~\ref{def:predicates-and-functs-P},
for some $r \in \{x,t\}$, $r\sigma$ is linear.
Also by hypothesis, $\sharelin_p(x,t)$ holds so that,
by Definition~\ref{def:predicates-and-functs-P},
if $r' = \{x,t\} \setdiff \{r\}$,
for all $z \in \vars(r\sigma) \inters \vars(r'\sigma)$,
$\occlin(z,r'\sigma)$ holds.
Therefore,
\begin{equation}
\label{eq:lem:hvars-correctness-share-linearly:r-r'-nonlinvars}
\vars(r\sigma) \inters \nonlinvars(r'\sigma) = \emptyset.
\end{equation}

By Lemma~\ref{lem:iff-application} and the congruence axioms,
$\sigma \union \{x = t\} \implies \{r\sigma = r'\sigma\}$.
Thus, as $\sigma \union \{x = t\}$ is satisfiable in $\RT$,
$\mgs(r\sigma = r'\sigma) \neq \emptyset$.
Thus, as $r\sigma$ is linear and
(\ref{eq:lem:hvars-correctness-share-linearly:r-r'-nonlinvars}) holds,
we can apply
Lemma~\ref{lem:share-linearly-mgs-properties}
(where $\bar{s} = r\sigma$ and  $\bar{t} = r'\sigma$)
so that
there exists $\mu \in \mgs(x\sigma = t\sigma)$
such that, for all
\(
  w \in \dom(\mu)
   \setdiff
        \bigl(\vars(x\sigma) \inters \vars(t\sigma)\bigr)
\),
\begin{equation}
\label{eq:lem:hvars-correctness-share-linearly:wmu-dommu-empty}
\vars(w\mu) \inters \dom(\mu) = \emptyset.
\end{equation}
Note that, by
(\ref{eq:lem:hvars-correctness-share-linearly:x-t-sigma-domsigma}),
\begin{equation}
\label{eq:lem:hvars-correctness-share-linearly:domnu-varsmu-empty}
\dom(\sigma) \inters \vars(\mu) = \emptyset.
\end{equation}

Let
\begin{align*}
  \nu  &\defeq \bigl\{\, z = z\sigma\mu \bigm| z \in \dom(\sigma) \,\bigr\}, \\
  \tau &\defeq \nu \union \mu.
\end{align*}
Then,
as $\sigma, \mu \in \RSubst$, it follows from
(\ref{eq:lem:hvars-correctness-share-linearly:domnu-varsmu-empty}) that
$\nu, \tau \in \Eqs$ have no identities or circular subsets
so that $\nu, \tau \in \RSubst$.
By Lemma~\ref{lem:iff-application},
$\tau \in \mgs\bigl(\sigma \union \{x = t\}\bigr)$.

Suppose $y \in \hvars(\sigma) \setdiff \sharesamevar_p(x,t)$.
We show that $y \in \hvars(\tau)$.
As $y \in \hvars(\sigma)$,
using Proposition~\ref{prop:hvars-rt-is-rsigma-VSubst},
$\rt(y,\sigma) = y\sigma$
and
\begin{equation}
\label{eq:lem:hvars-correctness-share-linearly:y-sigma-domsigma}
\vars(y\sigma) \inters \dom(\sigma) = \emptyset.
\end{equation}
As $y \notin \sharesamevar_{p}(x,t)$,
by Definition~\ref{def:predicates-and-functs-P},
\begin{equation}
\label{eq:lem:hvars-correctness-share-linearly:y-sigma-notin-inters}
  \vars(y\sigma) \inters \vars(x\sigma) \inters \vars(t\sigma)
   = \emptyset.
\end{equation}
Therefore, using
(\ref{eq:lem:hvars-correctness-share-linearly:y-sigma-notin-inters})
if $y \notin \dom(\sigma)$ and
(\ref{eq:lem:hvars-correctness-share-linearly:x-t-sigma-domsigma})
if $y \in \dom(\sigma)$, it follows that
\begin{equation}
\label{eq:lem:hvars-correctness-share-linearly:y-notin-inters}
  y \notin \vars(x\sigma) \inters \vars(t\sigma).
\end{equation}

We show that $\vars(y\tau) \inters \dom(\tau) = \emptyset$.
Now, if $y \notin \dom(\tau)$, the result holds trivially.
Suppose that $y \in \dom(\nu)$, then $y\tau = y\sigma\mu$.
Let $w$ be any variable in $\vars(y\sigma)$.
Then, by
(\ref{eq:lem:hvars-correctness-share-linearly:y-sigma-notin-inters}),
$w \notin \bigl(\vars(x\sigma) \inters \vars(t\sigma)\bigr)$ and,
by
(\ref{eq:lem:hvars-correctness-share-linearly:y-sigma-domsigma}),
$w \notin \dom(\sigma)$.
If $w \notin \dom(\mu)$, then
$w = w\mu \notin \dom(\tau)$.
If $w \in \dom(\mu)$, then $\vars(w\mu) \sseq \vars(\mu)$
so that,
by (\ref{eq:lem:hvars-correctness-share-linearly:domnu-varsmu-empty}),
we also have $\vars(w\mu) \inters \dom(\nu) = \emptyset$.
Moreover
(\ref{eq:lem:hvars-correctness-share-linearly:wmu-dommu-empty})
applies so that $\vars(w\mu) \inters \dom(\mu) = \emptyset$.
Therefore,
$\vars(w\mu) \inters \dom(\tau) = \emptyset$.
It follows that
$\vars(y\nu) \inters \dom(\tau) = \emptyset$.
Finally, suppose $y \in \dom(\mu)$.
Then $y\tau = y\mu$ and,
by (\ref{eq:lem:hvars-correctness-share-linearly:domnu-varsmu-empty}),
$\vars(y\mu) \inters \dom(\nu) = \emptyset$.
As (\ref{eq:lem:hvars-correctness-share-linearly:y-notin-inters}) holds,
(\ref{eq:lem:hvars-correctness-share-linearly:wmu-dommu-empty}) applies
where $w$ is replaced by $y$
so that $\vars(y\mu) \inters \dom(\mu) = \emptyset$.
Thus $\vars(y\mu) \inters \dom(\tau) = \emptyset$.

Therefore, using Definition~\ref{def:finiteness-op}, we have that
$y \in \hvars(\tau)$ as required.
\qed\end{pf}

\begin{proposition}
\label{prop:hvars-correctness-x-or-t-finite}
Let $p \in P$ and $(x \mapsto t) \in \Bind$,
where $\{x\} \union \vars(t) \sseq \VI$.
Let also $\sigma \in \concrP(p) \inters \VSubst$ and
suppose that $\{r,r'\} = \{x, t\}$,
$\vars(r) \sseq \hvars(\sigma)$ and $\lin_p(r)$ holds.
Then, for all $\tau \in \mgs\bigl(\sigma \union \{x = t\}\bigr)$, we have
\begin{equation}
\label{eq:lem:hvars-correctness-x-or-t-finite:to-prove}
\hvars(\sigma) \setdiff \sharewith_{p}(r) \sseq \hvars(\tau).
\end{equation}
\end{proposition}

\begin{pf}
If $\sigma \union \{x = t\}$ is not satisfiable,
the result is trivial.
We therefore assume, for the rest of the proof,
that $\sigma \union \{x = t\}$ is satisfiable in $\RT$.
It follows from Corollary~\ref{cor:all-mgs-have-same-hvars}
that we just have to show that there exists
$\tau \in \mgs\bigl(\sigma \union \{x = t\}\bigr)$
such that
(\ref{eq:lem:hvars-correctness-x-or-t-finite:to-prove}) holds.

By hypothesis, $\vars(r) \sseq \hvars(\sigma)$.
Hence,
by Proposition~\ref{prop:hvars-rt-is-rsigma-VSubst},
$\rt(r,\sigma) = r\sigma$ and
\begin{equation}
\label{eq:lem:hvars-correctness-x-or-t:r-sigma-domsigma}
\vars(r\sigma) \inters \dom(\sigma) = \emptyset.
\end{equation}
By hypothesis, $\lin_p(r)$ holds,
so that,
by Definition~\ref{def:predicates-and-functs-P},
$r\sigma$ is linear.

Let
\begin{align*}
  \{u_1, \ldots, u_l\}
     &\defeq
  \dom(\sigma) \inters \bigl(\vars(x\sigma) \union \vars(t\sigma)\bigr), \\
\bar{s} &\defeq (u_1, \ldots, u_l, r\sigma), \\
\bar{t} &\defeq (u_1\sigma, \ldots, u_l\sigma, r'\sigma).
\end{align*}
Since $r\sigma$ is linear,
it follows from
(\ref{eq:lem:hvars-correctness-x-or-t:r-sigma-domsigma})
that $\bar{s}$ is linear.
By Lemma~\ref{lem:iff-application} and the congruence axioms,
$\sigma \union \{x = t\} \implies \bar{s} = \bar{t}$.
Thus, as $\sigma \union \{x = t\}$ is satisfiable in $\RT$,
we have $\mgs(\bar{s} = \bar{t}) \neq \emptyset$.
Therefore, we can apply
Lemma~\ref{lem:semi-linearity-mgs-properties}
so that
there exists $\mu \in \mgs(\bar{s} = \bar{t})$
such that, for all
$w \in \dom(\mu) \setdiff \vars(\bar{s})$,
\begin{equation}
\label{eq:lem:hvars-correctness-x-or-t:wmu-dommu-empty}
\vars(w\mu) \inters \dom(\mu) = \emptyset.
\end{equation}
Note that, since $\sigma \in \VSubst$, for each $i = 1$, \dots,~$l$,
we have
\begin{align}
\notag
  \vars(u_i\sigma)
    &\sseq
      \vars(x\sigma) \union \vars(t\sigma). \\
\intertext{%
  Thus
}
\label{eq:lem:hvars-correctness-x-or-t:mu-in-union}
  \vars(\mu)
    &\sseq
      \vars(x\sigma) \union \vars(t\sigma).
\end{align}

Let
\begin{align*}
  \nu
    &\defeq
      \Bigl\{\,
        z = z\sigma\mu
      \Bigm|
        z \in \dom(\sigma)
          \setdiff \bigl(\vars(x\sigma) \union \vars(t\sigma)\bigr)
      \,\Bigr\}, \\
  \tau
    &\defeq
      \nu \union \mu.
\end{align*}
Then,
as $\sigma, \mu \in \RSubst$, it follows from
(\ref{eq:lem:hvars-correctness-x-or-t:mu-in-union}) that
$\nu, \tau \in \Eqs$ have no identities or circular subsets
so that $\nu, \tau \in \RSubst$.
By Lemma~\ref{lem:iff-application},
$\tau \in \mgs\bigl(\sigma \union \{x = t\}\bigr)$.

Suppose $y \in \hvars(\sigma) \setdiff \sharewith_{p}(r)$.
Then we show that $y \in \hvars(\tau)$.
As $y \in \hvars(\sigma)$,
by Proposition~\ref{prop:hvars-rt-is-rsigma-VSubst},
$\rt(y,\sigma) = y\sigma$ and
\begin{equation}
\label{eq:lem:hvars-correctness-x-or-t:y-sigma-domsigma}
\vars(y\sigma) \inters \dom(\sigma) = \emptyset.
\end{equation}
As $y \notin \sharewith_{p}(r)$,
by Definition~\ref{def:predicates-and-functs-P},
$y \notin \sharesamevar_{p}(y,r)$ so that,
using the same definition,
\begin{equation}
\label{eq:lem:hvars-correctness-x-or-t:y-sigma-notin-r-sigma}
  \vars(y\sigma) \inters \vars(r\sigma) = \emptyset.
\end{equation}
Therefore using
(\ref{eq:lem:hvars-correctness-x-or-t:y-sigma-notin-r-sigma})
if $y \notin \dom(\sigma)$ and
(\ref{eq:lem:hvars-correctness-x-or-t:r-sigma-domsigma})
if $y \in \dom(\sigma)$, it follows that
\begin{equation}
\label{eq:lem:hvars-correctness-x-or-t:y-notin-r-sigma}
  y \notin \vars(r\sigma).
\end{equation}

We show that $\vars(y\tau) \inters \dom(\tau) = \emptyset$.
Now, if $y \notin \dom(\tau)$, the result holds trivially.
Suppose that $y \in \dom(\nu)$.
Then $y\tau = y\sigma\mu$
and $y\in \dom(\sigma)$.
It follows from
 (\ref{eq:lem:hvars-correctness-x-or-t:y-sigma-domsigma})
and
 (\ref{eq:lem:hvars-correctness-x-or-t:y-sigma-notin-r-sigma}) that
$\vars(y\sigma) \inters \vars(\bar{s}) = \emptyset$.
Let $w$ be any variable in $\vars(y\sigma)$
so that $w \notin \vars(\bar{s})$.
By (\ref{eq:lem:hvars-correctness-x-or-t:y-sigma-domsigma}),
we have $w \notin \dom(\sigma)$.
If $w \notin \dom(\mu)$, then we have $w=w\mu \notin \dom(\tau)$.
If $w \in \dom(\mu)$, then $\vars(w\mu) \sseq \vars(\mu)$
so that,
by (\ref{eq:lem:hvars-correctness-x-or-t:mu-in-union}),
$\vars(w\mu) \inters \dom(\nu) = \emptyset$.
Moreover
(\ref{eq:lem:hvars-correctness-x-or-t:wmu-dommu-empty}) applies
so that $\vars(w\mu) \inters \dom(\mu) = \emptyset$.
Therefore,
$\vars(w\mu) \inters \dom(\tau) = \emptyset$.
It follows that
$\vars(y\nu) \inters \dom(\tau) = \emptyset$.
Finally, suppose $y \in \dom(\mu)$.
Then $y\tau = y\mu$ and,
by (\ref{eq:lem:hvars-correctness-x-or-t:mu-in-union}),
$\vars(y\mu) \inters \dom(\nu) = \emptyset$.
Since $\sigma \in \VSubst$ and $y \in \hvars(\sigma)$,
we have
\(
  y \notin
   \dom(\sigma) \inters \bigl(\vars(r\sigma) \union \vars(r'\sigma)\bigr)
\)
and hence
$y \notin \vars(\bar{s})$.
Therefore
(\ref{eq:lem:hvars-correctness-x-or-t:wmu-dommu-empty}) applies
and $\vars(y\mu) \inters \dom(\mu) = \emptyset$.
Thus $\vars(y\mu) \inters \dom(\tau) = \emptyset$.

Therefore, using Definition~\ref{def:finiteness-op}, we have that
$y \in \hvars(\tau)$ as required.
\qed\end{pf}

\begin{proposition}
\label{prop:amgu-correctness-non-affected-vars}
Let $p \in P$ and $(x \mapsto t) \in \Bind$,
where $\{x\} \union \vars(t) \sseq \VI$.
Let also $\sigma \in \concrP(p) \inters \VSubst$.
Then, for all
$\tau \in \mgs\bigl(\sigma \union \{x = t\}\bigr)$,
\begin{equation}
\label{eq:lem:amgu-correctness-non-affected-vars:to-prove}
  \hvars(\sigma)
     \setdiff
  \bigl(\sharewith_{p}(x)
          \union
        \sharewith_{p}(t)
  \bigr)
   \sseq \hvars(\tau).
\end{equation}
\end{proposition}

\begin{pf}
If $\sigma \union \{x = t\}$ is not satisfiable,
the result is trivial.
We therefore assume, for the rest of the proof,
that $\sigma \union \{x = t\}$ is satisfiable in $\RT$.
It follows from Corollary~\ref{cor:all-mgs-have-same-hvars}
that we just have to show that there exists
$\tau \in \mgs\bigl(\sigma \union \{x = t\}\bigr)$
such that
(\ref{eq:lem:amgu-correctness-non-affected-vars:to-prove}) holds.

Let
\begin{align*}
  \{u_1, \ldots, u_l\}
     &\defeq
  \dom(\sigma) \inters \bigl(\vars(x\sigma) \union \vars(t\sigma)\bigr), \\
\bar{s} &\defeq (u_1, \ldots, u_l, x\sigma), \\
\bar{t} &\defeq (u_1\sigma, \ldots, u_l\sigma, t\sigma).
\end{align*}
Note that, since $\sigma \in \VSubst$, for each $i = 1$, \dots,~$l$,
we have
\begin{align}
\notag
  \vars(u_i\sigma)
    &\sseq
      \vars(x\sigma) \union \vars(t\sigma).
\intertext{%
  Thus, for any $\mu \in \mgs(\bar{s} = \bar{t})$, we have
}
\label{eq:lem:hvars-correctness-non-affected-vars:mu-in-union}
  \vars(\mu)
    &\sseq
      \vars(x\sigma) \union \vars(t\sigma).
\end{align}

Let
\begin{align*}
  \nu
    &\defeq
      \Bigl\{\,
        z = z\sigma\mu
      \Bigm|
        z \in \dom(\sigma)
          \setdiff
            \bigl(\vars(x\sigma) \union \vars(t\sigma)\bigr)
      \,\Bigr\}, \\
  \tau
    &\defeq
      \nu \union \mu.
\end{align*}
Then,
as $\sigma, \mu \in \RSubst$, it follows from
(\ref{eq:lem:hvars-correctness-non-affected-vars:mu-in-union}) that
$\nu, \tau \in \Eqs$ have no identities or circular subsets
so that $\nu, \tau \in \RSubst$.
Thus, using Lemma~\ref{lem:iff-application} and the assumption that
$\sigma \union \{x = t\}$ is satisfiable in $\RT$,
$\tau \in \mgs\bigl(\sigma \union \{x = t\}\bigr)$.

Suppose that
\(
  y
    \in
      \hvars(\sigma)
        \setdiff \bigl(\sharewith_{p}(x) \union \sharewith_{p}(t) \bigr)
\).
We show that $y \in \hvars(\tau)$.
As $y \in \hvars(\sigma)$,
by Proposition~\ref{prop:hvars-rt-is-rsigma-VSubst},
$\rt(y,\sigma) = y\sigma$ and
\begin{equation}
\label{eq:lem:hvars-correctness-non-affected-vars:y-sigma-domsigma}
  \vars(y\sigma) \inters \dom(\sigma) = \emptyset.
\end{equation}
As
$y \notin \sharewith_{p}(x) \union \sharewith_{p}(t)$,
it follows from Definition~\ref{def:predicates-and-functs-P}
that
\[
  y \notin
   \sharesamevar_{p}(y,x) \union \sharesamevar_{p}(y,t)
\]
so that,
using the same definition with the result that $\rt(y,\sigma) = y\sigma$,
we obtain
\begin{equation}
\label{eq:lem:hvars-correctness-non-affected-vars:y-sigma-notin-union}
  \vars(y\sigma) \inters \bigl(\vars(x\sigma) \union \vars(t\sigma)\bigr)
       = \emptyset.
\end{equation}
Therefore, using
(\ref{eq:lem:hvars-correctness-non-affected-vars:y-sigma-notin-union})
if $y \notin \dom(\sigma)$ and
using the fact that $\sigma \in \VSubst$,
if $y \in \dom(\sigma)$, it follows that
\begin{equation}
\label{eq:lem:hvars-correctness-non-affected-vars:y-notin-union}
  y \notin \vars(x\sigma) \union \vars(t\sigma).
\end{equation}

We show that $\vars(y\tau) \inters \dom(\tau) = \emptyset$.
Now, if $y \notin \dom(\tau)$, the result holds trivially.
Suppose that $y \in \dom(\tau)$.
Then, by
(\ref{eq:lem:hvars-correctness-non-affected-vars:mu-in-union})
and
(\ref{eq:lem:hvars-correctness-non-affected-vars:y-notin-union}),
$y \notin \vars(\mu)$ so that $y \notin \dom(\mu)$
and $\vars(y\mu) \inters \dom(\mu) = \emptyset$.
Thus we must have $y \in \dom(\nu)$ and $y\tau = y\sigma$.
Then, by
(\ref{eq:lem:hvars-correctness-non-affected-vars:mu-in-union})
and
(\ref{eq:lem:hvars-correctness-non-affected-vars:y-sigma-notin-union}),
$\vars(y\sigma) \inters \dom(\mu) = \emptyset$.
Moreover, by
(\ref{eq:lem:hvars-correctness-non-affected-vars:y-sigma-domsigma}),
$\vars(y\sigma) \inters \dom(\sigma) = \emptyset$.
It follows that
$\vars(y\sigma) \inters \dom(\tau) = \emptyset$
and hence, as $y\sigma = y\tau$,
$\vars(y\tau) \inters \dom(\tau) = \emptyset$.

Therefore, using Definition~\ref{def:finiteness-op}, we have that
$y \in \hvars(\tau)$ as required.
\qed\end{pf}

\begin{pf*}{Proof of Theorem~\vref{thm:soundness-of-amguH}.}
By hypothesis, $\sigma \in \concrP(p)$.
By Theorem~\ref{thm:VSubst-generality},
there exists $\sigma' \in \VSubst$ such that
$\RT \entails \forall (\sigma \piff \sigma')$.
By Proposition~\ref{prop:project-hvars},
as $\sigma, \sigma'$ are satisfiable in $\RT$,
we have that $\hvars(\sigma) = \hvars(\sigma')$.
By Definition~\ref{def:P},
$\sigma \in \concrP(p)$
if and only if
$\sigma' \in \concrP(p)$.
We therefore safely assume that $\sigma \in \VSubst$.

By hypothesis, we have $\sigma \in \concrH(h)$.
Therefore, it follows from Definition~\ref{def:abstr-concr-H} that
$h \sseq \hvars(\sigma)$.
Similarly, by Definition~\ref{def:abstr-concr-H},
in order to prove $\tau \in \concrH(h')$,
we just need to show that $h' \sseq \hvars(\tau)$
where $h'$ is as defined in Definition~\ref{def:abs-funcs-H}.
There are eight cases that have to be considered.
\begin{enumerate}
%%
%% case 1.
\item
$\hterm_h(x) \land \ground_{p}(x)$ holds.

As $\hterm_h(x)$ holds,
by Definition~\ref{def:abs-funcs-H},
$x \in h$.
Hence,
by Definition~\ref{def:abstr-concr-H},
we have $x \in \hvars(\sigma)$.
As $\ground_{p}(x)$ holds,
by Definition~\ref{def:predicates-and-functs-P},
$\rt(x,\sigma) \in \GTerms$.
Therefore we can apply
Proposition~\ref{prop:finite-ground-propagates-finiteness},
where $r$ is replaced by $x$ and $r'$ by $t$, to conclude that
\[
  \hvars(\sigma) \union \vars(t) \sseq \hvars(\tau).
\]
%%
%% case 2.
\item
$\hterm_h(t) \land \ground_{p}(t)$ holds.

As $\hterm_h(t)$ holds,
by Definition~\ref{def:abs-funcs-H},
$\vars(t) \sseq h$.
Hence,
by Definition~\ref{def:abstr-concr-H},
$\vars(t) \sseq \hvars(\sigma)$.
As $\ground_{p}(t)$ holds,
by Definition~\ref{def:predicates-and-functs-P},
$\rt(t,\sigma) \in \GTerms$.
Therefore we can apply
Proposition~\ref{prop:finite-ground-propagates-finiteness},
where $r$ is replaced by $t$ and $r'$ by $x$, to conclude that
\[
  \hvars(\sigma) \union \{x\} \sseq \hvars(\tau).
\]
%%
%% case 3.
\item
\(
  \hterm_h(x) \land \hterm_h(t)
    \land
      \ind_p(x,t) \land \orlin_p(x,t)
\)
holds.

As $\hterm_h(x)$ and $\hterm_h(t)$ hold,
by Definition~\ref{def:abs-funcs-H},
$x \in h$ and $\vars(t) \sseq h$.
Hence,
by Definition~\ref{def:abstr-concr-H},
$x \in \hvars(\sigma)$ and $\vars(t) \sseq \hvars(\sigma)$.
Therefore we can apply
Proposition~\ref{prop:hvars-correctness-lin-ind} to conclude that
\[
  \hvars(\sigma) \sseq \hvars(\tau).
\]
%%
%% case 4.
\item
\(
  \hterm_h(x) \land \hterm_h(t)
    \land
      \gfree_p(x) \land \gfree_p(t)
\)
holds.

As $\hterm_h(x)$ and $\hterm_h(t)$ hold,
by Definition~\ref{def:abs-funcs-H},
$x \in h$ and $\vars(t) \sseq h$.
Hence,
by Definition~\ref{def:abstr-concr-H},
$x \in \hvars(\sigma)$ and $\vars(t) \sseq \hvars(\sigma)$.
Therefore we can apply
Proposition~\ref{prop:both-finite-preserve-finiteness-bis} to conclude that
\[
  \hvars(\sigma) \sseq \hvars(\tau).
\]
%%
%% case 5.
\item
\(
  \hterm_h(x) \land \hterm_h(t)
    \land
      \sharelin_p(x, t) \land \orlin_p(x,t)
\)
holds.

As $\hterm_h(x)$ and $\hterm_h(t)$ hold,
by Definition~\ref{def:abs-funcs-H},
$x \in h$ and $\vars(t) \sseq h$.
Hence,
by Definition~\ref{def:abstr-concr-H},
$x \in \hvars(\sigma)$ and $\vars(t) \sseq \hvars(\sigma)$.
Therefore we can apply
Proposition~\ref{prop:hvars-correctness-share-linearly} to conclude that
\[
  \hvars(\sigma) \setdiff \sharesamevar_p(x, t) \sseq \hvars(\tau).
\]
%%
%% case 6.
\item
\(
 \hterm_h(x) \land \lin_p(x)
\)
holds.

As $\hterm_h(x)$ holds,
by Definition~\ref{def:abs-funcs-H},
$x \in h$.
Hence,
by Definition~\ref{def:abstr-concr-H},
we have $x \in \hvars(\sigma)$.
Therefore we can apply
Proposition~\ref{prop:hvars-correctness-x-or-t-finite}
where $r$ is replaced by $x$ and $r'$ by $t$, to conclude that
\[
  \hvars(\sigma) \setdiff \sharewith_p(x) \sseq \hvars(\tau).
\]
%%
%% case 7.
\item
\(
  \hterm_h(t) \land \lin_p(t)
\)
holds.

As $\hterm_h(t)$ holds,
by Definition~\ref{def:abs-funcs-H},
$\vars(t) \sseq h$.
Hence,
by Definition~\ref{def:abstr-concr-H},
$\vars(t) \sseq \hvars(\sigma)$.
Therefore we can apply
Proposition~\ref{prop:hvars-correctness-x-or-t-finite}
where $r$ is replaced by $t$ and $r'$ by $x$, to conclude that
\[
  \hvars(\sigma) \setdiff \sharewith_p(t) \sseq \hvars(\tau).
\]
%%
%% case 8.
\item
For all $(x \mapsto t) \in \Bind$ where $\{x\} \union \vars(t) \sseq \VI$,
Proposition~\ref{prop:amgu-correctness-non-affected-vars} applies
so that
\[
  \hvars(\sigma)
    \setdiff \bigl(
               \sharewith_p(x)
                 \union
                   \sharewith_p(t)
             \bigr) \sseq \hvars(\tau).
\]
\end{enumerate}
\qed\end{pf*}

\begin{pf*}{Proof of Theorem~\vref{thm:soundness-of-projH}.}
Suppose that $\tau \in \projectD x \st \{ \sigma \}$.
We need to show that $\tau \in \concrH\bigl( \projH(h, x) \bigr)$.

Let $\overline{V} = \Vars \setdiff \VI$.
Then, by Definition~\ref{def:concrete-domain},
\(
  \RT
    \entails
      \forall
        \bigl(
          \exists \overline{V} \st (\tau \piff  \exists x \st \sigma)
        \bigr)
\).
Thus we have
\begin{equation}
\label{thm:soundness-of-projH:expanded-hypothesis}
  \RT
    \entails
      \forall
        \Bigl(
          \bigl(\exists \overline{V} \st \tau\bigr)
            \piff
              \bigl(\exists \overline{V} \union \{ x \} \st \sigma\bigr)
        \Bigr).
\end{equation}
Suppose $v \in \overline{V} \setdiff \vars(\sigma)$.
As we assumed that $\Vars$ is denumerable and that $\VI$ is finite,
such a $v$ will exist.  Moreover, as $x \in \VI$, we have $x \neq v$.
Let $\sigma' \in \RSubst$ be obtained from $\sigma$
by replacing every occurrence of $x$ by $v$.
Formally, if $\rho = \{x \mapsto v\}$, let
\[
  \sigma'
    \defeq
      \bigl\{\,
        y \mapsto y\sigma \rho
      \bigm|
        y \in \dom(\sigma) \setdiff \{ x \}
      \,\bigr\}
        \union \sigma'',
\]
where $\sigma'' = \{v \mapsto x\sigma \rho\}$
if $x \in \dom(\sigma)$ and $\emptyset$ otherwise.
Then $\sigma' \in \RSubst$ and
\[
  \RT
    \entails
      \forall
        \Bigl(
          \bigl(\exists \overline{V} \st \sigma'\bigr)
            \piff \bigl(\exists \overline{V} \union \{ x \} \st \sigma\bigr)
        \Bigr).
\]
Thus, by~(\ref{thm:soundness-of-projH:expanded-hypothesis}),
\(
  \RT
    \entails
      \forall
        \bigl(
          (\exists \overline{V} \st \tau)
            \piff (\exists \overline{V} \st \sigma')
        \bigr)
\).
Therefore, by Proposition~\ref{prop:project-hvars},
\begin{align}
\label{thm:soundness-of-projH:hvars tau=hvars-sigma'}
  \hvars(\tau) \inters \VI
    &= \hvars(\sigma') \inters \VI. \\
\intertext{
As $\sigma' \in \RSubst$ and $x \notin \dom(\sigma')$,
$\rt(x, \sigma') = x$ so that,
by Proposition~\ref{def:finiteness-op},
$x \in \hvars(\sigma')$.
Also, as $\sigma'$ is obtained from $\sigma$ by renaming $x$
to the new variable $v$,
$\hvars(\sigma') \Sseq \hvars(\sigma) \setdiff \{ v \}$.
Since $v \notin \VI$, we have
}
\notag
  \hvars(\sigma') \inters \VI
    &\Sseq \bigl(\hvars(\sigma) \union \{ x \}\bigr) \inters \VI. \\
\intertext{%
Therefore,
by~(\ref{thm:soundness-of-projH:hvars tau=hvars-sigma'}),
}
\label{thm:soundness-of-projH:hvars tau=hvars-sigma-union-x}
  \hvars(\tau) \inters \VI
    &\Sseq \bigl(\hvars(\sigma) \union \{ x \}\bigr) \inters \VI.
\end{align}

By hypothesis, $\sigma \in \concrH(h)$,
so that, by Definition~\ref{def:abstr-concr-H},
$\hvars(\sigma) \Sseq h$.
Therefore,
by~(\ref{thm:soundness-of-projH:hvars tau=hvars-sigma-union-x}),
$\hvars(\tau) \inters \VI \Sseq \bigl(h  \union \{ x \}\bigr) \inters \VI$.
Thus, by applying Definition~\ref{def:abstr-concr-H},
we can conclude that $\tau \in \concrH\bigl(h \union \{ x \}\bigr)$.
\qed\end{pf*}

\subsection{Finite-Tree Dependencies}
\label{sec:proofs-finite-tree-dependencies}
The proof of Theorem~\ref{thm:concrFD-properties} depends on the fact that
finite-tree dependencies only capture permanent information
and that the $\concrFD$ function is \emph{meet-preserving}.

\begin{proposition}
\label{prop:concrFD-is-entailment-closed}
Let $\sigma,\tau \in \RSubst$ and $\phi \in \Bfun$,
where $\sigma \in \concrFD(\phi)$ and $\tau \in \down \sigma$.
Then $\tau \in \concrFD(\phi)$.
\end{proposition}
\begin{pf}%%[of Proposition~\vref{prop:concrFD-is-entailment-closed}]
By the hypothesis, $\tau \in \down \sigma$, so that,
for each $\upsilon \in \down \tau$,
$\upsilon \in \down\sigma$.
Therefore, as $\sigma \in \concrFD(\phi)$,
it follows from Definition~\ref{def:concrFD} that,
for all $\upsilon \in \down \tau$,
$\phi\bigl(\hval(\upsilon)\bigr) = 1$
and hence $\tau \in \concrFD(\phi)$.
\qed\end{pf}

\begin{proposition}
\label{prop:concrFD-coadditive}
Let $\phi_1, \phi_2 \in \Bfun$.
Then
\[
  \concrFD(\phi_1 \land \phi_2)
    =  \concrFD(\phi_1) \inters \concrFD(\phi_2).
\]
\end{proposition}
\begin{pf}%%[of Proposition~\vref{prop:concrFD-coadditive}]
\begin{align*}
  \concrFD(\phi_1 \land \phi_2)
    &=
      \bigl\{\,
        \sigma \in \RSubst
      \bigm|
        \forall \tau \in \down\sigma
          \itc (\phi_1 \land \phi_2)\bigl(\hval(\tau)\bigr) = 1
      \,\bigr\} \\
    &=
      \bigl\{\,
        \sigma \in \RSubst
      \bigm|
        \forall \tau \in \down\sigma
          \itc
            \forall i \in \{ 1, 2 \}
              \itc \phi_i\bigl(\hval(\tau)\bigr) = 1
      \,\bigr\} \\
    &=
      \bigl\{\,
        \sigma \in \RSubst
      \bigm|
        \forall \tau \in \down\sigma
          \itc \phi_1\bigl(\hval(\tau)\bigr) = 1
      \,\bigr\} \\
    &\qquad\inters
      \bigl\{\,
        \sigma \in \RSubst
      \bigm|
        \forall \tau \in \down\sigma
          \itc \phi_2\bigl(\hval(\tau)\bigr) = 1
      \,\bigr\} \\
  &=
    \concrFD(\phi_1) \inters \concrFD(\phi_2).
\end{align*}
\qed\end{pf}

\begin{pf*}{Proof of Theorem~\vref{thm:concrFD-properties}.}
Assuming the hypothesis of the theorem,
we will prove each relation separately.

(\ref{eq:concrFD-properties:binding}).
Let $\sigma = \{ x \mapsto t \}$
and suppose that $\tau \in \down \sigma$.
Then, by Proposition~\ref{prop:down-entailment-equivalence},
$\RT \entails \forall\bigl(\tau \pimplies \sigma\bigr)$.
It follows from Lemma~\ref{lem:rt-equal-terms} that
$\rt(x, \tau) = \rt(t, \tau)$
and thus,
by Proposition~\ref{prop:rt-hvars-RSubst},
$x \in \hvars(\tau)$ if and only if $\vars(t) \sseq \hvars(\tau)$.
This is equivalent to
\(
  \bigl(x \piff \bigland \vars(t)\bigr)
    \bigl(\Bvalzero\bigl[1/\hvars(\tau)\bigr]\bigr)
      = 1
\)
and, by Definition~\ref{def:concrFD}, to
\(
  \bigl(x \piff \bigland \vars(t)\bigr)
     \bigl(\hval(\tau)\bigr) = 1
\).
As this holds for all $\tau \in \down \sigma$,
by Definition~\ref{def:concrFD},
$\sigma \in \concrFD\bigl(x \piff \bigland \vars(t)\bigr)$.

(\ref{eq:concrFD-properties:neg-x}).
Let $\sigma = \{ x \mapsto t \}$, where $x \in \vars(t)$.
By Definition~\ref{def:finiteness-op},
$x \notin \hvars(\sigma)$.
By case~(\ref{eq:down-gvars-hvars:hvars})
of Proposition~\ref{prop:down-gvars-hvars},
for all $\tau \in \down\sigma$,
we have $\hvars(\tau) \sseq \hvars(\sigma)$.
Thus $x \notin \hvars(\tau)$
and $(\neg x)\bigl(\hval(\tau)\bigr) = 1$.
Therefore, by Definition~\ref{def:concrFD},
$\sigma \in \concrFD(\neg x)$.

(\ref{eq:concrFD-properties:x}).
Let $\sigma \in \RSubst$ such that
$x \in \gvars(\sigma) \inters \hvars(\sigma)$.
By case~(\ref{eq:down-gvars-hvars:gvars-inters-hvars})
of Proposition~\ref{prop:down-gvars-hvars},
we have $x \in \hvars(\tau)$ for all $\tau \in \down\sigma$.
So $(x)\bigl(\hval(\tau)\bigr) = 1$.
Therefore, by Definition~\ref{def:concrFD},
$\sigma \in \concrFD(x)$.

(\ref{eq:concrFD-properties:and}).
Let $\sigma_1 \in \Sigma_1$ and $\sigma_2 \in \Sigma_2$.
Then, by hypothesis $\sigma_1 \in \concrFD(\phi_1)$
and $\sigma_2 \in \concrFD(\phi_2)$.
Let $\tau \in \mgs(\sigma_1 \union \sigma_2)$.
By definition of $\mgs$,
$\RT \entails \forall(\tau \pimplies \sigma_1)$ and
$\RT \entails \forall(\tau \pimplies \sigma_2)$.
Thus, by Proposition~\ref{prop:down-entailment-equivalence},
we have $\tau \in \down \sigma_1 \inters \down \sigma_2$.
Therefore, by Proposition~\ref{prop:concrFD-is-entailment-closed},
$\tau \in \concrFD(\phi_1) \inters \concrFD(\phi_2)$.
The result then follows by Proposition~\ref{prop:concrFD-coadditive}.

(\ref{eq:concrFD-properties:or}).
We have
\begin{align*}
  \concrFD(\phi_1 \lor \phi_2)
    &=
      \bigl\{\,
        \sigma \in \RSubst
      \bigm|
         \forall \tau \in \down\sigma
           \itc (\phi_1 \lor \phi_2)\bigl(\hval(\tau)\bigr) = 1
      \,\bigr\} \\
    &=
      \bigl\{\,
        \sigma \in \RSubst
      \bigm|
         \forall \tau \in \down\sigma
           \itc
             \exists i \in \{ 1, 2 \}
               \st \phi_i\bigl(\hval(\tau)\bigr) = 1
      \,\bigr\} \\
    &\Sseq
      \bigl\{\,
        \sigma \in \RSubst
      \bigm|
         \forall \tau \in \down\sigma
           \itc \phi_1\bigl(\hval(\tau)\bigr) = 1
      \,\bigr\} \\
    &\qquad\union
      \bigl\{\,
        \sigma \in \RSubst
      \bigm|
         \forall \tau \in \down\sigma
           \itc \phi_2\bigl(\hval(\tau)\bigr) = 1
      \,\bigr\} \\
  &=
    \concrFD(\phi_1) \union \concrFD(\phi_2) \\
  &\Sseq
    \Sigma_1 \union \Sigma_2.
\end{align*}

(\ref{eq:concrFD-properties:exists}).
Let $\sigma \in \Sigma$ and
let $\sigma' \in \projectD x \st \{ \sigma \}$.
We will show that $\sigma' \in \concrFD(\exists x \st \phi)$.

Let $\tau' \in \down \sigma'$.
Then there exists $\sigma'_1 \in \RSubst$ such that
$\RT \entails \forall\bigl(\tau' \piff (\sigma' \union \sigma_1')\bigr)$.
Let $\sigma_1 \in \projectD x \st \{\sigma_1'\}$ and
let $W \defeq (\Vars \setdiff \VI) \union \{x\}$.
Then, by Definition~\ref{def:concrete-domain}, it follows
$\RT \entails \forall\bigl(\exists W \st (\sigma' \piff \sigma)\bigr)$ and
$\RT \entails \forall\bigl(\exists W \st (\sigma_1' \piff \sigma_1)\bigr)$.
As a consequence
\[
  \RT
    \entails
      \forall
        \bigl(
          \exists W \st (\sigma' \union \sigma'_1)
            \piff
              \exists W \st (\sigma \union \sigma_1)
        \bigr).
\]
Therefore $\sigma \union \sigma_1$ is satisfiable in $\RT$
so that, for some $\tau \in \RSubst$,
\(
  \RT
    \entails
      \forall
        \bigl(
          \tau \piff (\sigma \union \sigma_1)
        \bigr)
\).
Thus
\(
  \RT
    \entails
      \forall(\exists W \st \tau \piff \exists W \st \tau')
\).
By Proposition~\ref{prop:project-hvars},
$\hvars(\tau') \setdiff W = \hvars(\tau) \setdiff W$ so that
\begin{equation}
\label{eq:concrFD-properties:exists:hvars-invariant}
  \bigl(\hvars(\tau') \inters \VI\bigr) \union \{x\}
    = (\hvars(\tau) \inters \VI) \union \{x\}.
\end{equation}
Let $c \defeq \hval(\tau)(x)$.
Then, since $\tau \in \down \sigma$ and,
by hypothesis, $\sigma \in \concrFD(\phi)$,
we have the following chain of implications:
\begin{align*}
  \phi\bigl(\hval(\tau)\bigr) &= 1
    &\law{by Defn.~\ref{def:concrFD}} \\
  \phi\bigl(\hval(\tau)[c/x]\bigr) &= 1
    &\law{by Defn.~\ref{def:boolean-valuations}} \\
  \phi\bigl(
        \Bvalzero \bigl[
                    1/\hvars(\tau) \inters \VI
                  \bigr][c/x]
      \bigr) &= 1
    &\law{by Defn.~\ref{def:concrFD}} \\
  \phi\bigl(
        \Bvalzero \bigl[
                    1/\bigl(\hvars(\tau) \inters \VI\bigr) \union \{x\}
                  \bigr][c/x]
      \bigr) &= 1
    &\law{by Defn.~\ref{def:boolean-valuations}} \\
  \phi\bigl(
        \Bvalzero \bigl[
                    1/\bigl(\hvars(\tau') \inters \VI \bigr) \union \{x\}
                  \bigr][c/x]
      \bigr) &= 1
    &\law{by~(\ref{eq:concrFD-properties:exists:hvars-invariant})} \\
  \phi\bigl(
        \Bvalzero \bigl[
                    1/\hvars(\tau') \inters \VI
                  \bigr][c/x]
      \bigr) &= 1
    &\law{by Defn.~\ref{def:boolean-valuations}} \\
  \phi\bigl(\hval(\tau')[c/x]\bigr) &= 1
    &\law{by Defn.~\ref{def:concrFD}} \\
  \phi[c/x]\bigl(\hval(\tau')\bigr) &= 1.
    &\law{by Defn.~\ref{def:boolean-functions}}
\end{align*}
From this last relation, since $\phi[c/x] \models \exists x \st \phi$,
it follows that
\[
  (\exists x \st \phi)\bigl(\hval(\tau')\bigr) = 1.
\]
As this holds for all $\tau' \in \down \sigma'$,
by Definition~\ref{def:concrFD},
$\sigma' \in \concrFD(\exists x \st \phi)$.
\qed\end{pf*}

\begin{pf*}{Proof of Theorem~\vref{thm:concrH-concrFD}.}
Since $h \sseq h'$,
by the monotonicity of $\concrH$ we have
$\concrH(h) \Sseq \concrH(h')$,
whence one of the inclusions:
$\concrH(h) \inters \concrFD(\phi) \Sseq \concrH(h') \inters \concrFD(\phi)$.

In order to establish the other inclusion,
we now prove that
$\sigma \in \concrH(h')$
assuming
$\sigma \in \concrH(h) \inters \concrFD(\phi)$.
To this end, by Definition~\ref{def:abstr-concr-H},
it is sufficient to prove that
$h' \sseq \hvars(\sigma)$.

Let $z \in h'$
and let $\psi = \bigl(\phi \land \bigland h\bigr)$,
so that, by hypothesis, $h' = \truev(\psi)$.
Therefore, we have $\psi \models z$.
Consider now
$\psi' = \bigl(\phi \land \bigland \hvars(\sigma)\bigr)$.
Since $\sigma \in \concrH(h)$,
by Definition~\ref{def:abstr-concr-H} we have
$h \sseq \hvars(\sigma)$,
so that $\psi' \models \psi$ and thus $\psi' \models z$.

Since $\sigma \in \concrFD(\phi)$,
by Definition~\ref{def:concrFD} we have
$\phi\bigl(\hval(\sigma)\bigr) = 1$.
Also note that
$\bigl(\bigland \hvars(\sigma)\bigr)\bigl(\hval(\sigma)\bigr) = 1$.
From these, by the definition of conjunction for Boolean
formulas, we obtain
$\psi'\bigl(\hval(\sigma)\bigr) = 1$.
Thus we can observe that
\begin{align*}
  \psi'\bigl(\hval(\sigma)\bigr) = 1
    &\iff
      (\psi' \land z)\bigl(\hval(\sigma)\bigr) = 1 \\
    &\implies
      z \in \hvars(\sigma).
\end{align*}
\qed\end{pf*}

\begin{pf*}{Proof of Theorem~\vref{thm:concrH-concrFD-consistency}.}
Suppose there exists
$\sigma \in \concrH(h) \inters \concrFD(\phi)$.
By Definition~\ref{def:concrFD},
since $\sigma \in \down\sigma$, we have
$\phi\bigl(\hval(\sigma)\bigr) = 1$;
moreover, we have
$\bigl( \bigland \hvars(\sigma) \bigr)\bigl(\hval(\sigma)\bigr) = 1$;
therefore, by the definition of conjunction for Boolean formulas,
we obtain
\[
  \Bigl( \phi \land \bigland h \Bigr)\bigl(\hval(\sigma)\bigr) = 1.
\]
As a consequence, we also have
\[
  \hvars(\sigma)
    \inters \falsev\Bigl( \phi \land \bigland h \Bigr)
      = \emptyset;
\]
by Definition~\ref{def:abstr-concr-H},
$h \sseq \hvars(\sigma)$,
so that we can conclude
$h \inters \falsev\bigl( \phi \land \bigland h \bigr) = \emptyset$.
\qed\end{pf*}

\subsection{Relation Between Groundness Dependencies
            and Finite-Tree Dependencies}
\label{sec:proofs-groundness-dependencies}

As was the case for finite-tree dependencies,
groundness dependencies only capture permanent information.
Moreover, the $\concrGD$ function is \emph{meet-preserving}.

\begin{proposition}
\label{prop:concrGD-is-entailment-closed}
Let $\sigma,\tau \in \RSubst$ and $\psi \in \Pos$,
where we have $\sigma \in \concrGD(\psi)$ and $\tau \in \down \sigma$.
Then $\tau \in \concrGD(\psi)$.
\end{proposition}
\begin{pf}%[of Proposition~\vref{prop:concrGD-is-entailment-closed}]
By the hypothesis, $\tau \in \down \sigma$, so that,
for each $\upsilon \in \down \tau$,
$\upsilon \in \down\sigma$.
Therefore, as $\sigma \in \concrGD(\psi)$,
it follows from Definition~\ref{def:concrGD} that,
for all $\upsilon \in \down \tau$,
$\psi\bigl(\gval(\upsilon)\bigr) = 1$
and hence $\tau \in \concrGD(\psi)$.
\qed\end{pf}

\begin{proposition}
\label{prop:concrGD-coadditive}
Let $\psi_1, \psi_2 \in \Pos$.
Then
\[
  \concrGD(\psi_1 \land \psi_2)
    =  \concrGD(\psi_1) \inters \concrGD(\psi_2).
\]
\end{proposition}

\begin{pf}%[of Proposition~\vref{prop:concrGD-coadditive}]
\begin{align*}
  \concrGD(\psi_1 \land \psi_2)
    &=
      \bigl\{\,
        \sigma \in \RSubst
      \bigm|
         \forall \tau \in \down\sigma
           \itc (\psi_1 \land \psi_2)\bigl(\gval(\tau)\bigr) = 1
      \,\bigr\} \\
    &=
      \sset
        {\sigma \in \RSubst}
        {\forall \tau \in \down\sigma
           \itc
             \forall i \in \{1,2\}
               \itc \\
         \qquad
                 \psi_i\bigl(\gval(\tau)\bigr) = 1
        } \\
    &=
      \bigl\{\,
        \sigma \in \RSubst
      \bigm|
         \forall \tau \in \down\sigma
           \itc \psi_1\bigl(\gval(\tau)\bigr) = 1
      \,\bigr\} \\
    &\qquad\inters
      \bigl\{\,
        \sigma \in \RSubst
      \bigm|
         \forall \tau \in \down\sigma
           \itc \psi_2\bigl(\gval(\tau)\bigr) = 1
      \,\bigr\} \\
  &=
    \concrGD(\psi_1) \inters \concrGD(\psi_2).
\end{align*}
\qed\end{pf}

Since non-ground terms can be made cyclic by instantiating their variables,
those terms detected as definitely finite on $\Bfun$
are also definitely ground.
\begin{proposition}
\label{prop:x-concrFD-sseq-concrGD}
Let $x \in \VI$.
Then $\concrFD(x) \sseq \concrGD(x)$.
\end{proposition}

\begin{pf}%[of Proposition~\vref{prop:x-concrFD-sseq-concrGD}]
Suppose that $\sigma \in \concrFD(x)$.
Then, by Definition~\ref{def:concrFD},
$(x)\bigl(\hval(\tau)\bigr) = 1$ for all $\tau \in \down \sigma$,
so that $x \in \hvars(\tau)$;
in particular, $x \in \hvars(\sigma)$.
We prove $x \in \gvars(\sigma)$ by contradiction.
That is, we show that if $x \in \hvars(\sigma) \setdiff \gvars(\sigma)$,
then there exists $\tau \in \down \sigma$ for which $x \notin \hvars(\tau)$.

Suppose that $x \in \hvars(\sigma) \setdiff \gvars(\sigma)$.
Then, by Propositions~\ref{prop:rt-hvars-RSubst} and
\ref{prop:RSubst-gvars},
$\rt(x, \sigma) \in \HTerms \setdiff \GTerms$.
Hence, by Proposition~\ref{prop:rt-vars-HTerms},
there exists $i \in \Nset$ such that $\rt(x, \sigma) = x\sigma^i$
and there exists $y \in \vars(x\sigma^i) \setdiff \dom(\sigma)$.
As we assumed that $\Sig$ contains a
function symbol of non-zero arity,
there exists $t \in \HTerms \setdiff \{ y \}$
for which $\{ y \} = \vars(t)$.
It follows that $\sigma' = \{ y \mapsto t \} \in \RSubst$
and, by Definition~\ref{def:finiteness-op},
$y \notin \hvars(\sigma')$.
Since $y \notin \dom(\sigma)$, by Lemma~\ref{lem:add-binding-new},
$\tau = \sigma \union \sigma' \in \RSubst$.
Since $\tau \in \down \sigma'$ then,
by case~(\ref{eq:down-gvars-hvars:hvars})
of Proposition~\ref{prop:down-gvars-hvars},
we have $y \notin \hvars(\tau)$.

By Lemma~\ref{lem:application}, we have
$\RT \entails \forall\bigl(\sigma \pimplies (x = x\sigma^i)\bigr)$.
Thus, since we also have $\tau \in \down \sigma$, we obtain
$\RT \entails \forall\bigl(\tau \pimplies (x = x\sigma^i)\bigr)$.
By applying Lemma~\ref{lem:rt-equal-terms},
we have that $\rt(x, \tau) = \rt(x\sigma^i, \tau)$ and thus,
by Proposition~\ref{prop:rt-hvars-RSubst},
we obtain $x \in \hvars(\tau)$ if and only if
$\vars(x\sigma^i) \sseq \hvars(\tau)$.
However, as observed before, we know that
$y \in \vars(x\sigma^i) \setdiff \hvars(\tau)$,
so that we also have $x \notin \hvars(\tau)$.

Therefore $x \in \gvars(\sigma) \inters \hvars(\sigma)$ and,
by case~(\ref{eq:down-gvars-hvars:gvars-inters-hvars})
of Proposition~\ref{prop:down-gvars-hvars},
for all $\tau \in \down \sigma$,
$x \in \gvars(\tau) \inters \hvars(\tau)$.
As a consequence, for all $\tau \in \down \sigma$,
$(x)\bigl(\gval(\tau)\bigr) = 1$,
so that, by Definition~\ref{def:concrGD},
we can conclude that $\sigma \in \concrGD(x)$.
\qed\end{pf}

\begin{pf*}{Proof of Theorem~\vref{thm:concrH-concrFD-concrGD}.}
\par\indent

Proof of (\ref{eq:concrH-concrFD-->>concrGD}).
Since $\psi \land \psi' \models \psi$, the inclusion
\[
  \concrH(h) \inters \concrFD(\phi) \inters \concrGD(\psi)
    \Sseq
      \concrH(h) \inters \concrFD(\phi) \inters \concrGD(\psi \land \psi')
\]
follows by the monotonicity of $\concrGD$.

We now prove the reverse inclusion.
Let us assume
$\sigma \in \concrH(h) \inters \concrFD(\phi) \inters \concrGD(\psi)$.
By Proposition~\ref{prop:concrGD-coadditive} we have that
$\concrGD(\psi \land \psi') = \concrGD(\psi) \inters \concrGD(\psi')$.
Therefore it is enough to show that
$\sigma \in \concrGD(\psi')$.
By hypothesis, $\psi' = \pos(\exists \VI \setdiff h \st \phi)$.
Moreover, by Definition~\ref{def:concrFD}, $h \sseq \hvars(\sigma)$.
Thus, to prove the result, we will show, by contradiction, that
\(
  \sigma
    \in
      \concrGD\Bigl(
                \pos\bigl(\exists \VI \setdiff \hvars(\sigma) \st \phi \bigr)
              \Bigr)
\).

Suppose therefore that
\(
  \sigma
    \notin
      \concrGD\Bigl(
                \pos\bigl(\exists \VI \setdiff \hvars(\sigma) \st \phi \bigr)
              \Bigr)
\).
Then there exists $\tau \in \down \sigma$
such that
\begin{equation}
\label{eq:hconcr-fdepconcr-gndconcr:improve-gnd:gval=0}
  \pos\bigl(
        \exists \VI \setdiff \hvars(\sigma) \st \phi
      \bigr)
    \bigl(\gval(\tau)\bigr) = 0.
\end{equation}

Let $z \in \hvars(\sigma) \inters \VI$.
By
Proposition~\ref{prop:rt-hvars-RSubst},
$\rt(z, \sigma) \in \HTerms$.
By Proposition~\ref{prop:rt-vars-HTerms},
there exists $i \in \Nset$
such that $\rt(z, \sigma) = z\sigma^i$ and
$\vars(z\sigma^i) \inters \dom(\sigma) = \emptyset$.
Therefore, by Definition~\ref{def:finiteness-op},
$\vars(z\sigma^i) \sseq \hvars(\sigma)$.
Thus, we have
\begin{equation}
\label{eq:hconcr-fdepconcr-gndconcr:improve-gnd:z-in-hvars-sigma}
  \vars(z\sigma^i) \sseq \hvars(\sigma) \setdiff \dom(\sigma).
\end{equation}
By Lemma~\ref{lem:application},
as $\tau \in \down \sigma$,
$\RT \entails \forall\bigl(\tau \pimplies (z = z\sigma^i)\bigr)$.
By Lemma~\ref{lem:rt-equal-terms}, we have
$\rt(z, \tau) = \rt(z\sigma^i, \tau)$
so that, by Proposition~\ref{prop:RSubst-gvars},
\begin{equation}
\label{eq:hconcr-fdepconcr-gndconcr:improve-gnd:z-in-gvars-tau}
  z \in \gvars(\tau)
    \iff
      \vars(z\sigma^i) \sseq \gvars(\tau).
\end{equation}

Take $t \in \GTerms \inters \HTerms$ and let
\begin{align*}
%%\label{eq:hconcr-fdepconcr-gndconcr:improve-gnd:upsilon1}
  \upsilon_1
    &\defeq
      \Bigl\{\,
        y \mapsto t
      \Bigm|
        y \in \bigl( \hvars(\sigma) \inters \gvars(\tau) \bigr)
                \setdiff \dom(\sigma)
      \,\Bigr\}. \\
\intertext{%
As we assumed that $\Sig$ contains
a function symbol of non-zero arity,
for each $y \in \Vars$
there exists $t_y \in \HTerms \setdiff \{ y \}$
such that $\vars(t_y) = \{ y \}$. Thus let
}
%%\label{eq:hconcr-fdepconcr-gndconcr:improve-gnd:upsilon2}
  \upsilon_2
    &\defeq
      \sset{
        y \mapsto t_y
      }{
        y \in \bigl( \VI \union \vars(\sigma) \bigr) \inters \hvars(\sigma) \\
        y \notin \gvars(\tau) \union \dom(\sigma)
      }.
\end{align*}
Note that $\upsilon_1$, $\upsilon_2 \in \RSubst$,
$\vars(\upsilon_1) \inters \vars(\upsilon_2) = \emptyset$
and $\vars(\upsilon_i) \inters \dom(\sigma) = \emptyset$, for $i = 1$,~$2$.
Thus, by Lemma~\ref{lem:add-binding-new},
$\tau' \defeq (\sigma \union \upsilon_1 \union \upsilon_2) \in \RSubst$
is satisfiable in $\RT$.

We now show that
\begin{equation}
\label{eq:hconcr-fdepconcr-gndconcr:improve-gnd:contradict-equation}
  z \in \gvars(\tau)
    \iff
      z \in \hvars(\tau').
\end{equation}

Assume first that $z \in \gvars(\tau)$.
Then, by~(\ref{eq:hconcr-fdepconcr-gndconcr:improve-gnd:z-in-gvars-tau}),
we have $\vars(z\sigma^i) \sseq \gvars(\tau)$.
From this, since
also~(\ref{eq:hconcr-fdepconcr-gndconcr:improve-gnd:z-in-hvars-sigma})
holds, we obtain $\vars(z\sigma^i) \sseq \dom(\upsilon_1)$ so that,
by Definitions~\ref{def:groundness-op} and~\ref{def:finiteness-op},
$\vars(z\sigma^i) \sseq \gvars(\upsilon_1) \inters \hvars(\upsilon_1)$.
Since $\tau' \in \down \upsilon_1$,
by case~(\ref{eq:down-gvars-hvars:gvars-inters-hvars}) of
Proposition~\ref{prop:down-gvars-hvars},
$\vars(z\sigma^i) \sseq \gvars(\tau') \inters \hvars(\tau')$.
Thus, by Propositions~\ref{prop:rt-hvars-RSubst}
and~\ref{prop:RSubst-gvars},
$\rt(z\sigma^i, \tau') \in \GTerms \inters \HTerms$.
Now $\tau' \in \down \sigma$ so that, by Lemma~\ref{lem:application},
$\RT \entails \forall\bigl(\tau' \pimplies  (z = z\sigma^i)\bigr)$.
By Lemma~\ref{lem:rt-equal-terms},
$\rt(z\sigma^i, \tau') = \rt(z, \tau') \in \GTerms \inters \HTerms$
so that, by Proposition~\ref{prop:rt-hvars-RSubst} and
Proposition~\ref{prop:RSubst-gvars},
$z \in \hvars(\tau')$.

We prove the other direction by contraposition,
assuming that $z \notin \gvars(\tau)$.
By~(\ref{eq:hconcr-fdepconcr-gndconcr:improve-gnd:z-in-gvars-tau}),
there exists $y \in \vars(z\sigma^i) \setdiff \gvars(\tau)$.
Also note that $y \in \VI \union \vars(\sigma)$ and,
by~(\ref{eq:hconcr-fdepconcr-gndconcr:improve-gnd:z-in-hvars-sigma}),
$y \notin \dom(\sigma)$ so that $y \in \dom(\upsilon_2)$.
By Definition~\ref{def:finiteness-op},
we have $y \notin \hvars(\upsilon_2)$ and,
since $\tau' \in \down \upsilon_2$,
by case~(\ref{eq:down-gvars-hvars:hvars}) of
Proposition~\ref{prop:down-gvars-hvars},
$y \notin \hvars(\tau')$.
Thus, by Proposition~\ref{prop:rt-hvars-RSubst},
we have that $\rt(z\sigma^i, \tau') \notin \HTerms$.
Moreover, as
$\RT \entails \forall\bigl(\tau' \pimplies (z = z\sigma^i)\bigr)$,
by Lemma~\ref{lem:rt-equal-terms} we have
$\rt(z\sigma^i, \tau') = \rt(z, \tau') \notin \HTerms$
and therefore, by Proposition~\ref{prop:rt-hvars-RSubst},
$z \notin \hvars(\tau')$.

Since $z$ was an arbitrary variable in $\hvars(\sigma) \inters \VI$,
it follows
from~(\ref{eq:hconcr-fdepconcr-gndconcr:improve-gnd:gval=0})
and~(\ref{eq:hconcr-fdepconcr-gndconcr:improve-gnd:contradict-equation})
that,
\begin{equation}
\label{eq:hconcr-fdepconcr-gndconcr:improve-gnd:hval=0}
  \pos\bigl(
        \exists \VI \setdiff \hvars(\sigma) \st \phi
      \bigr)
    \bigl(\hval(\tau')\bigr) = 0.
\end{equation}
We have by hypothesis that $\sigma \in \concrFD(\phi)$,
so that, as $\tau' \in \down \sigma$,
by Definition~\ref{def:concrFD} we have
$\phi\bigl(\hval(\tau')\bigr) = 1$.
Therefore, since
\(
  \phi
    \models
      \pos\bigl( \exists \VI \setdiff \hvars(\sigma) \st \phi \bigr)
\),
we obtain
\(
  \pos\bigl(
        \exists \VI \setdiff \hvars(\sigma) \st \phi
      \bigr)
    \bigl(\hval(\tau')\bigr)
      = 1
\),
which contradicts (\ref{eq:hconcr-fdepconcr-gndconcr:improve-gnd:hval=0}).

Proof of (\ref{eq:concrH-concrGD-->>concrFD}).
Since $\phi \land \phi' \models \phi$, the inclusion
\[
  \concrH(h) \inters \concrFD(\phi) \inters \concrGD(\psi)
    \Sseq
      \concrH(h) \inters \concrFD(\phi \land \phi') \inters \concrGD(\psi)
\]
follows by the monotonicity of $\concrFD$.

We now prove the reverse inclusion.
Assume that
$\sigma \in \concrH(h) \inters \concrFD(\phi) \inters \concrGD(\psi)$.
By Proposition~\ref{prop:concrFD-coadditive} we have that
$\concrFD(\phi \land \phi') = \concrFD(\phi) \inters \concrFD(\phi')$.
Therefore it is enough to show that
$\sigma \in \concrFD(\phi')$.
By hypothesis, $\phi' = \exists \VI \setdiff h \st \psi$.
Moreover, by Definition~\ref{def:abstr-concr-H}, $h \sseq \hvars(\sigma)$.
Thus, to prove the result, we will show, by contradiction, that
\(
  \sigma
    \in
      \concrFD\bigl(
                  \exists \VI \setdiff \hvars(\sigma) \st \psi
                \bigr)
\).

Suppose therefore that
\(
  \sigma
    \notin
      \concrFD\bigl(
                  \exists \VI \setdiff \hvars(\sigma) \st \psi
                \bigr)
\).
Then there exists $\tau \in \down \sigma$
such that
\begin{equation}
\label{eq:concrH-concrFD-concrGD:hval=0}
  \bigl(\exists \VI \setdiff \hvars(\sigma) \st \psi\bigr)
       \bigl(\hval(\tau)\bigr) = 0.
\end{equation}

Take $t \in \GTerms \inters \HTerms$ and let
\begin{equation}
\label{eq:concrH-concrFD-concrGD:upsilon}
  \upsilon
    \defeq
      \Bigl\{\,
        y \mapsto t
      \Bigm|
        y \in \vars(\sigma)
                \inters
                  \bigl(
                    \hvars(\tau) \setdiff \dom(\sigma)
                  \bigr)
      \,\Bigr\}.
\end{equation}
By Lemma~\ref{lem:add-binding-new},
$\tau' \defeq \sigma \union \upsilon \in \RSubst$
is satisfiable in $\RT$.

Let $z$ be any variable in $\hvars(\sigma)$.
By Proposition~\ref{prop:rt-hvars-RSubst},
$\rt(z,\sigma) \in \HTerms$.
Then, by Proposition~\ref{prop:rt-vars-HTerms},
there must exists $i \in \Nset$
such that $\rt(z, \sigma) = z\sigma^i$ and
$\vars(z\sigma^i) \inters \dom(\sigma) = \emptyset$.
Therefore, by Definition~\ref{def:finiteness-op},
$\vars(z\sigma^i) \sseq \hvars(\sigma)$.
Thus, we have
\begin{equation}
\label{eq:concrH-concrFD-concrGD:z-in-hvars-sigma}
  \vars(z\sigma^i) \sseq \hvars(\sigma) \setdiff \dom(\sigma).
\end{equation}
By Lemma~\ref{lem:application},
as $\tau \in \down \sigma$,
$\RT \entails \forall\bigl(\tau \pimplies  (z = z\sigma^i)\bigr)$.
By Lemma~\ref{lem:rt-equal-terms}, we have
$\rt(z, \tau) = \rt(z\sigma^i, \tau)$
so that, by Proposition~\ref{prop:rt-hvars-RSubst},
\begin{equation}
\label{eq:concrH-concrFD-concrGD:z-in-hvars-tau}
  z \in \hvars(\tau)
    \iff
      \vars(z\sigma^i) \sseq \hvars(\tau).
\end{equation}

We now show that
\begin{equation}
\label{eq:concrH-concrFD-concrGD:contradict-equation}
  \hvars(\tau) = \hvars(\sigma) \inters \gvars(\tau').
\end{equation}
Since $\tau \in \down \sigma$,
it follows from case~(\ref{eq:down-gvars-hvars:hvars}) of
Proposition~\ref{prop:down-gvars-hvars} that
$\hvars(\tau) \sseq \hvars(\sigma)$.
Thus, as $z \in \hvars(\sigma)$,
either $z \in \hvars(\tau)$ or $z \in \hvars(\sigma) \setdiff \hvars(\tau)$.
We consider these cases separately.

First, assume that $z \in \hvars(\tau)$.
Then, by~(\ref{eq:concrH-concrFD-concrGD:z-in-hvars-tau}),
$\vars(z\sigma^i) \sseq \hvars(\tau)$.
Also, by case~(\ref{eq:down-gvars-hvars:hvars}) of
Proposition~\ref{prop:down-gvars-hvars},
we have $z \in \hvars(\sigma)$, so that
we can apply~(\ref{eq:concrH-concrFD-concrGD:z-in-hvars-sigma})
to derive $\vars(z\sigma^i) \inters \dom(\sigma) = \emptyset$.
Therefore, $\vars(z\sigma^i) \sseq \dom(\upsilon)$ and,
by Definitions~\ref{def:groundness-op} and~\ref{def:finiteness-op},
$\vars(z\sigma^i) \sseq \gvars(\upsilon) \inters \hvars(\upsilon)$.
Since $\tau' \in \down \upsilon$,
by case~(\ref{eq:down-gvars-hvars:gvars-inters-hvars}) of
Proposition~\ref{prop:down-gvars-hvars}, we have
$\vars(z\sigma^i) \sseq \gvars(\tau') \inters \hvars(\tau')$.
Thus, by Propositions~\ref{prop:rt-hvars-RSubst}
and~\ref{prop:RSubst-gvars},
$\rt(z\sigma^i, \tau') \in \GTerms \inters \HTerms$.
Now $\tau' \in \down \sigma$ so that, by Lemma~\ref{lem:application},
we have $\RT \entails \forall\bigl(\tau' \pimplies  (z = z\sigma^i)\bigr)$.
Thus, by Lemma~\ref{lem:rt-equal-terms},
$\rt(z\sigma^i, \tau') = \rt(z, \tau') \in \GTerms \inters \HTerms$
so that, by
Propositions~\ref{prop:rt-hvars-RSubst} and  \ref{prop:RSubst-gvars},
$z \in \hvars(\tau') \inters \gvars(\tau')$.
Hence, by case~(\ref{eq:down-gvars-hvars:hvars})
of Proposition~\ref{prop:down-gvars-hvars},
we can conclude
$z \in \hvars(\sigma) \inters \gvars(\tau')$.
Thus $\hvars(\tau) \sseq \hvars(\sigma) \inters \gvars(\tau')$.

Secondly, assume that $z \in \hvars(\sigma) \setdiff \hvars(\tau)$.
Since $z \notin \hvars(\tau)$,
by~(\ref{eq:concrH-concrFD-concrGD:z-in-hvars-tau}),
there exists $y \in \vars(z\sigma^i) \setdiff \hvars(\tau)$.
Also, since $z \in \hvars(\sigma)$,
by~(\ref{eq:concrH-concrFD-concrGD:z-in-hvars-sigma}),
we have $y \in \hvars(\sigma) \setdiff \dom(\sigma)$ so that,
by Definition~\ref{def:groundness-op},
we have $y \notin \gvars(\sigma)$.
By (\ref{eq:concrH-concrFD-concrGD:upsilon}),
since $y \notin \dom(\sigma) \union \hvars(\tau)$,
we have $y \notin \dom(\upsilon)$ so that $y \notin \gvars(\tau')$.
Thus, by Proposition~\ref{prop:RSubst-gvars},
we have $\rt(z\sigma^i, \tau') \notin \GTerms$.
Moreover, since we have
$\RT \entails \forall\bigl(\tau' \pimplies  (z = z\sigma^i)\bigr)$,
we obtain, by Lemma~\ref{lem:rt-equal-terms},
$\rt(z\sigma^i, \tau') = \rt(z, \tau') \notin \GTerms$
and thus, by
Proposition~\ref{prop:RSubst-gvars},
$z \notin \gvars(\tau')$.
Thus $\hvars(\tau) \Sseq \hvars(\sigma) \inters \gvars(\tau')$.

It follows
from (\ref{eq:concrH-concrFD-concrGD:hval=0})
and (\ref{eq:concrH-concrFD-concrGD:contradict-equation})
that,
\begin{align}
\label{eq:concrH-concrFD-concrGD:gval=0}
 \bigl(
    \exists \VI \setdiff \hvars(\sigma) \st \psi
 \bigr)
   \bigl(\gval(\tau')\bigr)
     &= 0.
\intertext{%
We have by hypothesis that $\sigma \in \concrGD(\psi)$,
so that, as $\tau' \in \down \sigma$,
by Definition~\ref{def:concrGD} we have
$\psi\bigl(\gval(\tau')\bigr) = 1$.
Therefore, as
$\psi \models \exists \VI \setdiff \hvars(\sigma) \st \psi$,
}
\notag
  \bigl(
    \exists \VI \setdiff \hvars(\sigma) \st \psi
  \bigr)
    \bigl(\gval(\tau')\bigr)
      &= 1.
\end{align}
which contradicts (\ref{eq:concrH-concrFD-concrGD:gval=0}).
\qed\end{pf*}

\begin{pf*}{Proof of Theorem~\vref{thm:concrFD-->>concrGD}.}
Since $\psi \land \bigland \truev(\phi) \models \psi$,
the inclusion
\begin{align*}
  \concrFD(\phi) \inters \concrGD(\psi)
    &\Sseq \concrFD(\phi)
             \inters \concrGD\Bigl(\psi \land \bigland \truev(\phi)\Bigl) \\
\intertext{%
follows by the monotonicity of $\concrGD$.
To prove the inclusion
}
  \concrFD(\phi) \inters \concrGD(\psi)
    &\sseq
      \concrFD(\phi)
        \inters \concrGD\Bigl(\psi \land \bigland \truev(\phi)\Bigl)
\end{align*}
we will show that
$\concrFD(\phi) \sseq \concrGD\Bigl(\bigland \truev(\phi)\Bigl)$.
The thesis will thus follow by Proposition~\ref{prop:concrGD-coadditive}.
We have
\begin{align*}
  \concrFD(\phi)
    &\sseq
      \concrFD\Bigl(\bigland \truev(\phi)\Bigl)
&\law{since $\phi \models \bigland \truev(\phi)$} \\
    &=
      \biginters
        \bigl\{\, \concrFD(x) \bigm| x \in \truev(\phi) \,\bigr\}
&\law{by Proposition~\ref{prop:concrFD-coadditive}} \\
    &\sseq
      \biginters
        \bigl\{\, \concrGD(x) \bigm| x \in \truev(\phi) \,\bigr\}
&\law{by Proposition~\ref{prop:x-concrFD-sseq-concrGD}} \\
    &=
      \concrGD\Bigl(\bigland \truev(\phi)\Bigl).
&\law{by Proposition~\ref{prop:concrGD-coadditive}}
\end{align*}
\qed\end{pf*}

Part of the proof of Theorem~\ref{thm:concrH-concrFD-concrGD-subsumed}
relies on the following lemma.

\begin{lemma}
\label{lem:exists-pos}
Let $h \in H$ and $\phi \in \Bfun$ be such that
$\concrH(h) \inters \concrFD(\phi) \neq \emptyset$.
Then $(\exists \VI \setminus h \st \phi) \in \Pos$.
\end{lemma}
\begin{pf}
By hypothesis, there exists
$\sigma \in \concrH(h) \inters \concrFD(\phi)$ so that,
by Definitions~\ref{def:abstr-concr-H} and~\ref{def:concrFD},
we have $h \sseq \hvars(\sigma)$ and
\(
  \bigland \hvars(\sigma)
    \models
      \exists \VI \setminus \hvars(\sigma) \st \phi
\).
%$\phi \neq \bot$.

Towards a contradiction, suppose that
$(\exists \VI \setminus h \st \phi) \notin \Pos$, i.e.,
\begin{align*}
& (\exists \VI \setminus h \st \phi)(\Bvalone) = 0. \\
\intertext{%
Since existential quantification is an extensive operator on $\Bfun$
and $h \sseq \hvars(\sigma)$, we obtain
\(
  \exists \VI \setminus \hvars(\sigma) \st \phi
    \models
      \exists \VI \setminus h \st \phi
\),
so that
}
& \bigl(\exists \VI \setminus \hvars(\sigma) \st \phi\bigr)(\Bvalone) = 0. \\
\intertext{%
Moreover, since
\(
  \bigland \hvars(\sigma)
    \models
      \exists \VI \setminus \hvars(\sigma) \st \phi
\),
we have
}
& \Bigl( \bigland \hvars(\sigma) \Bigr)(\Bvalone) = 0.
\end{align*}
which is a contradiction.
% because $\bigland \hvars(\sigma) \in \Pos$.
Therefore, $(\exists \VI \setminus h \st \phi) \in \Pos$
\qed\end{pf}

\begin{pf*}{Proof of Theorem~\vref{thm:concrH-concrFD-concrGD-subsumed}.}
Let us assume the hypotheses and prove each statement in turn.

Consider first the case where $i=1$,
which corresponds to the application of the abstract disjunction operator.
Then, for the finiteness component $h_1$ we have:
\begin{align*}
  h_1
    &=
      h \inters h' \\
    &\Sseq
      \truev\Bigl(\phi \land \bigland h\Bigr)
        \inters
      \truev\Bigl(\phi' \land \bigland h'\Bigr) \\
    &\Sseq
      \truev\Bigl(\phi \land \bigland (h \inters h')\Bigr)
        \inters
      \truev\Bigl(\phi' \land \bigland (h \inters h')\Bigr) \\
    &=
      \truev\Bigl(\phi \land \bigland (h \inters h')
                    \lor
                  \phi' \land \bigland (h \inters h')
            \Bigr) \\
    &=
      \truev\Bigl((\phi \lor \phi') \land \bigland (h \inters h')\Bigr) \\
    &=
      \truev\Bigl(\phi_1 \land \bigland h_1\Bigr). \\
\intertext{%
For the finite-tree dependencies component $\phi_1$, we have:
}
  \phi_1
    &=
      \phi \lor \phi' \\
    &\models
      (\exists \VI \setdiff h \st \psi)
        \lor
          (\exists \VI \setdiff h' \st \psi') \\
    &\models
      \bigl(\exists \VI \setdiff (h \inters h') \st \psi \bigl)
        \lor
          \bigl(\exists \VI \setdiff (h \inters h') \st \psi' \bigl) \\
    &=
      \exists \VI \setdiff (h \inters h') \st \psi \lor \psi' \\
    &=
      \exists \VI \setdiff h_1 \st \psi_1. \\
\intertext{%%
For the groundness dependencies component $\psi_1$ we have:
}
  \psi_1
    &=
      \psi \lor \psi' \\
    &\models
      \pos(\exists \VI \setdiff h \st \phi)
        \lor
          \pos(\exists \VI \setdiff h' \st \phi') \\
    &=
      \Bigl(
        (\exists \VI \setdiff h \st \phi)
          \lor \bigland \VI 
      \Bigr)
        \lor
          \Bigl(
            (\exists \VI \setdiff h' \st \phi')
              \lor \bigland \VI
          \Bigr) \\
    &=
      (\exists \VI \setdiff h \st \phi)
        \lor (\exists \VI \setdiff h' \st \phi')
          \lor \bigland \VI \\
    &=
      \pos\bigl(
            (\exists \VI \setdiff h \st \phi)
              \lor
            (\exists \VI \setdiff h' \st \phi')
          \bigr) \\
    &\models
      \pos\Bigl(
            \bigl(\exists \VI \setdiff (h \inters h') \st \phi \bigl)
              \lor
            \bigl(\exists \VI \setdiff (h \inters h') \st \phi' \bigl)
          \Bigr) \\
     &=
      \pos\bigl(
            \exists \VI \setdiff (h \inters h') \st \phi \lor \phi'
          \bigr) \\
    &=
      \pos(\exists \VI \setdiff h_1 \st \phi_1). \\
\intertext{%%
Consider now the case where $i=2$,
which corresponds to the application of the abstract projection operator.
Then, for the finiteness component $h_2$ we have:
}
  h_2
    &=
      h \union \{x\} \\
    &\Sseq
      \truev\Bigl(\phi \land \bigland h\Bigr) \union \{x\} \\
    &\Sseq
      \truev\Bigl((\exists x \st \phi) \land \bigland h\Bigr) \union \{x\} \\
    &=
      \truev\Bigl((\exists x \st \phi)
                    \land
                  \bigland \bigl(h \union \{x\}\bigr)
            \Bigr) \\
    &=
      \truev\Bigl(\phi_2 \land \bigland h_2\Bigr). \\
\intertext{%%
For the finite-tree dependencies component $\phi_2$ we have:
}
  \phi_2
    &=
      \exists x \st \phi \\
    &\models
      \exists x \st \exists \VI \setdiff h \st \psi \\
    &=
      \exists \VI \setdiff h \st \exists x \st \psi \\
    &=
      \exists \VI \setdiff \bigl(h \union \{x\}\bigr)
          \st \exists x \st \psi \\
    &=
      \exists \VI \setdiff h_2 \st \psi_2. \\
\intertext{%%
By hypothesis,
$\concrH(h) \inters \concrFD(\phi) \neq \emptyset$
so that we also have
\(
    \concrH\bigl(h \union \{x\}\bigr) \inters \concrFD(\exists x \st \phi)
       \neq \emptyset
\).
Thus, 
for the groundness dependencies component $\psi_2$ we have:
}
  \psi_2
    &=
      \exists x \st \psi \\
    &\models
      \exists x \st \pos(\exists \VI \setdiff h \st \phi) \\
    &=
      \exists x \st \exists \VI \setdiff h \st \phi
 &&\text{[by Lemma~\ref{lem:exists-pos}]} \\
    &=
      \exists \VI \setdiff h \st \exists x \st \phi \\
    &=
      \exists \VI \setdiff \bigl(h \union \{x\} \bigr)
        \st \exists x \st \phi \\
    &=
      \pos\Bigl(
            \exists \VI \setdiff \bigl(h \union \{x\} \bigr)
              \st \exists x \st \phi
          \Bigr)
 &&\text{[by Lemma~\ref{lem:exists-pos}]} \\
    &=
      \pos(\exists \VI \setdiff h_2 \st \phi_2).
\end{align*}
\qed\end{pf*}


\begin{thebibliography}{10}
\expandafter\ifx\csname url\endcsname\relax
  \def\url#1{\texttt{#1}}\fi
\expandafter\ifx\csname urlprefix\endcsname\relax\def\urlprefix{URL }\fi

\bibitem{Colmerauer82}
A.~Colmerauer, {Prolog} and infinite trees, in: K.~L. Clark, S.~{\AA}.
  T{\"a}rnlund (Eds.), Logic Programming, APIC Studies in Data Processing,
  Vol.~16, Academic Press, New York, 1982, pp. 231--251.

\bibitem{Colmerauer90}
A.~Colmerauer, An introduction to {Prolog-III}, Communications of the ACM
  33~(7) (1990) 69--90.

\bibitem{SICStusManual-3_9}
Swedish Institute of Computer Science, Intelligent Systems Laboratory,
  {SICStus} {Prolog} User's Manual, release 3.9 Edition (2002).

\bibitem{SmolkaT94}
G.~Smolka, R.~Treinen, Records for logic programming, Journal of Logic
  Programming 18~(3) (1994) 229--258.

\bibitem{YAPManual-4_3_20}
V.~{Santos Costa}, L.~Damas, R.~Reis, R.~Azevedo, YAP User's Manual,
  Universidade do Porto, version 4.3.20 Edition (2001).

\bibitem{EggertC83}
P.~R. Eggert, K.~P. Chow, Logic programming, graphics and infinite terms, Tech.
  Rep. UCSB DoCS TR 83-02, Department of Computer Science, University of
  California at Santa Barbara (1983).

\bibitem{GiannesiniC84}
F.~Giannesini, J.~Cohen, Parser generation and grammar manipulation using
  {Prolog's} infinite trees, Journal of Logic Programming 3 (1984) 253--265.

\bibitem{CousotC95a}
P.~Cousot, R.~Cousot, Formal language, grammar and set-constraint-based program
  analysis by abstract interpretation, in: Proceedings of the Seventh ACM
  Conference on Functional Programming Languages and Computer Architecture, ACM
  Press, La Jolla, California, 1995, pp. 170--181.

\bibitem{JanssensB92}
G.~Janssens, M.~Bruynooghe, Deriving descriptions of possible values of program
  variables by means of abstract interpretation, Journal of Logic Programming
  13~(2{\&}3) (1992) 205--258.

\bibitem{VanHentenryckCLC95a}
P.~{Van Hentenryck}, A.~Cortesi, B.~{Le Charlier}, Type analysis of {Prolog}
  using type graphs, Journal of Logic Programming 22~(3) (1995) 179--209,.

\bibitem{Filgueiras84}
M.~Filgueiras, A {Prolog} interpreter working with infinite terms, in: Campbell
   \cite{Campbell84}, pp. 250--258.

\bibitem{HaridiS84}
S.~Haridi, D.~Sahlin, Efficient implementation of unification of cyclic
  structures, in: Campbell  \cite{Campbell84}, pp. 234--249.

\bibitem{Carro04}
M.~Carro, An application of rational trees in a logic programming interpreter
  for a procedural language, Tech. Rep.~{\tt arXiv:cs.DS/0403028}, School of
  Computer Science, Technical University of Madrid (UPM), available from
  \url{http://arxiv.org/} (2004).

\bibitem{Mukai91th}
K.~Mukai, Constraint logic programming and the unification of information,
  Ph.D. thesis, Department of Computer Science, Faculty of Engineering, Tokio
  Institute of Technology (1991).

\bibitem{PollardS94}
C.~Pollard, I.~A. Sag, Head-Driven Phrase Structure Grammar, University of
  Chicago Press, Chicago, 1994.

\bibitem{Carpenter92}
B.~Carpenter, The Logic of Typed Feature Structures with Applications to
  Unification-based Grammars, Logic Programming and Constraint Resolution,
  Vol.~32 of Cambridge Tracts in Theoretical Computer Science, Cambridge
  University Press, New York, 1992.

\bibitem{Erbach95}
G.~Erbach, {ProFIT}: {Prolog} with {F}eatures, {I}nheritance and {T}emplates,
  in: Proceedings of the 7th Conference of the European Chapter of the
  Association for Computational Linguistics, Dublin, Ireland, 1995, pp.
  180--187.

\bibitem{CodishT97}
M.~Codish, C.~Taboch, A semantic basis for termination analysis of logic
  programs and its realization using symbolic norm constraints, in: M.~Hanus,
  J.~Heering, K.~Meinke (Eds.), Algebraic and Logic Programming, 6th
  International Joint Conference, Vol. 1298 of Lecture Notes in Computer
  Science, Springer-Verlag, Berlin, Southampton, U.K., 1997, pp. 31--45.

\bibitem{CodishT99}
M.~Codish, C.~Taboch, A semantic basis for the termination analysis of logic
  programs, Journal of Logic Programming 41~(1) (1999) 103--123.

\bibitem{LindenstraussSS97}
N.~Lindenstrauss, Y.~Sagiv, A.~Serebrenik, {TermiLog}: A system for checking
  termination of queries to logic programs, in: O.~Grumberg (Ed.), Computer
  Aided Verification: Proceedings of the 9th International Conference, Vol.
  1250 of Lecture Notes in Computer Science, Springer-Verlag, Berlin, Haifa,
  Israel, 1997, pp. 444--447.

\bibitem{MesnardR05TPLP}
F.~Mesnard, R.~Bagnara, {cTI}: A constraint-based termination inference tool
  for {ISO-Prolog}, Theory and Practice of Logic Programming 5~(1{\&}2), to
  appear.

\bibitem{Stark96}
R.~F. St{\"a}rk, Total correctness of pure {Prolog} programs: A formal
  approach, in: R.~Dyckhoff, H.~Herre, P.~Schroeder-Heister (Eds.), Extensions
  of Logic Programming: Proceedings of the 5th International Workshop, Vol.
  1050 of Lecture Notes in Computer Science, Springer-Verlag, Berlin, Leipzig,
  Germany, 1996, pp. 237--254.

\bibitem{Stark98}
R.~F. St{\"a}rk, The theoretical foundations of {LPTP} (a {L}ogic {P}rogram
  {T}heorem {P}rover), Journal of Logic Programming 36~(3) (1998) 241--269.

\bibitem{CortesiLCR97}
A.~Cortesi, B.~{Le Charlier}, S.~Rossi, Specification-based automatic
  verification of {Prolog} programs, in: J.~P. Gallagher (Ed.), Logic Program
  Synthesis and Transformation: Proceedings of the 6th International Workshop,
  Vol. 1207 of Lecture Notes in Computer Science, Springer-Verlag, Berlin,
  Stockholm, Sweden, 1997, pp. 38--57.

\bibitem{DebrayL93}
S.~Debray, N.-W. Lin, Cost analysis of logic programs, ACM Transactions on
  Programming Languages and Systems 15~(5) (1993) 826--875.

\bibitem{HermenegildoBPL99}
M.~V. Hermenegildo, F.~Bueno, G.~Puebla, P.~L\'opez, Program analysis,
  debugging, and optimization using the ciao system preprocessor, in: D.~{De
  Schreye} (Ed.), Logic Programming: The 1999 International Conference, MIT
  Press Series in Logic Programming, The MIT Press, Las Cruces, New Mexico,
  1999, pp. 52--66.

\bibitem{ISO-Prolog-part-1}
ISO/IEC, {ISO/IEC 13211-1: 1995 Information technology --- Programming
  languages --- {Prolog} --- Part 1: General core}, International Standard
  Organization (1995).

\bibitem{BagnaraZH00}
R.~Bagnara, E.~Zaffanella, P.~M. Hill, Enhanced sharing analysis techniques: A
  comprehensive evaluation, in: M.~Gabbrielli, F.~Pfenning (Eds.), Proceedings
  of the 2nd International ACM SIGPLAN Conference on Principles and Practice of
  Declarative Programming, Association for Computing Machinery, Montreal,
  Canada, 2000, pp. 103--114.

\bibitem{BagnaraZH05TPLP}
R.~Bagnara, E.~Zaffanella, P.~M. Hill, Enhanced sharing analysis techniques: A
  comprehensive evaluation, Theory and Practice of Logic Programming
  5~(1{\&}2), to appear.

\bibitem{CortesiLCVH00}
A.~Cortesi, B.~{Le Charlier}, P.~{Van Hentenryck}, Combinations of abstract
  domains for logic programming: Open product and generic pattern construction,
  Science of Computer Programming 38~(1--3) (2000) 27--71.

\bibitem{CousotC92LP}
P.~Cousot, R.~Cousot, Abstract interpretation and applications to logic
  programs, Journal of Logic Programming 13~(2{\&}3) (1992) 103--179.

\bibitem{HillZB04TPLP}
P.~M. Hill, E.~Zaffanella, R.~Bagnara, A correct, precise and efficient
  integration of set-sharing, freeness and linearity for the analysis of finite
  and rational tree languages, Theory and Practice of Logic Programming 4~(3)
  (2004) 289--323, to appear.

\bibitem{Zaffanella01th}
E.~Zaffanella, Correctness, precision and efficiency in the sharing analysis of
  real logic languages, Ph.D. thesis, School of Computing, University of Leeds,
  Leeds, U.K., available at \url{http://www.cs.unipr.it/~zaffanella/} (2001).

\bibitem{BagnaraGHZ01}
R.~Bagnara, R.~Gori, P.~M. Hill, E.~Zaffanella, Finite-tree analysis for
  constraint logic-based languages, in: P.~Cousot (Ed.), Static Analysis: 8th
  International Symposium, SAS 2001, Vol. 2126 of Lecture Notes in Computer
  Science, Springer-Verlag, Berlin, Paris, France, 2001, pp. 165--184.

\bibitem{BagnaraZGH01}
R.~Bagnara, E.~Zaffanella, R.~Gori, P.~M. Hill, Boolean functions for
  finite-tree dependencies, in: R.~Nieuwenhuis, A.~Voronkov (Eds.), Proceedings
  of the 8th International Conference on Logic for Programming, Artificial
  Intelligence and Reasoning (LPAR 2001), Vol. 2250 of Lecture Notes in
  Artificial Intelligence, Springer-Verlag, Berlin, Havana, Cuba, 2001, pp.
  579--594.

\bibitem{BerarducciVZ93}
A.~Berarducci, M.~{Venturini Zilli}, Generalizations of unification, Journal of
  Symbolic Computation 15 (1993) 479--491.

\bibitem{King00}
A.~King, Pair-sharing over rational trees, Journal of Logic Programming
  46~(1--2) (2000) 139--155.

\bibitem{Colmerauer84}
A.~Colmerauer, Equations and inequations on finite and infinite trees, in:
  Proceedings of the International Conference on Fifth Generation Computer
  Systems (FGCS'84), ICOT, Tokyo, Japan, 1984, pp. 85--99.

\bibitem{JaffarLM87}
J.~Jaffar, J.-L. Lassez, M.~J. Maher, {Prolog-II} as an instance of the logic
  programming scheme, in: M.~Wirsing (Ed.), Formal Descriptions of Programming
  Concepts {III}, North-Holland, Amsterdam, 1987, pp. 275--299.

\bibitem{Keisu94th}
T.~Keisu, Tree constraints, Ph.D. thesis, The Royal Institute of Technology,
  Stockholm, Sweden, also available in the SICS Dissertation Series:
  SICS/D--16--SE (May 1994).

\bibitem{Maher88}
M.~J. Maher, Complete axiomatizations of the algebras of finite, rational and
  infinite trees, in: Proceedings, Third Annual Symposium on Logic in Computer
  Science, IEEE Computer Society Press, Edinburgh, Scotland, 1988, pp.
  348--357.

\bibitem{ArmstrongMSS98}
T.~Armstrong, K.~Marriott, P.~Schachte, H.~S{\o}ndergaard, Two classes of
  {Boolean} functions for dependency analysis, Science of Computer Programming
  31~(1) (1998) 3--45.

\bibitem{MarriottS89}
K.~Marriott, H.~S{\o}ndergaard., Notes for a tutorial on abstract
  interpretation of logic programs, North American Conference on Logic
  Programming, Cleveland, Ohio, USA (1989).

\bibitem{CortesiFW91}
A.~Cortesi, G.~Fil\'e, W.~Winsborough, \emph{Prop} revisited: Propositional
  formula as abstract domain for groundness analysis, in: Proceedings, Sixth
  Annual IEEE Symposium on Logic in Computer Science, IEEE Computer Society
  Press, Amsterdam, The Netherlands, 1991, pp. 322--327.

\bibitem{MarriottS93}
K.~Marriott, H.~S{\o}ndergaard, Precise and efficient groundness analysis for
  logic programs, ACM Letters on Programming Languages and Systems 2~(1--4)
  (1993) 181--196.

\bibitem{Schroeder1877}
E.~Schr{\"o}der, Der Operationskreis des Logikkalkuls, B. G. Teubner, Leibzig,
  1877.

\bibitem{CousotC77}
P.~Cousot, R.~Cousot, Abstract interpretation: {A} unified lattice model for
  static analysis of programs by construction or approximation of fixpoints,
  in: Proceedings of the Fourth Annual ACM Symposium on Principles of
  Programming Languages, ACM Press, New York, 1977, pp. 238--252.

\bibitem{CousotC92fr}
P.~Cousot, R.~Cousot, Abstract interpretation frameworks, Journal of Logic and
  Computation 2~(4) (1992) 511--547.

\bibitem{BagnaraHZ00}
R.~Bagnara, P.~M. Hill, E.~Zaffanella, Efficient structural information
  analysis for real {CLP} languages, in: M.~Parigot, A.~Voronkov (Eds.),
  Proceedings of the 7th International Conference on Logic for Programming and
  Automated Reasoning (LPAR 2000), Vol. 1955 of Lecture Notes in Artificial
  Intelligence, Springer-Verlag, Berlin, R\'eunion Island, France, 2000, pp.
  189--206.

\bibitem{CrnogoracKS96}
L.~Crnogorac, A.~D. Kelly, H.~S{\o}ndergaard, A comparison of three occur-check
  analysers, in: R.~Cousot, D.~A. Schmidt (Eds.), Static Analysis: Proceedings
  of the 3rd International Symposium, Vol. 1145 of Lecture Notes in Computer
  Science, Springer-Verlag, Berlin, Aachen, Germany, 1996, pp. 159--173.

\bibitem{Sondergaard86}
H.~S{\o}ndergaard, An application of abstract interpretation of logic programs:
  Occur check reduction, in: B.~Robinet, R.~Wilhelm (Eds.), Proceedings of the
  1986 European Symposium on Programming, Vol. 213 of Lecture Notes in Computer
  Science, Springer-Verlag, Berlin, Saarbr{\"u}cken, Federal Republic of
  Germany, 1986, pp. 327--338.

\bibitem{BruynoogheCM94}
M.~Bruynooghe, M.~Codish, A.~Mulkers, Abstract unification for a composite
  domain deriving sharing and freeness properties of program variables, in:
  F.~S. {de Boer}, M.~Gabbrielli (Eds.), Verification and Analysis of Logic
  Languages, Proceedings of the W2 Post-Conference Workshop, International
  Conference on Logic Programming, Santa Margherita Ligure, Italy, 1994, pp.
  213--230.

\bibitem{HansW92}
W.~Hans, S.~Winkler, Aliasing and groundness analysis of logic programs through
  abstract interpretation and its safety, Tech. Rep. 92--27, Technical
  University of Aachen (RWTH Aachen) (1992).

\bibitem{HillBZ02TPLP}
P.~M. Hill, R.~Bagnara, E.~Zaffanella, Soundness, idempotence and commutativity
  of set-sharing, Theory and Practice of Logic Programming 2~(2) (2002)
  155--201.

\bibitem{JacobsL89}
D.~Jacobs, A.~Langen, Accurate and efficient approximation of variable aliasing
  in logic programs, in: E.~L. Lusk, R.~A. Overbeek (Eds.), Logic Programming:
  Proceedings of the North American Conference, MIT Press Series in Logic
  Programming, The MIT Press, Cleveland, Ohio, USA, 1989, pp. 154--165.

\bibitem{CortesiF99}
A.~Cortesi, G.~Fil\'e, Sharing is optimal, Journal of Logic Programming 38~(3)
  (1999) 371--386.

\bibitem{HillZB01TR}
P.~M. Hill, E.~Zaffanella, R.~Bagnara, A correct, precise and efficient
  integration of set-sharing, freeness and linearity for the analysis of finite
  and rational tree languages, Quaderno 273, Dipartimento di Matematica,
  Universit\`a di Parma, Italy, available at
  \url{http://www.cs.unipr.it/Publications/}. Also published as technical
  report No.~2001.22, School of Computing, University of Leeds, U.K. (2001).

\bibitem{Dart91}
P.~W. Dart, On derived dependencies and connected databases, Journal of Logic
  Programming 11~(1{\&}2) (1991) 163--188.

\bibitem{BagnaraS99}
R.~Bagnara, P.~Schachte, Factorizing equivalent variable pairs in {ROBDD}-based
  implementations of \textit{Pos}, in: A.~M. Haeberer (Ed.), Proceedings of the
  ``Seventh International Conference on Algebraic Methodology and Software
  Technology (AMAST'98)'', Vol. 1548 of Lecture Notes in Computer Science,
  Springer-Verlag, Berlin, Amazonia, Brazil, 1999, pp. 471--485.

\bibitem{Bryant92}
R.~E. Bryant, Symbolic boolean manipulation with ordered binary-decision
  diagrams, ACM Computing Surveys 24~(3) (1992) 293--318.

\bibitem{JacobsL92}
D.~Jacobs, A.~Langen, Static analysis of logic programs for independent {AND}
  parallelism, Journal of Logic Programming 13~(2{\&}3) (1992) 291--314.

\bibitem{CodishSS99}
M.~Codish, H.~S{\o}ndergaard, P.~J. Stuckey, Sharing and groundness
  dependencies in logic programs, ACM Transactions on Programming Languages and
  Systems 21~(5) (1999) 948--976.

\bibitem{CortesiFW98}
A.~Cortesi, G.~Fil\'e, W.~Winsborough, The quotient of an abstract
  interpretation for comparing static analyses, Theoretical Computer Science
  202~(1{\&}2) (1998) 163--192.

\bibitem{Bagnara97th}
R.~Bagnara, Data-flow analysis for constraint logic-based languages, Ph.D.
  thesis, Dipartimento di Informatica, Universit\`a di Pisa, Pisa, Italy,
  printed as Report TD-1/97 (Mar. 1997).

\bibitem{Ramakrishnan88}
R.~Ramakrishnan, {Magic Templates}: A spellbinding approach to logic programs,
  in: R.~A. Kowalski, K.~A. Bowen (Eds.), Logic Programming: Proceedings of the
  Fifth International Conference and Symposium on Logic Programming, MIT Press
  Series in Logic Programming, The MIT Press, Seattle, USA, 1988, pp. 140--159.

\bibitem{Bourdoncle93}
F.~Bourdoncle, Efficient chaotic iteration strategies with widenings, in:
  D.~Bj{\o}rner, M.~Broy, I.~V. Pottosin (Eds.), Proceedings of the
  International Conference on ``Formal Methods in Programming and Their
  Applications'', Vol. 735 of Lecture Notes in Computer Science,
  Springer-Verlag, Berlin, Academgorodok, Novosibirsk, Russia, 1993, pp.
  128--141.

\bibitem{Bourdoncle93th}
F.~Bourdoncle, S\'emantiques des langages imp\'eratifs d'ordre sup\'erieur et
  interpr\'etation abstraite, PRL Research Report~22, DEC Paris Research
  Laboratory (1993).

\bibitem{Campbell84}
J.~A. Campbell (Ed.), Implementations of {Prolog}, Ellis Horwood/Halsted
  Press/Wiley, 1984.

\bibitem{BagnaraHZ02TCS}
R.~Bagnara, P.~M. Hill, E.~Zaffanella, Set-sharing is redundant for
  pair-sharing, Theoretical Computer Science 277~(1-2) (2002) 3--46.

\bibitem{ZaffanellaHB02TPLP}
E.~Zaffanella, P.~M. Hill, R.~Bagnara, Decomposing non-redundant sharing by
  complementation, Theory and Practice of Logic Programming 2~(2) (2002)
  233--261.

\bibitem{CodishDY91}
M.~Codish, D.~Dams, E.~Yardeni, Derivation and safety of an abstract
  unification algorithm for groundness and aliasing analysis, in: K.~Furukawa
  (Ed.), Logic Programming: Proceedings of the Eighth International Conference
  on Logic Programming, MIT Press Series in Logic Programming, The MIT Press,
  Paris, France, 1991, pp. 79--93.

\bibitem{Scozzari00}
F.~Scozzari, Abstract domains for sharing analysis by optimal semantics, in:
  J.~Palsberg (Ed.), Static Analysis: 7th International Symposium, SAS 2000,
  Vol. 1824 of Lecture Notes in Computer Science, Springer-Verlag, Berlin,
  Santa Barbara, CA, USA, 2000, pp. 397--412.

\bibitem{Clark78}
K.~L. Clark, Negation as failure, in: H.~Gallaire, J.~Minker (Eds.), Logic and
  Databases, Plenum Press, Toulouse, France, 1978, pp. 293--322.

\bibitem{HillBZ98b}
P.~M. Hill, R.~Bagnara, E.~Zaffanella, The correctness of set-sharing, in:
  G.~Levi (Ed.), Static Analysis: Proceedings of the 5th International
  Symposium, Vol. 1503 of Lecture Notes in Computer Science, Springer-Verlag,
  Berlin, Pisa, Italy, 1998, pp. 99--114.

\end{thebibliography}
\end{document}